\documentclass[fleqn,usenatbib]{mnras}

\usepackage{newtxtext,newtxmath}

\usepackage[T1]{fontenc}

\DeclareRobustCommand{\VAN}[3]{#2}
\let\VANthebibliography\thebibliography
\def\thebibliography{\DeclareRobustCommand{\VAN}[3]{##3}\VANthebibliography}

\usepackage{graphicx}	
\usepackage{amsmath}	
\usepackage{multirow}
\usepackage{orcidlink}  
\usepackage{stfloats}   
\usepackage{hyperref}   
\usepackage{pdflscape}
\newcommand{\citedysmalpy}{\citep{Davies2004_dysmalpy, Davies2004b_dysmalpy, Cresci2009_dysmalpy, Davies2011, Wuyts2016_KMOS3d, Lang2017_stacking_rc, Genzel2017, Ubler2018_ha_co_rc, Price2021_rc41, Lee2025_dysmalpy}}

\newcommand{\kms}{\, \rm km\,s^{-1}}
\newcommand{\Msun}{\, \rm M_\odot}

\title[\texttt{RotCurves}: Rotation Curve fitting tool]{\texttt{RotCurves}: A \texttt{PYTHON} package for efficient modelling and fitting of galactic rotation curves at high-z}

\author[Nestor Shachar et al.]{
A. Nestor Shachar\,\orcidlink{0000-0003-1785-1357}$^{1}$
A. Sternberg\,\orcidlink{0000-0001-5065-9530}$^{2,1,3}$,
S. H. Price\,\orcidlink{0000-0002-0108-4176}$^{4}$, 
N. M. F{\"o}rster Schreiber\,\orcidlink{0000-0003-4264-3381}$^{3}$,
R. Genzel\,\orcidlink{0000-0002-2767-9653}$^{3}$,\newauthor
L. J. Tacconi\,\orcidlink{0000-0002-1485-9401}$^{3}$,
H. {\"U}bler\, \orcidlink{0000-0003-4891-0794}$^{3}$
C. Barfety\,\orcidlink{0000-0002-1952-3966}$^{3}$,
A. Burkert\,\orcidlink{0000-0001-6879-9822}$^{5}$,
J. Chen\,\orcidlink{0000-0003-3921-3313}$^{3}$,
R. Davies\,\orcidlink{0000-0003-4949-7217}$^{3}$,
F. Eisenhauer$^{3}$,\newauthor
J. M. Espejo Salcedo\,\orcidlink{0000-0001-6703-4676}$^{3}$,
R. Herrera-Camus\,\orcidlink{0000-0002-2775-0595}$^{6}$,
J. B. Jolly\,\orcidlink{0000-0002-3405-5646}$^{3}$
L. L. Lee\,\orcidlink{0000-0001-7457-4371}$^{3}$,
T. Naab\,\orcidlink{0000-0002-7314-2558}$^{7}$,
S. Pastras\,\orcidlink{0009-0009-0472-6080}$^{3,7}$,\newauthor
C. Pulsoni\,\orcidlink{0000-0002-1428-1558}$^{3}$,
T. T. Shimizu\,\orcidlink{0000-0002-2125-4670}$^{3}$,
G. Tozzi\,\orcidlink{0000-0003-4226-7777}$^{3}$,
\\
$^{1}$School of Physics and Astronomy, Tel Aviv University, Tel Aviv 6997801, Israel\\
$^{2}$Center for Computational Astrophysics, Flatiron Institute. 162 5th Avenue, New York, NY, 10010, USA\\
$^{3}$Max-Planck-Institut für Extraterrestrische Physik (MPE), Giessenbachstr.1, 85748 Garching, Germany\\
$^{4}$Space Telescope Science Institute, 3700 San Martin Drive, Baltimore, Maryland 21218, USA\\
$^{5}$Universitäts-Sternwarte Ludwig-Maximilians-Universität (USM), Scheinerstr. 1, München, D-81679, Germany\\
$^{6}$Departamento de Astronomía, Universidad de Concepción, Barrio Universitario, Concepción, Chile\\
$^{7}$Max Planck Institute for Astrophysik (MPA), Karl-Schwarzschild-Str 1, D-85748 Garching, Germany
}

\date{Accepted 2026 January 11. Received 2026 January 11; in original form 2025 November 13}

\pubyear{\the\year{}}

\begin{document}
\label{firstpage}
\pagerange{\pageref{firstpage}--\pageref{lastpage}}
\maketitle

\begin{abstract}
Rotation curves are a fundamental tool in the study of galaxies across cosmic time, and with the advent of large integral field unit (IFU) kinematic surveys there is an increasing need for efficient and flexible modelling tools. We present \texttt{RotCurves}, a parametric forward-modeling tool designed for rotation curve analysis at high-z, correcting for ``beam smearing" by projecting and convolving the beam PSF in the plane of the galaxy. We benchmark \texttt{RotCurves} against the established parametric code \texttt{dysmalpy} using synthetic observations. The typical runtime with \texttt{RotCurves} is a few $\sim10$ms, a factor $\approx 250$ faster than \texttt{dysmalpy} for a single realization. For well-resolved systems (PSF FWHM $< R_{\rm eff}$), the mock observed rotation and dispersion curves agree to within 5\% up to $3R_{\rm eff}$, 
whereas in marginally resolved systems (PSF FWHM $\gtrsim 1.5 R_{\rm eff}$) discrepancies increase to up to 15\%. Using a built-in MCMC fitting procedure, \texttt{RotCurves} recovers well the intrinsic model parameters across a wide range of galaxy properties and accounting for realistic noise patterns. Systematic biases emerge for the effective radius and for low disk masses ($M_{\rm disk} \lesssim 3 \times  10^{9} \Msun$). We show excellent parameter recovery at high signal-to-noise ratios (S/N$\gtrsim 25$), with increasing deviations in parameter recovery at lower S/N. \texttt{RotCurves} is best suited for inclinations of $10^\circ < i < 80^\circ$. \texttt{RotCurves} is built as an exploratory tool for rapid testing of mass model assumptions, parameter studies and for efficiently processing large samples of observational data from large IFU surveys. The code is publicly available on \href{https://github.com/nestoramit/RotCurves}{github}.
\end{abstract}


\section{Introduction} \label{sec:intro}
Galaxy rotation curves have long been a powerful tool to study the structure and evolution of galaxies. As direct probes of the gravitational potentials, they are key for understanding galactic structure, evolution, dynamics, and the nature of the dark matter halos. Since the pioneering work of \cite{Burbidge1959_rc, Rubin1970_rc, Bosma1981_rc, vanDerKruit1978_rc} and many others, the study of rotation curves has evolved from single-object spectroscopy in the Local Universe to large integral field unit (IFU) surveys and millimeter interferometry of distance objects across cosmic time \citep{Courteau1997_rc, Courteau2015_rc, deBlok2008_things, ForsterSchreiber2009_SINS, Wisnioski2015_kmos3d, Stott2016_kross, Harrison2017_kross, Genzel2017, Genzel2020_rc41, Tiley2019_kges, Bouche2022_MUSE, NestorShachar2023_rc100, Puglisi2023_KURVS, Lee2025_cristal}. Studies at higher redshifts have shown that most ``main sequence" star-forming galaxies (SFGs) at redshifts $z\sim 1-3$ are rotating turbulent disks \citep{ForsterSchreiber2009_SINS, ForsterSchreiber2020_review, Wuyts2016_KMOS3d, Genzel2017, Genzel2020_rc41, Lang2017_stacking_rc, Wisnioski2019_kmos3d, Bouche2022_MUSE}. Recent kinematic studies from ground-based mm interferometric and JWST IFU data indicate that the prevalence of rotating, turbulent gas disks SFGs extends out to $z \sim 6$ \citep[e.g.][and references within]{Lee2025_cristal}, with tentative evidence for rotation as early as $z\sim 14$ \citep{Scholtz2025_z14rc}. Order rotation points out to a more regulated scenario of galaxy evolution over frequent violent major-mergers. More recently, \emph{JWST}/NIRCam have established that star forming galaxies are predominantly disks, with some evidence for the existence of spiral structure, up to redshifts of $z \sim 6$ \citep{Ferreira2023_jwst_disks, Kartaltepe2023_jwst_disks, Huertas-Company2024_jwst_disks, Kuhn2024_jwst_disks, Tohill2024_jwst_disks, Smethurst2025_jwst_disks, EspejoSalcedo2025_morphology, Geron2025_jwst_features, Chugunov2025_jwst_features, vanDerWel2025_disks}.
Analysis of rotation curves is essential for the study of galaxies across cosmic time, even at the earliest stages of galaxy formation. 

In this paper we present our rotation curve modeling tool \texttt{RotCurves}.

\subsection{Spectroscopic surveys}\label{sec:surveys}
The advent of large-scale ground-based surveys, especially with the introduction of IFU and adaptive-optics modules, has revolutionized our ability to study galaxy kinematics with unprecedented detail and statistical power. In the Local Universe, The HI Nearby Galaxy Survey \citep[THINGS:][]{Walter2008_things, deBlok2008_things}, followed by LITTLE THINGS targeting dwarf galaxies \citep{Hunter2012AJ_little_things}, mapped the neutral hydrogen through 21cm emission with extremely high spatial ($\sim 100$pc) and spectral ($\sim 5\kms$) resolutions. In the optical, the MaNGA survey targeted an unprecedented number of $\sim$10,000 galaxies \citep{Bundy2015_manga, Abdurrouf2022_manga}, while other IFU-surveys provided a higher spatial and spectral resolutions for large samples, such as SAMI \citep{Croom2021_sami_survey} and CALIFA \citep{Sanchez2016_califa, Sanchez2023_califa}. Subsequent studies have focused on resolving the kinematics of nearby galaxies at high spatial resolution, for example using samples from MaNGA \citep{Yoon2021_manga_rc}, the DISKMASS survey \citep{Martinsson2013_diskmass}, the Extended Disk Galaxy Explore Science (EDGES) survey \citep{Richards2016_edges}, the SPARC sample \citep[175 galaxies][]{Lelli2016_sparc, Lelli2017_sparc}, and others \citep{Cappellari2013_atlas3d, Gomez-lopez2019_rc_herschel, Lang2020_alma_phangs, DiTeodoro2021, DiTeodoro2023}. 

At higher redshifts, the small apparent sizes of galaxies together with surface brightness dimming, which scales as $\left(1+z\right)^4$, are a major challenge for resolved kinematic studies of galaxies, both spatially and spectrally. Ground-based observations from large telescopes were of great importance both in the near-infrared (NIR), tracing rest-optical emission lines, and in mm-radio interferometry, with little atmospheric absorption. The SINS/zC-SINF survey of H$\alpha$ kinematics with SINFONI at the Very Large Telescope targeted $\sim 80$ main sequence massive SFGs at $z \sim 1.5 - 2.5$ in seeing-limited mode, including 35 for which very deep, high-resolution adaptive optics-assisted observations were also obtained \citep[PSF FWHM $\sim 0.15"$][]{ForsterSchreiber2009_SINS, ForsterSchreiber2018_sinf, mancini2011_zcosmos}. Other studies have increased the number of galaxies using both VLT/SINFONI \citep{Epinat2009_sinfoni, Gnerucci2011_sinfoni} and KECK/OSIRIS \citep{Law2009_keck}, in both seeing-limited and adaptive-optics assisted modes. In other cases, the magnification caused by gravitational lensing can resolve galaxies down to the $\sim 100$pc scales and provide highly detailed velocity maps \citep{Livermore2015_lensed, Leethochawalit2016_lensed, Liu2023_lensed}.

Multi-target IFU instruments have been invaluable in increasing statistics, targeting multiple galaxies at once. In the near-IR, the VLT/KMOS allowed large surveys to collectively target thousands of galaxies at $z \sim 1-3$, with surveys such as KMOS$^{\rm 3D}$ \citep[$\sim$700 galaxies,][]{Wisnioski2019_kmos3d}, KROSS \citep[$\sim$800 galaxies,][]{Stott2016_kross, Harrison2017_kross} and KGES \citep[$\sim$300 galaxies,][]{Tiley2019_kges, Tiley2021_kges}.
Others have used KMOS to obtain deep observation on smaller samples, \citep[e.g., KURVS, targeting 22 low mass galaxies at $z \sim 1$, ][]{Puglisi2023_KURVS}. In the optical, VLT/MUSE have been used to resolve kinematics for $z \lesssim 1$ galaxies, such as the MUSE Hubble Ultra Deep Field \citep{Contini2016_muse, Guerou2017_muse} or the MXDF \citep[][]{Bouche2022_MUSE}.

In mm interferometry, ALMA and NOEMA have played a key role by tracing CO molecular gas kinematics up to at $z \sim 3$ \citep[][Jolly et al., in prep.]{Genzel2013_phibss, Tacconi2013_phibss, Tacconi2018_phibss, Chen2017_alma, Ubler2018_ha_co_rc, Tadaki2018_alma_co, Molina2019_co, Rizzo2021_alma, Rizzo2023_dispersion, Pastras2025_noema3d}, most notably with the ALMA-ALPAK survey \citep{Rizzo2023_dispersion} and NOEMA$^{\rm 3D}$ (\citealt{Pastras2025_noema3d}; Jolly et al., in prep.). More recently, the strong [CII] line have been used in kinematic surveys at higher redshifts $z \sim 3-6$, such as the ALPINE survey \citep{Jones2021_alpine_alma}, the CRISTAL survey \citep{HerreraCamus2025_cristal, Lee2025_cristal}, REBELS \citep{Rowland2024_rebels} and others \citep{Smit2018_alma_cii, Rizzo2021_alma, Akins2025_jwst}

The space telescope \emph{JWST} is recently being used to probe the high redshift range in H$\alpha$ (and [OIII]) kinematics, with surveys such as JADES \citep{deGraaff2024_jwst_jades}, GA-NIFS \citep{Ubler2024_GN20, Bertola2025_jwst_ganifs, Marconcini2024_jwst_ganifs} and others \citep{Barisic2025jwst_msa3d, Danhaive2025_dispersion, Fujimoto2025_cosmicgrapes}.


\subsection{Key Past results}\label{sec:intro_key}
Kinematic studies have been key in understanding how the dynamics of galaxies affect their evolution. A fundamental empirical scaling relation was found between the stellar (or baryonic) mass of a disk galaxy and its rotation velocity, $M \propto V_{\rm rot}^4$ (the ``Tully-Fisher" relation) \citep{Tully1977, McGaugh2000_TFR, Lelli2016_TFR, Lelli2019_TFR, Ubler2017_TFR}, and has been shown to hold up to $z \sim 3$, with application both in calibrating cosmological simulations \citep{Vogelsberger2014_illustris, Schaye2015_eagle, Sommerville2015} and as a cosmological distance measure \citep{Freedman2001_TFR_cosmology, Schombert2020_TFH0, Boubel2024_TFH0}. 

The velocity dispersion $\sigma_0$ of the gas in the interstellar medium (ISM) increases with redshift, for both the ionized and molecular gas phases \citep{Ubler2019_dispersion, ForsterSchreiber2020_review, deGraff2024_dispersion, Lee2025_cristal, Danhaive2025_dispersion, Wisnioski2025_dispersion}. Consequently, at $z \sim 2-3$ galaxies are typically less rotationally supported, as defined by the ratio of rotation velocity to their velocity dispersion, $V / \sigma_0$, which can become as low as $\approx 2-4$  \citep{ForsterSchreiber2020_review}. Such conditions lead to gravitational fragmentation of the disk via ``Toomre" instability, forming $\sim$kpc sized star forming clumps, which have been frequently observed (\citealt{Toomre1964, Burkert2010, Genzel2011_clumps}; F{\"o}rster Schreiber et al., in prep.). However, other studies, focused mostly on the cold gas phase, find velocity dispersions are much smaller and galaxies retain high rotation support of $V / \sigma_0 \sim 10$ up to $z \sim 4-5$ \citep{Rizzo2023_dispersion, Rizzo2024_dispersion, RomanOliveira2024_dispersion}. This tension is yet to be resolved; yet it is commonly agreed that the ionized gas is more turbulent than the molecular phase \citep[e.g.,][]{Girard2021_dispersion_phases_difference}.

The characteristic flat outer disk rotation curves, observed in many local spiral galaxies, is one of the key evidences for the existence of dark matter in galaxies \citep{Rubin1970_rc, Gunn1980_dm, vanAlbada1985_rcdm, deBlok2008_things}. Dark matter amounts to most of the total mass for massive spirals in the local universe, $\sim 40$-$80\%$ of the mass budget within the half-mass radius, and completely dominates the mass budget in dwarf galaxies \citep{Oh2011_little_things, Oh2015_little_things, Lelli2016_sparc}. Yet, this ubiquitous ''flatnss" indicates that the dark matter distribution has to be finely tuned to the baryonic content of the galaxy, a problem that became known as the ``disk-halo conspiracy" \citep[e.g.][]{vanAlbada1986_disks}. 

The distribution of dark matter is commonly described using empirical and theoretical density profiles. The most widely adopted is the Navarro-Frenk-White (NFW) profile, a universal density profile arising from collisionless N-body simulations with an inner cusp scaling as $\rho \propto r^{-1}$ \citep{Navarro1996, Navarro1997_nfw, Navarro2004}. However, other forms of dark matter profiles have been suggested, with varying inner power law slopes of the form $\rho \propto r^{-\alpha}$. These are typically split into cusps ($\alpha \gtrsim 1$; such as the NFW profile) or cores ($\alpha \approx 0$). A commonly used cored density profile is the ``Burkert" profile \citep{Burkert2015_cores}, suggest to fit local dwarf galaxies. Other profiles are parametrized in such a way to have either a cusp or a core, depending on the choice of parameters, such as the generalized-NFW \citep{Zhao1996}, Einasto \citep{Einasto1965}, or the Dekel-Zhao profile \citep{Freundlich2020_dekelzhao_profile}.

Including the effects of the baryons on the DM leads to complicating dynamics that can change the distribution of DM in the halo from the results of the idealized N-body simulations. On the one hand, the accumulation of baryons in the center of the halo causes the halo to further contract (``adiabatic contraction"), resulting in an inner DM cusp that is steeper than the NFW, $\rho \propto r^{-\alpha}$ with $\alpha > 1$ \citep[e.g.][]{Blumenthal1986}. On the other hand, feedback processes and mergers inject energy that can lead to an expansion of the halo and form an inner core, i.e. $\alpha \approx 0$, a process most efficient in massive dwarf galaxies \citep{Governato2012, Read2019_dmcores, Lazar2020_cored_einasto_FIRE}. Such DM cores have been observed in local dwarf galaxies \citep{deBlok2010_corecusp, Oh2015_little_things}, and some kinematic studies suggest they can also occur in massive galaxies at $z \sim 1-2$ \citep{Genzel2020_rc41, Bouche2022_MUSE, NestorShachar2023_rc100}.

\subsection{Rotation curve modelling at high-z}\label{sec:intro_rctools_highz}
At higher redshifts, both surface brightness dimming and limited angular resolutions have a significant impact on the kinematic maps. While the first can be overcome by long exposures \citep[e.g.][]{Genzel2017}, the latter is harder to overcome as it is limited by instrumental capabilities and atmospheric turbulence. This leads to ``beam smearing", an artificial smoothing of velocity gradients and local kinematic features as light is convolved from large physical areas over the source. Beam smearing creates significant biases in the derived rotation curves if not properly accounted for in the modeling process, and has long been a challenge for analyzing rotation curves \citep[e.g.,][]{EspejoSalcedo2025_ang}.

To address these challenges, the community has developed sophisticated modelling tools that can be applied directly in observational space, thereby avoiding the systematic biases introduced by traditional 2D kinematic map extraction. Commonly used 3D tools are $^{\rm 3D}$\texttt{Barolo}, \texttt{GalPak}$^{\rm 3D}$ and \texttt{dysmalpy}. Other tools, such as \texttt{2DBAT}, employ a 2D modelling approach but are intended mainly for highly-resolved systems.

\texttt{DysmalPy}\footnote{https://www.mpe.mpg.de/resources/ir/dysmalpy} (DYnamical Simulation and Modelling ALgorithm in PYthon), inspired by the earlier \texttt{DISDYN} program, is a forward modeling tool specifically designed for analyzing galaxy kinematics from modern IFU data cubes \citedysmalpy. It uses analytical mass components to construct an axisymmetric model of a rotating galaxy, generating a full 4D hypercube containing a spectrum at each pixel. The galaxy is then rotated to match the sky projection, and is convolved over a finite beam to recover the line-of-sight spectra, fully incorporating all relevant observational and projection effects. It uses a Bayesian inference method (MCMC) to fit observational data.

\texttt{GalPaK}$^{\rm 3D,\ }$\footnote{https://galpak3d.univ-lyon1.fr/} (Galaxy Parameters and Kinematics) is another parametric tool that is using Bayesian inference methods to fit observed data cubes \citep{Bouche2015_galpak3d, Bouche2021_galpak, Bacon2015_galpak, Mason2017galpak, Szakacs2021galpak, Ciocan2022_galpak, Puglisi2023_KURVS, Birkin2024_KAOSS}. However, instead of assuming the shape of the mass distribution, it uses analytics shapes for the rotation (e.g., $v_\mathrm{rot} \propto \tan^{-1}(r)$) and the emission profile (e.g., Sérsic profile), which are used to recover the kinematic and morphological properties. It was developed to work with 3D data cubes, recovering both geometric and kinematic properties.

$^{\rm3D}$\texttt{Barolo}\footnote{https://editeodoro.github.io/Bbarolo/} (3D-Based Analysis of Rotating Object via Line Observations) is a non-parametric tool developed to recover rotation velocities \cite{DiTeodoro2015_3dbarolo, DiTeodoro2016_barolo, Iorio2017_galpak, Rizzo2021_alma, Rizzo2023_dispersion, Deg2022_barolo, Westmeier2022_barolo, RomanOliveira2024_dispersion}. It uses tilted-ring models to recover the intrinsic rotation pattern directly from a data cube, while taking into account beam smearing and projection effects. It does not rely on assumptions about the mass distribution and allows flexibility in the shape of the inferred kinematic profiles.

While powerful, these 3D algorithms can become computationally expensive. It requires the construction of a full 3D cube  (in the case of $^{\rm 3D}$\texttt{Barolo} and \texttt{Galpak}$^{\rm 3D}$) or a 4D hypercube (in \texttt{dysmalpy}), which is then collapsed and convolved over the sky orientation. This amounts to a total of $n_x \times n_y \times n_z (\times n_{\lambda})$ points, at each implementation of the model. In a Bayesian inference framework, this has to be calculated repeatedly at each step. While this is useful to handle the degrees of freedom when simultaneously fitting multiple parameters, it can hamper the ability to test multiple forms of assumptions (i.e., priors, profile shapes, systematics, parameters co-dependency), in a time where galaxy samples are increasing in size.

Other kinematic tools enable 2D modeling, such as \texttt{2DBAT}\footnote{https://github.com/seheonoh/2dbat} \cite{Oh2018_2DBAT, Oh2020_2dbat_soft} and \texttt{diskfit}\footnote{https://kspekkens.github.io/DiskFit/} \cite{deNaray2012_diskfit_soft, Sellwood2015_diskfit}, yet they are designed for highly-resolved systems in the Local Universe. They focus on accurate recovery of the geometrical properties of the galaxy, including the existence of warps, and detailed kinematic features arising from bars, spiral arms, or non-circular motions \citep{Oh2022_2dbat, Oh2025_2dbat_b, Elagali2018_2dbat, Kim2020_2dbat, Zhang2020_2dbat, diFolco2020_diskfit, Belete2021_diskfit, Wong2021_2dbat, Bisaria2022_diskfit, Oh2025_2dbat_a, Liu2025_diskfit}. Both tools are non-parametric, with \texttt{2DBAT} using a tilted-ring approach and \texttt{diskfit} a global profile with no pre-assumed form. As these features require highly resolved observations, with high S/N, these tools have no beam smearing correction and are less relevant in the high-z regime, where such corrections become critical.

In this paper, we present \texttt{RotCurves}\footnote{https://github.com/nestoramit/RotCurves}: a new modeling tool designed to fit galactic rotation curves using a parametric, forward modelling approach, allowing flexibility and fast computational time. This is accomplished by implementing a 2D approximation of the galactic midplane, which is convolved given the instrumental PSF and sky orientation. Instead of rotating the galaxy to align with the beam PSF (which would have to be done in 3D), we use the projection of the beam as it intersects the midplane of the galaxy for a given inclination angle. Our code is intended to recover the physical properties of a galaxy, with the effects of beam-smearing, but assuming its inclination and PA are already known. RotCurves is not intended to directly fit these geometric properties, but relies on having them estimated beforehand. We demonstrate the accuracy of our tool with synthetic datasets created with \texttt{dysmalpy}, discussing the effects on the various parameters. 

The paper is structured as follows. In \S~\ref{sec:basic_equations} we describe the process of creating a mock rotation curve with \texttt{RotCurves}: the definition for the mass distribution and velocity profiles (\ref{sec:basic_equations - mass_distributions}, pressure support corrections (\ref{sec:pressure_support_equations}), the beam smearing process using our beam projection method (\ref{sec:sky_projection}), and the setup for the built-in MCMC fitting procedure (\ref{sec:mcmc_fitting}). In \S~\ref{sec:benchmark} we compare our results versus model generated with \texttt{dysmalpy}, comparing the beam smearing method for a fiducial model of a $z=2$ galaxy (\ref{sec:benchmark - beam smearing}), the performance for various model parameters (\ref{sec:fiducial_parameter_study}), the MCMC fitter (\ref{sec:benchmark - mcmc fitter}), and the effects of simulated noise on the recovery of the physical parameters (\ref{sec:mock_observation}) In \S~\ref{sec:discussion} we summarize and discuss the results.

Throughout the paper we adopt a flat $\Lambda$CDM cosmology, with $\Omega_{\mathrm{m}}=0.3$, $\Omega_{\mathrm{\Lambda}}=0.7$ and $H_0 = 70\, \mathrm{km\, s^{-1}\, Mpc^{-1}}$.

\section{Model outline} \label{sec:basic_equations}
We construct an idealized model of a galaxy as an axisymmetric system in centrifugal equilibrium 
with all gaseous and stellar mass components in circular orbits. The rotation velocity is a function of the radius $r$, given by $V_{\rm circ}^2 = -r \frac{d \phi (r)}{dr}$, where $\phi$ is the gravitational potential. As the gravitational potential can be expressed as a linear combination of different mass distributions, this expression can be treated as a sum of the squared circular velocity of each of these components, $V_{circ}^2 = \sum_i V_{i,circ}^2$.  
We calculate the intrinsic velocity field within the midplane of the galaxy, so we only need two coordinates, $\left( x_{\rm gal}, y_{\rm gal} \right)$, with the corresponding radius $r_{\rm gal} = \sqrt{x_{\rm gal}^2 + y_{\rm gal}^2}$ and angle with respect to the $x$-axis $\theta_{\rm gal} = \arctan \left(y_{\rm gal}/x_{\rm gal} \right)$. Given a composition of mass distributions, the circular velocity at each radius $V_{\rm circ}^2 \left( r_{\rm gal} \right)$ is calculated analytically. We sample the grid uniformly in the $xy$-plane, with a squared pixel size with sides $\Delta x_{\rm gal} = \Delta y_{\rm gal}$. If only the intrinsic velocity field is of interest, the pixel resolution can be set arbitrarily 
but when fitting to observations it is best to match the model resolution to the instrumental pixel size. 

\subsection{Mass Distributions} \label{sec:basic_equations - mass_distributions}
We calculate the intrinsic circular velocity in the midplane of the galaxy in the presence of the various mass components included. Baryons are considered to be purely axisymmetric with no warping, split into two distinct components of a bulge and a disk, and the dark matter halo is considered to be spherically symmetric. 
In the following subsections we describe the formulation used for the different density profiles, and also their mass and velocity profiles. Each mass distribution has a characteristic scale-length $r_{\rm s}$ and total mass $M$. All functions are given by the normalized radius:
\begin{equation}\label{eq:normalized_radius}
    x \equiv  r / r_{\rm s}
\end{equation}
In general, we describe the enclosed mass within radius $x$ using a scale mass $M_{\rm s}$ and a dimensionless function $f_{\rm M} (x)$ such that:
\begin{equation}\label{eq:enclosed_mass_general}
    M(<x)= M_{\rm s} \times  f_{\rm M}(x) \ \ \ .
\end{equation}
The formula for the dimensionless function $f_{\rm M}(x)$ is discussed later, and differs for cylindrical and spherical cases. 

The effective radius, or half mass radius, is the radius encompassing half the total mass of the profile, and is set by solving the implicit equation:
\begin{equation}
    f_{\rm M}(x_{\rm eff}) = \frac{1}{2} f_{\rm M}(x_{\rm max})
\end{equation}
where $x_{\rm max}$ is the edge of the distribution, and is typically $x_{\rm max} \rightarrow \infty$. The circular velocity profile:
\begin{equation}\label{eq:velocity_general}
\begin{split}    
    &V_{\rm circ} (x) = V_{\rm s} \times f_{\rm V}(x) \\
    &V_{\rm s} = \sqrt {\frac{G M_{\rm s}}{r_{\rm s}} } \ \ \  ,
\end{split}
\end{equation}
with the characteristic velocity $V_{\rm s}$ dimensionless function $f_{\rm V} (x)$, and $G$ being the gravitational constant.

\subsubsection{Baryonic components}\label{sec:basic_equations - baryons}
Axisymmetric mass distributions are best described by their local surface densities $\Sigma (r)$. We denote the central surface density as $\Sigma_0$ and define a dimensionless density function in terms of the normalized radius defined in Equation~\ref{eq:normalized_radius}:
\begin{equation}\label{eq:baryonic_surface_density}
    \Sigma (x) = \Sigma_0 \times f_\Sigma(x)
\end{equation}
where $f_{\Sigma}(0)=1$ and $f_{\Sigma} (x_{\rm max}) = 0$.
The scale mass $M_{\rm s}$ is given by
\begin{equation}\label{eq:baryonic_enclosed_mass}
    M_{\rm s} = 2 \pi \Sigma_0 r_{\rm s}^2 \ \ \ ,
\end{equation}
and the dimensionless mass function $f_{\rm M}(x)$
\begin{equation}
    f_{\rm M}(x) = \int_0^x t f_{\Sigma}(t)dt\ \ \ ,
\end{equation}
with $\Sigma_0$ normalized so that the total mass equals $M$
\begin{equation}\label{eq:baryonic_surface_density_normalization}
    \Sigma_0 = \frac{M}{2 \pi r_{\rm s}^2 f_M(x_{\rm max})}\ \ \ ,
\end{equation}
and $x_{\rm max} \rightarrow \infty$ by default. 

The circular velocity for razor-thin mass distributions have to be calculated explicitly by the integrations \citep[see][Eq. 2.158-2.160]{Binney_and_Tremaine2008}:
\begin{equation}\label{eq:vcirc_binney_tremaine}
\begin{split}    
    &f_{\rm V}^2 (x) = -x \int_0^{\infty} {J_1(kr) k S(k) dk} \\
    &S(k) = \int_0^{\infty} {J_0 (ku) f_{\Sigma} (u) udu} \ \ \ ,
\end{split}
\end{equation}
where $J_1$ and $J_2$ are the Bessel functions.

If the distribution is razor-thin, Equations~\ref{eq:baryonic_surface_density}-\ref{eq:baryonic_surface_density_normalization} are an exact description. However, in real galactic disks there is a finite thickness measured by the ratio, $q_0$, between the scale height and the scale length. In the Local Universe, disks of Milky Way like galaxies are close to razor-thin ($q_0 \approx 0.05$) while at higher-$z$ disks tend to be more oblate ($q_0 \approx 0.2)$ \citep{Lian2024_thickdisk, Tsukui2025_thickdisk}. In the more general case, the surface density is 
\begin{equation}\label{eq:baryonic_surface_density_z}
    \Sigma(x) = 2 \int_{0}^{\infty} \rho(x, z)dz 
\end{equation}
where $z$ is the vertical coordinate, and $\rho(x,z)$ is the local volume density. The circular velocity can be calculate by decomposing the local density and integrating over iso-density surfaces. We Follow the method provided by \cite{Noordermeer2008} to decompose Sérsic profiles, which are the most commonly type of profile for a disk or a bulge.

It is useful to define a radially dependent ``virial" coefficient, $\kappa (x)$, to quantify the local velocity deviations from Keplerian motions, i.e., the circular velocities for the masses but enclosed within spheres:
\begin{equation}\label{eq:virial_coefficient_kappa}
    \kappa (x) = \frac{f_{\rm V}^2(x)}{f_{\rm M}(x) / x} \ \ \ .
\end{equation}
In essence, $\kappa(x)$ encodes the effect the geometry of the mass distribution has on the potential compared to the simple spherical case, for which $\kappa (r) = 1$ \citep[see also][for a similar definition]{Price2022_sersic}. At large radii, we expect this parameter to approach unity $\kappa (x \rightarrow \infty) = 1$.

The light distribution of the baryon component follows the surface density with a constant mass-to-light ratio
\begin{equation} \label{eq:flux_distribution}
    I(r) = \Upsilon\, \times\, \Sigma(r)\ \ \ ,
\end{equation}
in arbitrary units, as only the relative flux is important for the process of the beam smearing. By default we set $\Upsilon = 1$ for baryonic components. If no light is emitted with respect to the tracer, $\Upsilon = 0$. For example, if considering a combination of disk and bulge in H$\alpha$ emission, it is reasonable to assume the disk overshadows the bulge $\Upsilon_\mathrm{disk} \gg \Upsilon_\mathrm{bulge}$. It is also possible to construct various combinations in which the M/L does not trace the mass completely, such as an emission ring above a faint exponential disk \cite[see][]{NestorShachar2025}.

While the physical distribution of various tracers might differ from that of the stellar disk, these difference are mitigated by the beam smearing in low-resolution studies. At high-z the common tracers are the ionized and molecular phases of the ISM, which although linked to sites of star-formation are typically widespread enough throughout the disk so that, when smeared, can become quite smooth \citep[e.g., ][]{Ubler2018_ha_co_rc, Jones2025}. The assumption of a constant M/L has been used successfully in multiple kinematic studies \citep{Genzel2017, Genzel2020_rc41, Price2021_rc41, NestorShachar2023_rc100, HerreraCamus2022_dispersion_highz, Lee2025_cristal}.

\begin{table*}
    \centering
    \begin{tabular}{c|cccccc}
    & & & intrinsic & & & \\
    Profile& Parameters& $r_{\rm s}$& axis ratio & $f_{\Sigma}(x)$& $R_{\rm eff} / r_{\rm s}$& $M / M_{\rm s}$ \\ \hline \hline
    
    Exponential disk & $M, R_{\rm d}$ & $R_{\rm d}$& $0$  &$\exp{\left\{-x \right\} }$
& $1.68$ & $1$ \\
    
    Thick S{\'e}rsic disk & $M, R_{\rm d}, n$ & $R_{\rm d}$& $q_0$  &$\exp{\left\{-x^{1/n}\right\} }$
& $b_n^n$ & $\Gamma (2n)$ \\
    
    Gaussian Ring$^{a}$& $M, R_{\rm p}, \sigma_{r}$& $R_{\rm p}$ & $0$ &$\exp {\left\{- A (x-1)^2 \right\}}$& $\approx 1 + \frac{1}{4} h^{-5/4}$ & $\frac{1}{2A} \left( e^{-A} + \sqrt{\pi A} + \sqrt{A} \gamma(\frac{1}{2}, A) \right)$\\
    & & &  && & $\approx \left( h-0.07 \right)^{-1}$ 
    \end{tabular}
    \caption{Dimensionless properties of the surface density profiles included, as a function of the normalized radius $x = r/r_{\rm s}$. $^{a}$For the gaussian Ring profile we define: $A = R_{\rm p}^2 / 2 \sigma_{\rm r}^2$ and $h = \sqrt{A} / 2\sqrt{\ln 2}$.}
    \label{tab:baryons_dimless_properties}
\end{table*}

\begin{table*}
    \centering
    \begin{tabular}{c|cc}
    Profile& $f_{\rm M} (x)$  &$f_{\rm V}^2 (x)$\\ \hline \hline
    
    Exponential disk & $1 - e^{-x} \left( 1+x \right)$  &$\frac{1}{2} x^2 \left[ I_0(\frac{x}{2}) K_0(\frac{x}{2}) - I_1(\frac{x}{2}) K_1(\frac{x}{2})  \right]$\\ \hline
    
    Thick S{\'e}rsic disk & $\gamma {(2n, x^{1/n})}$   &$\frac{2 b_n^{n+1}}{\pi n^2} \times \int_0^x { \frac{m^2}{\sqrt{x^2-m^2 (1-q_0^2)^2}} \left[ \int_m^{\infty} { \frac{e^{-b_n t^{1/n}} t^{1/n - 1}}{t^2 - m^2} dt} \right] dm}$\\ \hline 
    
    Gaussian Ring$^{a}$ & $ \frac{1}{2A} \left[e^{-A} - e^{-A\left(x-1\right)^2} \right] + $ & $ 2A \times  \int_0^x { \frac{m}{\sqrt{x^2-m^2}} \left[ \int_0^{\infty} { e^{ - A (x-1)^2} \frac{m (t-1)}{\sqrt{t^2 - m^2}} dt} \right] dm}$ \\
    & $ + \frac{1}{2 \sqrt{A}} \left[ \gamma \left( \frac{1}{2},A \right) + \gamma \left( \frac{1}{2},A \left(x-1 \right)^2 \right) \rm{sgn}(x-1) \right] $  &\\ 
    \end{tabular}
    \caption{Table \ref{tab:baryons_dimless_properties} cont.}
    \label{tab:baryons_dimless_properties_cont}
\end{table*}

\subsubsection{Dark Matter Halos}\label{sec:basic_equations - dm}
We assume spherically symmetric dark matter halos uniquely defined by redshift $z$, virial mass $M_{\rm vir}$ and concentration parameter $c$ (along with possible additional parameters in the density profile). For a local density $\rho_{\rm DM}(r)$ and scale length ($r_{\rm s}$), we define similarly to Equation~\ref{eq:baryonic_surface_density} the dimensionless density function
\begin{equation}\label{eq:dm_density}
    \rho(x) = \rho_0 \times f_{\rho}(x)\ \ \ ,
\end{equation}
the scale mass
\begin{equation}
    M_{\rm s} = \frac{4 \pi}{3} \rho_0 r_{\rm s}^3\ \ \ ,
\end{equation}
and the formula for the dimensionless enclosed mass within radius $x$
\begin{equation}\label{eq:dm_enclosed_mass}
    f_{\rm M}(x) = 3 \int_0^x t^2 f_{\rho}(t)dt\ \ \ .
\end{equation}
Any halo extends to the virial radius $R_{\rm vir}$, the radius within which the average matter density is $\Delta$ times the critical density of the Universe at redshift $z$, $\rho_{\rm crit}=\frac{3H^2(z)}{8\pi G}$. The enclosed mass within $R_{\rm vir}$ is then the virial mass of the halo, 
and is by definition:
\begin{equation}\label{eq:dm_virial_mass}
    M_{\rm vir} = \frac{4 \pi}{3} \Delta \rho_{\rm crit}(z) R_{\rm vir}^3 \ \ \ .
\end{equation}
Typical values for the virial overdensity are $\Delta=(100, 177, 200)$, and we use $\Delta=200$ as the default value \cite[see for example][]{Bullock2001_virial}.
The concentration parameter 
\begin{equation}\label{eq:dm_conc_parameter}
    c = R_{\rm vir} / r_{\rm s}
\end{equation}
is the ratio of the virial radius to the scale radius for the dark-matter density distribution, and this ratio quantifies the ``compactness" of the halo.
Using Equations~\ref{eq:enclosed_mass_general}, \ref{eq:dm_enclosed_mass} and \ref{eq:dm_virial_mass} we can write the central density in terms of $z, M_{\rm vir}, c$ as
\begin{equation}\label{eq:dm_halo_central_density}
    \rho_0 = \frac{c^3}{f_{\rm M}(c)} \Delta \rho_{\rm crit}(z) \ \ \ .
\end{equation}
For a spherical mass distribution, regardless of the specific profile, the circular velocity at radius $r$ is the Keplerian $V_{circ}^2 = GM(<r) / r$, and the dimensionless velocity function is:
\begin{equation}\label{eq:dm_circ_velocity}
    f_{\rm V} (x) = \frac{f_{\rm M} (x)}{x}\ \ \ .
\end{equation}

As baryons accumulate in the center of the halo (i.e., a galaxy forms), additional force acting on the inner halo would lead it to contract. Assuming galaxy assembly is slow and continuous, the contraction can be treated as adiabatic. Following \citealt{Mo1998}, the final (contracted) halo profile can be solved by requiring conservation of angular momentum for each DM particle, assuming pure circular orbits and no shell crossings:
\begin{equation}\label{eq:adaiabatic_contraction1}
    x M_f(<x)  = x_i M_i(<x_i)
\end{equation}
where $x_i$ is the initial radius and $x$ the radius after contraction, in a system including the baryons. The total DM mass remains the same within $x$ as is within the initial $x_i$, so that equation~\ref{eq:adaiabatic_contraction1} becomes:
\begin{equation}\label{eq:adaiabatic_contraction2}
    x \left( M_{\rm baryons}(< x) + M_i(< x_i) \right) = x_i M_i(<x_i)\ \ \ .
\end{equation}
We can solve for the ``contraction factor" $\eta(x) \equiv x / x_i$, for which the last equation can be rearranged as:
\begin{equation}\label{eq:adaiabatic_contraction3}
    \eta^{-1} (x) = 1 + \frac{M_{\rm baryons}(< x)}{M_i(< x / \eta(x))}  \ \ \ .
\end{equation}
The last equation can be solved numerically given the distribution of baryons. This contraction factor must be smaller than unity at all radii and approach it as $r \rightarrow R_{\rm vir}$, as the outer halo is expected to be less affected by the centrally concentrated baryons. Using the the contraction factor, the final density distribution is found by plugging the initial radius $x / \eta$:
\begin{equation}
    \rho_{\rm cont}(x) = \rho_i (x / \eta(x))
\end{equation}
with $\rho_i$ the initial halo profile before contraction, i.e., as given by equation~\ref{eq:dm_density}.
\begin{table*}
    \centering
    \begin{tabular}{c|cccc}
    Profile& Parameters & $r_{\rm s}$ & $f_{\rho} (x)$& $f_{\rm M} (x)$  \\ \hline \hline 
    
    NFW&$M, R_{\rm vir}, c$&$R_{vir} / c$& $\frac{1}{x \left(1 + x\right)^2 }$& $3 \left[ \ln{\left(1+x \right)} - \frac{x}{1+x} \right]$ \\ 
    
    $\alpha$-NFW&$M, R_{\rm vir}, c, \alpha$&$R_{vir} / c$& $\frac{1}{x^\alpha \left(1 + x\right)^{3-\alpha} }$& $\frac{3}{3-\alpha} \left[ x^{3-\alpha} \ _2F_1(3-\alpha, 3-\alpha; 4-\alpha;-x) \right]$\\ 
    
    Burkert&$M, R_{\rm vir}, c$&$R_{\rm vir}/c$& $\frac{1}{(1+x)\left( 1+x^2 \right) }$ & $\frac{3}{2} \left[ \frac{1}{2} \ln{\left(1+x^2 \right)} + \ln{(1+x)} - \arctan{(x)} \right]$ \\ 

    Einasto & $M, R_{\rm vir}, c, n$ & $R_{\rm vir}/c$ & $\exp {\left\{ -x^{1/n} \right\}}$ & $3n\ \gamma \left(3n,x^{1/n} \right)$ \\

    Dekel-Zhao & $M, R_{\rm vir}, c$ & $R_{\rm vir}/c$ & $\frac{3-a}{3} \left( 1 + \frac{3-\bar g}{3-a} x^{1/b} \right) \frac{1}{x^{a} \left(1 + x^{1/b} \right)^{1+b(\bar g -a)}}$ & $\frac{x^3}{x^{a} \left(1 + x^{1/b} \right)^{b(\bar g -a)}}$
    \end{tabular}
    \caption{Density profile for the dark matter. We assume spherically symmetric distributions, described with the normalized radius $x = r/r_{\rm s}$ (see Equation~\ref{eq:normalized_radius}). The dimensionless circular velocity for all profiles is $f_{\rm V} (x) = \sqrt{f_{\rm x}(x)/x}$ (see Equation~\ref{eq:velocity_general}.}
    \label{tab:dm_dimless_properties}
\end{table*}

\subsubsection{Light distributions}
The flux distribution 

\subsection{Pressure Support} \label{sec:pressure_support_equations}
Pressure gradients introduce additional forces that act on the individual mass elements, which can become dynamically significant \citep{Burkert2010_pressure_support}. Typically, the ISM gas has velocity dispersions ranging from $\sim 20\, \rm{km/s}$ at $z=0$, to $\sim 60\, \rm{km/s}$ at redshifts $z\approx 3$, and can become dynamically important as it introduces a pressure term that changes the equilibrium velocity.

The hydrostatic equilibrium equation for gas in a turbulent disk is \citep{Burkert2010}:
\begin{equation}
    \frac{V_{\rm rot}^2(r)}{r} = F_g(r)+\frac{1}{\rho(r)} \frac{dP(r)}{dr}
\end{equation}
where $V_{\rm rot}(r)$ is the equilibrium circular velocity, $F_g(r)$ is the gravitational force at $r$, and $\rho(r)$ is the gas density. The pressure $P = \rho \left( \sigma^2 + c_s^2 \right)$ has a kinetic component originating from the isotropic velocity dispersion $\sigma$ and a thermal component dependent on the sound speed $c_s$, which can be neglected as typically $c_s \ll \sigma$. If we denote $V_{\rm circ}$ as the circular velocity due to gravity alone (i.e., for $P=0$), the pressure-corrected (observable) rotational velocity $V_{\rm rot}$ can be written as:
\begin{equation} \label{eq:Burkert2010_general}
    V_{\rm rot}^2 \left( x \right) = V_{\rm circ}^2 \left( x \right) + \sigma^2 \frac{d\ln \left( f_\rho \sigma_0^2 \right)}{d \ln x}
\end{equation}
where we have used Eq.~(\ref{eq:dm_density}) and switched to the normalized radius $x = r / r_{\rm s}$. The importance of the pressure term is typically quantified using the velocity-to-dispersion ratio, $V / \sigma_0$. For high values the pressure gradients are negligible and the rotation is purely gravitationally supported (a ``cold disk"). For ratios below $\lesssim 2-3$ the disk is marginally stable and pressure gradient are important (a ``hot disk", \citealt{ForsterSchreiber2020_review}). Furthermore, at low $V/\sigma$ the disk can become gravitationally unstable and form clumps at the Toomre scale \citep{Toomre1964}. While these clumps produce asymmetries in the overall  mass distribution, the clump's masses ($10^{8}-10^{9}\, \rm{M_\odot}$) are generally too small to strongly affect the global gravitational potential. Other effects, such as inward migration due to torques exerted on the clumps, can create deviations from circular motions \citep{Romeo2010_clump_formation, Dekel2022_clump_migration}. Such inflows have been observed in several clumpy $z \sim 1-2$ galaxies, with velocities of $50-100\, \rm{km/s}$ ($\approx 10-30 \%$ of the circular velocity) (\citealt{Genzel2023_inflows, Pastras2025_noema3d}; Jolly et al., in prep.). We ignore any such inflow motions in the modelling, as these higher-order kinematic corrections are smaller along the kinematic major-axis.

Two additional approximations are typically applied to Equation~\ref{eq:Burkert2010_general}. One possibility is to assume that the velocity dispersion is independent of radius, $\sigma = \rm const. \equiv \sigma_0$, giving \citep{Burkert2010_pressure_support}:
\begin{equation}\label{eq:Burkert2010_const_sigma}
    V_{\rm rot}^2 \left( x \right) = V_{\rm circ}^2 \left( x \right) + \sigma_0^2 \frac{d\ln f_\rho(x)}{d \ln x}
\end{equation}
A second possibility is to assume vertical hydrostatic equilibrium (HSE), for which the local midplane density is proportional to the total surface density $\Sigma$ and the vertical velocity dispersion $\sigma_z$, namely, $\rho = \frac{\pi G \Sigma^2}{2\sigma_z^2}$ \citep[a ``Spitzer disk",][]{Spitzer1942_HSE, Binney_and_Tremaine2008}. In the case of an isotropic velocity dispersion, the radial and vertical velocity dispersion are equal $\sigma_z = \sigma_r \equiv \sigma $, giving
\begin{equation}\label{eq:Burkert2010_HSE}
    V_{\rm rot}^2 \left( x \right) = V_{\rm circ}^2 \left( x \right) + 2 \sigma^2 \frac{d\ln f_\Sigma (x)}{d \ln x} \ \ \ ,
\end{equation}
where $f_\Sigma(x)$ is the surface density at radius $x$. Interestingly, equation~\ref{eq:Burkert2010_HSE} is also valid for radially changing velocity dispersions, as long as it can be locally treated as isotropic. Assuming HSE is a convenient assumption when dealing with observations, as it allows to directly use a surface density profile (e.g., an exponential disk) without knowing the intrinsic local 3D density. For a S{\'e}rsic disk, $\ln{ f_\Sigma} = - \left(r/R_{\rm d} \right)^{1/n}$, and Eq.~(\ref{eq:Burkert2010_HSE}) gives \citep{Burkert2016_pressure_support}: 
\begin{equation}\label{eq:Burkert2010_HSE_sersic}
    V_{\rm rot}^2 \left( r \right) = V_{\rm circ}^2 \left( r \right) - 2 b_n \sigma_0^2 \left( r / R_{\rm eff} \right)^{1/n}
\end{equation}
where $R_{\rm eff}$ is the disk half mass radius, and $b_n \approx 2n - \frac{1}{3}$ is the solution to $\gamma(b_n,2n)=\frac{1}{2}\Gamma(2n)$ with $\gamma$ and $\Gamma$ being the incomplete and complete gamma function, respectively \citep{Ciotti1999_sersic}.

\subsection{Sky projection}\label{sec:sky_projection}
\begin{figure}
    \centering
    \includegraphics[width=0.9\linewidth]{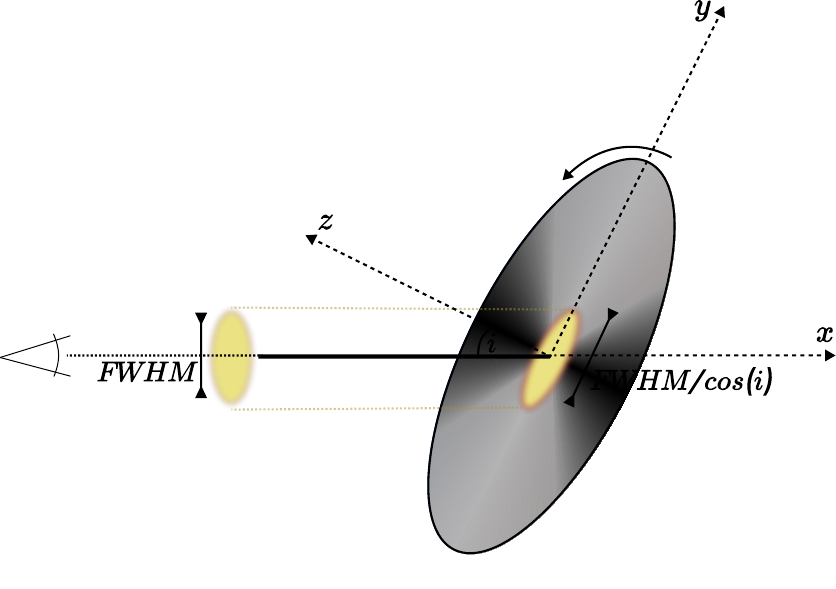}
    \caption{Illustration of the 2D beam smearing projection for an inclination angle $i$. As the galaxy becomes more edge-on ($i \rightarrow  90^\circ$), it intersects larger areas on the disk plane. The projected ellipse has an axis-ratio of $a/b = 1 / \cos{i}$ (for $i \neq 90^\circ$).}
    \label{fig:beam_projection_cartoon}
\end{figure}

After generating an intrinsic velocity profile in the galactic plane, we want to simulate the observed velocity map for a given sky orientation and finite beam resolution. This is typically done by rotating the coordinates in the frame of the galaxy to sky coordinates, given the inclination angle, then summing over the line-of-sight to create the two dimensional velocity map. This is followed by convolution with an instrumental point spread function (PSF). Treatment of this beam smearing effect is crucial, as large PSF sizes can cause dramatic effects on the resulting velocity map, and often can be very time consuming since the number of pixel scales rapidly as $\propto n^3$. Hence, instead of rotating the galaxy coordinates to match the sky coordinates, in \texttt{RotCurves} we replace the PSF with an ``effective beam", as a function of the inclination angle while still working in galactic coordinates. We illustrate this in Figure~\ref{fig:beam_projection_cartoon}. For example, a circular PSF through a face-on galaxy imprints a circle. For a more edge-on galaxy the beam intersects the galactic plane at an angle and it imprints a prolate ellipse. If the galaxy is axisymmetric, we can arbitrarily set the $x_{\rm gal}$ axis as the rotation axis for the inclination, so that the change is only applied to the $y_{\rm gal}$ axis. Specifically, if we consider a circular Gaussian PSF with a standard deviation $\sigma_{\rm PSF}$, it will become a bi-variate Gaussian with $\left( \sigma_x, \sigma_y \right) = \left( \sigma_{\rm PSF}, \sigma_{\rm PSF}/\cos \left( i \right) \right)$. This way a face-on projection ($i=0$) results in a circular shape, while the edge-on case ($i=90$) results in a cylinder. 

The observed rotation velocity is then calculated as a weighted average over the area of the ``effective beam". In addition to the PSF, the strength of the emission itself has to be taken into account. As kinematic observations typically use emission line tracers, it will not be isotropic and will tend to favor certain areas in the galaxy. For example, 21cm hydrogen emission arises from the neutral ISM covering most of the stellar disk, while H$\alpha$ emission is emitted from hot $\sim 10^4 \rm K$ clouds around young OB stars, tracing recent star formation. Therefore, ``light weighting" is included in this step, with a functional form for the emission, $I(r)$. By default, we assume a constant mass-to-light (M/L) ratio following the density profile of the disk. In addition, only line-of-sight velocities contribute observationally, introducing an additional $\cos \theta$ factor. 

In total, the observed velocity at a given radius $r$ is just the weighted average:
\begin{align}\label{eq:Vobs_psf_convolution}
    V_{\rm obs} & \equiv \langle  V_{\rm circ} (r) \rangle = \nonumber \\ 
    & \frac {\sum_{x_,y}\ \mathrm{PSF} \left(x_{\rm gal}, y_{\rm gal} \right) I\left(r_{\rm gal} \right) V_{\rm circ} \left(r_{\rm gal} \right) \cos \theta} {\sum_{x_,y}\ \mathrm{PSF} \left(x_{\rm gal}, y_{\rm gal} \right) I\left(r_{\rm gal} \right)}
\end{align}
and the velocity dispersion at a given radius $r$ is estimated as the weighted standard deviation over the ``effective" PSF:
\begin{equation}\label{eq:sigma_psf_convolution}
    \sigma_{\rm obs} = \sqrt{\left( <V_{\rm circ}^2> - V_{\rm obs}^2 \right)} \ \ \ .
\end{equation}
summing over all $(x_{\rm gal}, y_{\rm gal})$ within the FWHM of the PSF, where $r_{\rm gal} = \sqrt{x_{\rm gal}^2 + y_{\rm gal}^2 }$ and $\cos \theta = \arctan {\left( y_{\rm gal} / x_{\rm gal} \right)}$.

\subsection{MCMC Fitting}\label{sec:mcmc_fitting}
The sky-projected velocity map can be fitted to an observed velocity map using a Monte-Carlo Markov-Chain (MCMC) Bayesian algorithm implemented with the python package \texttt{emcee} \citep{ForemanMackey2013_emcee} together with \texttt{corner} to analyze the posterior distribution \citep{corner}. We can then simultaneously fit multiple free parameters and estimate their distribution given constraints on both observational data and priors (via scaling relations, other observational constraints, etc.). It does so by generating a Markov chain in which each new proposed state is accepted or rejected based on a probabilistic criterion, maximizing the log-probability function. Using many iterations, the walkers explore the posterior distribution of the free parameters and provide realistic estimates of the uncertainties, given the walkers are not correlated and the chain has converged. The best fit values are estimated as the median or the maximum of the posteriors, and even though these might not optimally maximize the log-likelihood they provide good estimates.

For a set of $n$ free parameters, the MCMC algorithm constructs a sequence of parameter states $\vec{\theta} = (\theta_1, \theta_2, ..., \theta_n)$, each sampled from an initial (prior) distribution $\mathcal{P}(\theta_i)$. Given $\vec{\theta}$, a galaxy model is constructed along with its projected, mock-observed rotation curve, as discussed in the previous paragraphs. At each step, the MCMC algorithm is attempting to minimize the log-probability function, which is the sum of the log-prior of each free parameter and the log-likelihood as defined by a least-squares:
\begin{align}\label{eq:mcmc_loss}
    \ln \mathcal{L} = &- \frac{1}{2} \sum_{i=1}^m \left( \frac{v_{\rm data,i} - v_{\rm rot,i} (\vec{\theta})}{v_{\rm err,i}} \right) ^{2} \\
     & +\sum_{j=1}^{n} \mathcal{P}(\theta_j)
\end{align}
with $m$ data points, and where $\vec{\theta}$ are the $n$ model parameters.

The priors are set for each parameter independently, and are drawn either from a uniform or a Gaussian probability distributions, both bounded by minimum and maximum values:
\begin{equation}
    \ln \mathcal{P(\theta)} =
    \begin{cases}
        -\infty, & \theta < \theta_{\rm min}\, , \, \theta > \theta_{\rm max}\\
        0, & \mathrm{uniform} \\
        - \frac{1}{2} \left( \frac{\theta-\theta _0}{\sigma_{\theta}} \right)^2, & \mathrm{Gaussian}
    \end{cases}
\end{equation}
with $\theta_0$ the Gaussian center (by default the initial value) and $\sigma_{\theta}$ the standard deviation.

Each walker then makes a random step, repeating the process and attempting to maximize the log-likelihood while still allowing for occasional ``bad" steps that would help if the walker is stuck at local minima. We find that a combination of different steps in best to ensure rapid exploration of the parameter space and break correlated steps: each step has a 60\% chance of being a \texttt{emcee.moves.StretchMove} with $a_{\rm stretch} = 2$, 30\% chance of being a \texttt{emcee.moves.DEMove} and 10\% chance for a \texttt{emcee.moves.KDEMove}.


To properly sample the posterior space, a large number of individual ``walkers" are required with a total number of iterations (or steps). We estimate the auto-correlation time $\tau$ of the walkers for each parameter using the built-in method \texttt{sampler.get\_autocorr\_time}, and choose the maximum value as our estimate. The total number of steps is chosen to be much larger to ensure independent sampling, $n_{\rm steps} \approx  50\tau$, of which the first $n_{\rm burn} \approx 5 \tau$ are discarded as burn-in steps. With this number of steps, we set the number of walkers to be $n_{\rm walkers} = 100$ ensuring we have over $\sim 3000$ independent samples ($> 1000$) in the posterior distribution.

The best-fit values for each parameter are estimated as the median of the distribution, and the lower (upper) uncertainty is defined as the $16^{th}$ ($84^{th}$) percentiles of the posterior. If the posterior distribution is highly asymmetric, which can often be the case, the median might be significantly different than the value of maximum probability, and the maximum a-posteriori (MAP) might be a better estimator for the best-fit value. 


\section{benchmark} \label{sec:benchmark}
We use the widely-used, publicly available, parametric code \texttt{Dysmalpy} to compare the results of our beam smearing process. \texttt{Dysmalpy} has been used in the literature to analyze rotation curves at high redshift multiple times with great success \citep[][Espejo Salcedo et al., in prep.]{Genzel2006, Wuyts2016_KMOS3d, Lang2017_stacking_rc, Genzel2017, Ubler2018_ha_co_rc, Ubler2024_GN20, Ubler2024_stellar, Price2021_rc41, Liu2023_lesned, Posses2024, Telikova2025_cristal, Fei2025}. The code is described in detail in \cite{Price2021_rc41} and \cite{Lee2025_dysmalpy}. Essentially, \texttt{dysmalpy} constructs a model of an idealized, axisymmetric rotating galaxy, comprised of multiple mass components, each with a defined functional form (e.g., a Sérsic profile) and a set of parameters (such as mass, effective radius, etc.). Given a combination of mass components, a full 3D cube of the galaxy is constructed, where each pixel contains a Gaussian distribution of velocities centered around the rotational velocity, $V_{\rm rot}$, with a standard deviation of the intrinsic velocity dispersion of the gas, $\sigma_0$. The beam smearing is done for this 4D-hypercube, i.e. $(x_{\rm gal},y_{\rm gal}, z_{\rm gal}, v)$, by rotating it to sky coordinates and calculating the line-of-sight emission profile, taking into account the line-spread-function (LSF), point-spread-function (PSF) and the flux-weighting (or $\rm M/L$ ratio).

Using \texttt{dysmalpy} allows us to isolate the effect of the beam smearing correction alone, as we can define an intrinsically identical model for the galaxy and use similar pixel scales and box sizes to directly compare the recovered physical parameter of the galaxy. In contrast, other kinematic modelling tools (mentioned in \S~\ref{sec:intro_rctools_highz}) do not assume an underlying mass profile but use either non-parametric tilted-ring approaches or purely mathematical forms (e.g., $v_\mathrm{rot} \propto tan^{-1} (r)$).This is important for our setup, as such these underlying differences can lead to systematics beyond the scope of this comparison \citep[e.g.,][]{Lee2025_dysmalpy, Yttergren2025_code_comparison}. The implementation of the beam smearing correction has been the subject of some debate in recent years, and the ``correct" answer is very elusive as the ``true" kinematic values of a galaxy remain unknown. \cite{Lee2025_dysmalpy} used a sample of observationally-motivated mock-observed galaxies to demonstrate that all three codes are in excellent agreement for the recovered velocity profile in the high signal-to-noise regime, with less than a $<10 \%$ difference in the recovered properties up to twice the effective radius.

We mention that \texttt{2DBAT} in particular, with its 2D approach and similar Bayesian inference framework, make a good candidate for comparison. However, \texttt{2DBAT} does not include the crucial beam smearing correction needed for interpreting high-z observations. \texttt{2DBAT} also assumes that the halo dominates the gravitational potential at all radii, making it more suitable for highly dark-matter dominated systems \citep{Oh2018_2DBAT}. At Cosmic Noon, there are many galaxies with very low contributions of dark matter on galactic scales, and \texttt{RotCurves} is built to support a wide variety in velocity profiles \citep{Genzel2017, Genzel2020_rc41, NestorShachar2023_rc100, Lee2025_cristal, Danhaive2025_fdm}. Furthermore, the strength of \texttt{2DBAT} is in the recovery of detailed non-axisymmetric features through varying PA and inclination with the tilted-ring approach, which differ fundamentally from the assumptions made in \texttt{RotCurves}, in which the galaxy is axisymmetric and the PA and inclination do not vary radially. In this paper we focus on how \texttt{RotCurves} recover of the physical parameters of the galaxy, and not its geometric properties, making such comparisons beyond the scope of this current work.

Therefore, we limit our comparison to \texttt{dysmalpy} alone, as a more apples-to-apples approach for testing and comparing our parametric forward modelling code.

\begin{table}\label{tab:runtimes}
\centering
\begin{tabular}{c|ccc}
Code & $\rm n_{pix}$ & pixel size [mas]& runtime [ms]\\ \hline \hline 
\texttt{RotCurves}& 95& 50& 20\\
\texttt{dysmalpy} & 95& 50& 3638 \end{tabular}
\caption{Average runtimes of the beam-smearing correction for the fiducial model, calculated from $n=100$ procedures ($\lesssim 1\%$ statistical error) on a single CPU. \texttt{RotCurves} generates a ($\mathrm{n_{pix}}, \mathrm{n_{pix}}$) grid whereas \texttt{dysmalpy} generates a ($\mathrm{n_{pix}}, \mathrm{n_{pix}}, \mathrm{n_{pix}}, \mathrm{n_{spec}}$) hyper-cube, with $\mathrm{n_{spec}}=200$. The beam FWHM is 10 pixels.}
\end{table}
\begin{figure}
    \centering
    \includegraphics[width=0.8\columnwidth]{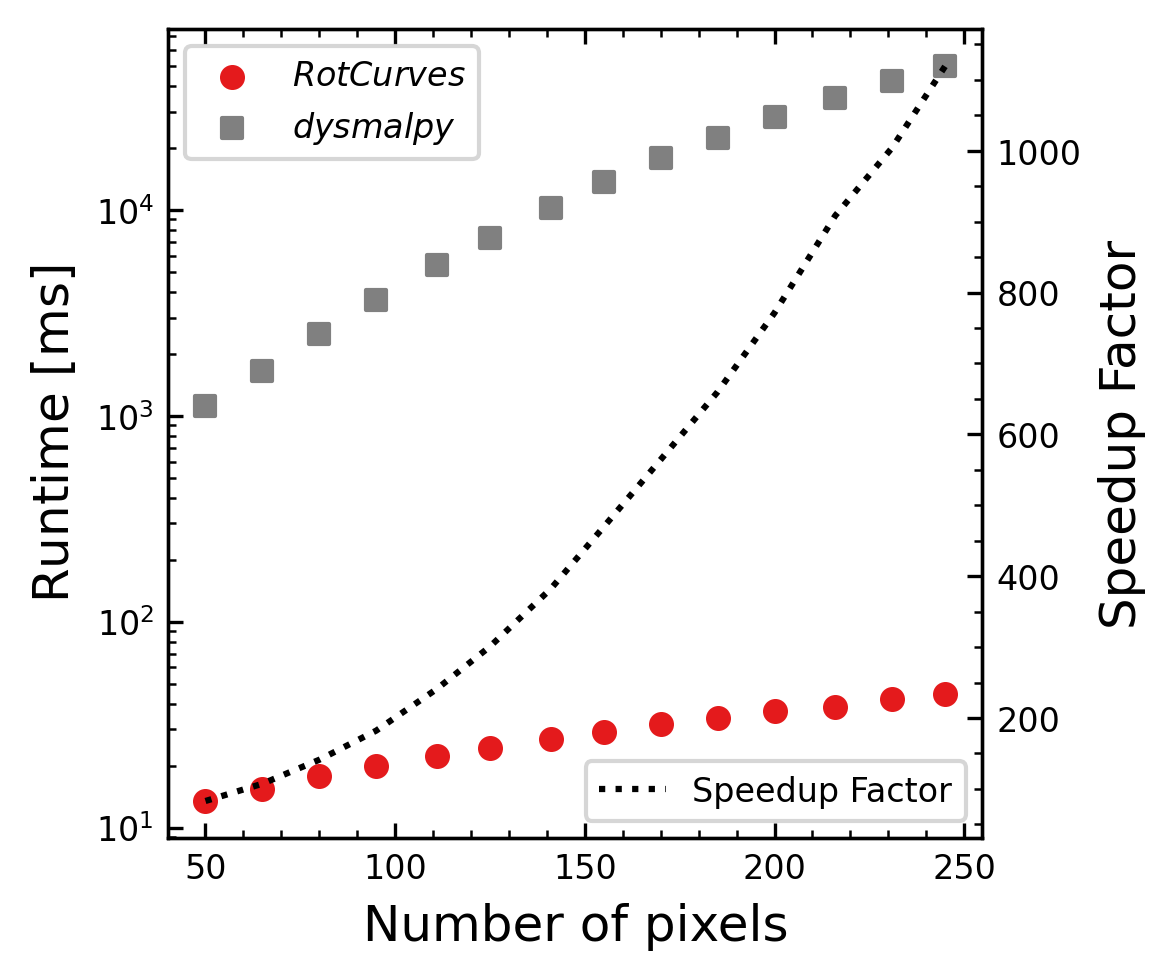}
    \caption{Average runtime of \texttt{RotCurves} (red circles) and \texttt{dysmalpy} (grey squares) as a function of the number of pixels on one side of the grid. Runtimes are calculated as the average CPU time from 100 computations. The speedup factor is shown in a dashed line, corresponding to the right side of the figure. \texttt{RotCurves} keeps under $\lesssim100$ms even when resolving finely sampled grids, and is $\approx 300-2500$ faster than \texttt{dysmalpy}.}
    \label{fig:runtimes}
\end{figure}

\subsection{Beam smearing} \label{sec:benchmark - beam smearing}
The procedure for beam smearing is much more detailed in \texttt{dysmalpy} with its full 4D-hypercube, so we expect any differences in the mock-observed rotation curves to increase as the inclination becomes more edge-on (high $i$), or as the PSF FWHM increases. 

For comparisons, we construct a fiducial model consisting of an NFW dark matter halo, an exponential disk, and a bulge, with values typical for massive, star-forming galaxies at $z=2$, and with a constant velocity dispersion for the gas (see Table~\ref{tab:fiducial_parameters}). Motivated by large H$\alpha$ IFU surveys of massive SFGs \citep[e.g.][]{ForsterSchreiber2009_SINS, Wisnioski2015_kmos3d} we set the mass of the disk to $M_{\rm disk} = 5 \times 10^{10}\  \rm M_\odot$, comprising both stellar and gas content, with an effective radius of $R_{\rm eff} = 5\, \rm kpc$. We assume an intrinsic axis-ratio of $q_0 = 0.2$ to account for disk thickness \citep{HamiltonCampos2023_thickdisk}.
We set the central bulge mass to $M_{\rm bulge}=3 \times 10^{10}\, M_\odot$ (bulge-to-total ratio of 0.38), with a de-Vaucouleurs profile ($n_s=4$) and an effective radius of $R_{\rm eff} = 1\, \rm kpc$. We assume an NFW density profile for the dark matter halo, with concentration parameter $c=4.5$ appropriate for redshift $z=2$ \citep[e.g.][]{Dutton2014,Genzel2020_rc41, NestorShachar2023_rc100, Puglisi2023_KURVS}. For this halo the dark-matter fraction at $R_{\rm eff}$ is $f_{\rm DM}=0.28$.

\begin{table}\label{tab:fiducial_parameters}
    \centering
    \begin{tabular}{c|cc}
         Component&  Parameter&Fiducial Value\\ \hline \hline
         Cosmology & Redshift & $2$ \\
         &  kpc-to-arcsec & $8.4$ \\
         Disk&  Mass $[M_\odot]$&$3 \times 10^{10}$\\
         &  $R_{\rm eff} [\rm kpc]$&$4$  \\
         &  $n_s$ &$1$  \\
 & $q_0$&$0.2$  \\
         &  $\Upsilon$&1\\
         Bulge&  Mass $[M_\odot]$ &$10^{10}$\\
         &  $R_{\rm eff} [\rm kpc]$ &$1$ \\
         &  $n_s$ &$4$ \\
 & $q_0$ &$1$ \\
         &  $\Upsilon$&$10^{-2}$\\
 NFW Halo& $f_{\rm DM} (R_{\rm eff})$&$0.28$ \\
 & Mass [$M_\odot$] & $11.8$ \\
 & concentration&$4.5$ \\
 Pressure Support& $V_{\rm circ}(R_{\rm eff}) / \sigma_0$&$4$ \\
        & $\sigma_0$ [km/s] & 56 \\
 Geometry& inclination &$60$ \\
 Beam& $\mathrm{FWHM / R_{\rm eff}}$ & $0.5$\\
 &  $\mathrm{FWHM}\ ["]$ &$0.25$ \\
 &  pixel size $[mas]$ & 50
    \end{tabular}
    \caption{Parameters for the fiducial model including an exponential disk, a de-Vauceulours bulge and an NFW halo, motivated by observational properties of massive SFGs at $ z\sim 2$ \citep{Genzel2020_rc41, NestorShachar2023_rc100, Ubler2019_dispersion} and by cosmological scaling relations \citep{Dutton2014, Moster2018}. The mass includes both stellar and gas contents, assumed to have a similar distribution.}
\end{table}

Table~\ref{tab:runtimes} shows the average runtimes of each code as a function of the number of pixels and the size of the beam. \texttt{RotCurves} generates a mock-observed rotation curve at a scale of $\sim10$ms, while \texttt{dysmalpy}  will generate the same rotation curve at a scale of $\sim 10$s, the exact numbers depending on the number of pixels. Figure~\ref{fig:runtimes} shows that the runtime for \texttt{RotCurves} increases slowly with the number of pixels and is $\lesssim100$ms even for a large number of pixels. On reasonable pixels scales and FOV sizes, \texttt{RotCurves} performs $\approx 200-300$ times faster than \texttt{dysmalpy}, improving further as the grid increases in size. This boost in performance allows running high-resolution and large-scale models much faster, increasing the ability to test multiple scenarios. Runtime increases with increasing number of pixels, yet it increases slower for \texttt{RotCurves} than \texttt{dysmalpy}. 

\begin{figure}
    \centering
    \includegraphics[width=\columnwidth]{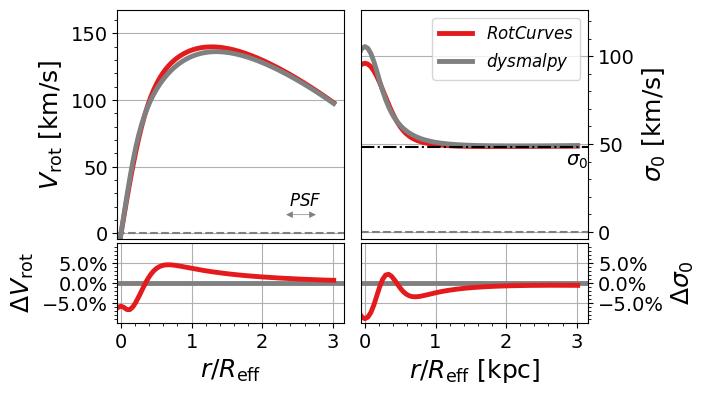}    
    \caption{Mock-observed rotation (left) and dispersion (right) curves for \texttt{RotCurves} (red) and \texttt{dysmalpy} (black). The bottom panels show the residuals relative to \texttt{dysmalpy}. A PSF FWHM of $0.25"$ ($0.5 R_{\rm eff}$) is considered, comparable to ground-based high-z spectroscopy. The residuals are $\lesssim 5\%$ for both rotation and dispersion, with largest deviations at $\approx 0.5R_{\rm eff}$. The root mean square error (in km/s) for the full curves is given in the top-left corner.}
    \label{fig:fiducial_rc_comparison}
\end{figure}

Comparing the two beam-smeared rotation curves in Figure~\ref{fig:fiducial_rc_comparison}, we find very good agreement between \texttt{RotCurves} and \texttt{dysmalpy}. Up to 3 times the effective radius, the difference does not exceed $5\%$, in both rotation velocity and velocity dispersion. Beyond the effective radius, at $r\approx 1-3 R_{\rm eff}$, the RC is highly accurate with deviations as small as $\lesssim 3\%$. The inner part is more susceptible to differences, as the intrinsic large velocity gradients make the vertical scale more important. Yet even at $r < R_{\rm eff}$ deviations do not exceed $5 \%$. In \texttt{dysmalpy} the beam collects light from particles off the midplane, and due to the inclination angle they will have larger radial distances that alter the line velocity centroids and broaden the line. Further out, and especially after the turnover radius, the off-midplane contribution will have both higher and lower velocities (i.e., closer and further radii). For a dropping rotation curve this leads to a lower centroid velocities, and vice versa for a rising curve. 

Such effects are not included in \texttt{RotCurves} as it considers only midplane motions, and the velocities may be over-estimated, and dispersions under-estimated, compared to \texttt{dysmalpy}. However, as shown for our fiducial model, these differences are only at the $\sim$few percent level. 
A major contribution to the differences, mainly at large radii, is that \texttt{dysmalpy} assumes an exponentially dropping light profile in the z-direction, with the scale height tied to the effective radius of the disk, $h_z \sim q_0 R_d \sim 0.1 R_{\rm eff}$. Therefore, this correction becomes important only when velocity gradients are large enough over $\sim 3h_z$, which is typically the case around the area of the bulge, but it is not the case further out in the disk. For the fiducial model, $h_z \sim 400\, \rm pc =0.1 R_{\rm eff}$. 

\begin{figure*}
    \centering
    \includegraphics[width=0.85\linewidth]{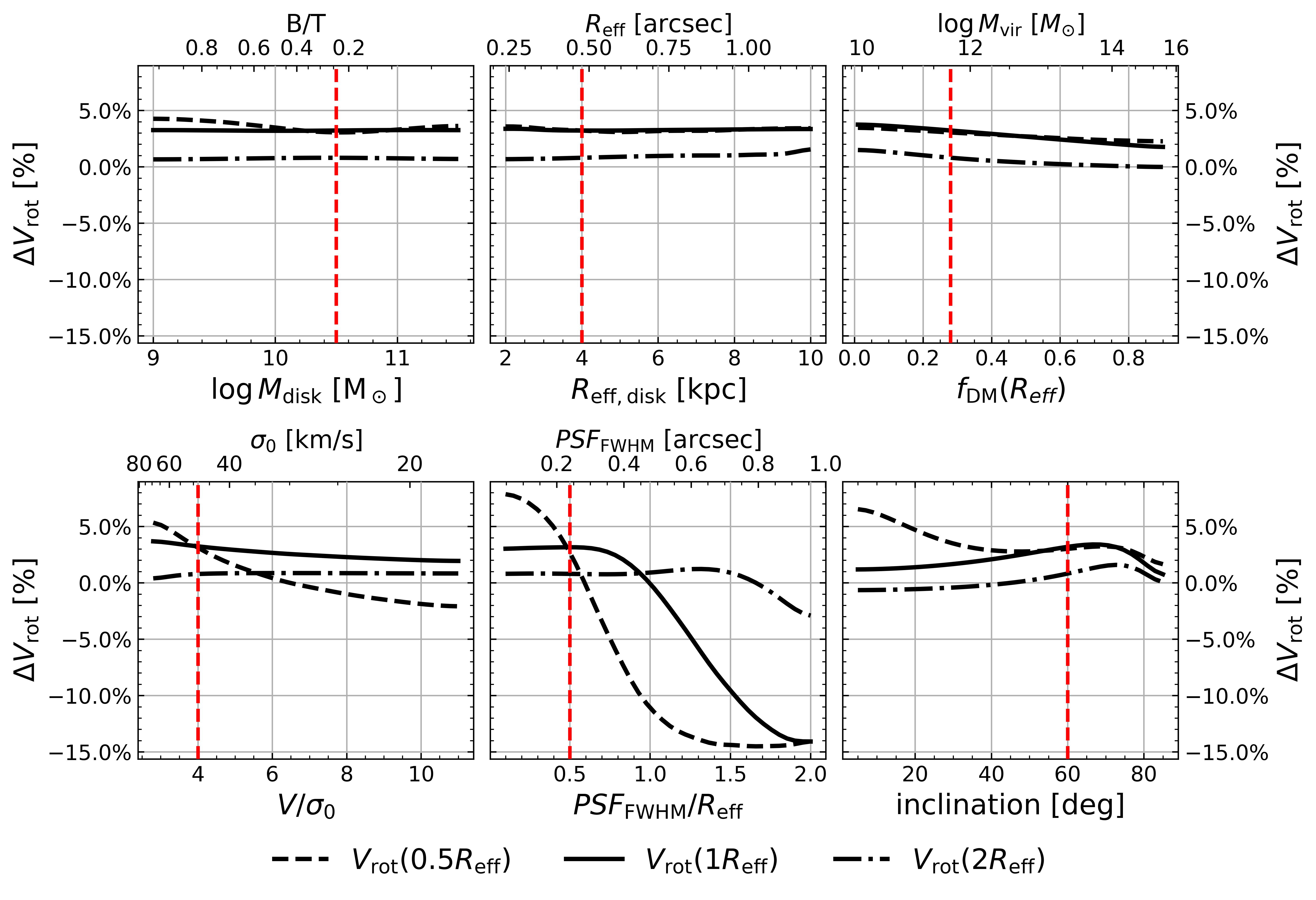} \\
    \includegraphics[width=0.85\linewidth]{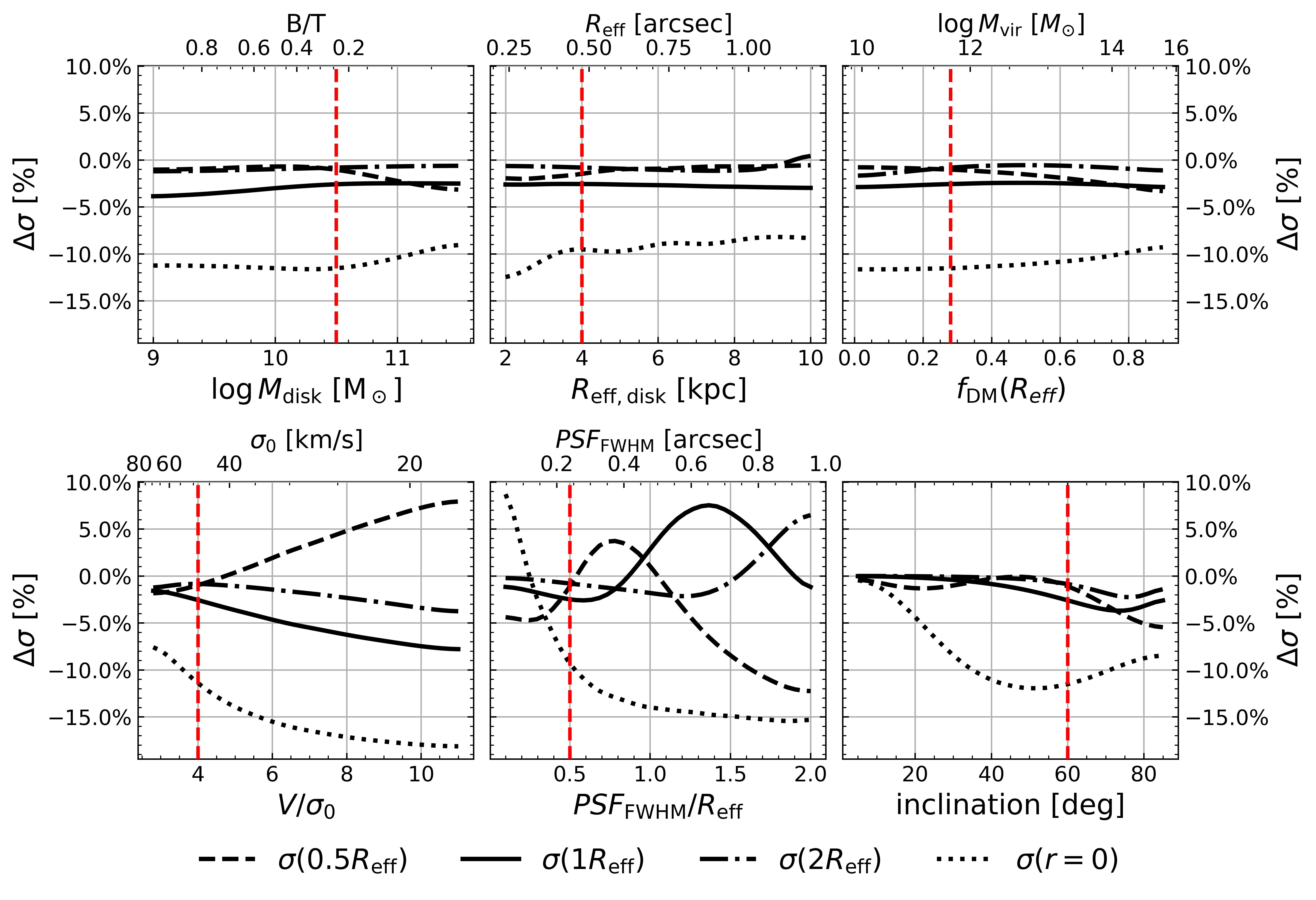}
    \caption{Comparison of the beam-smeared curves generated by \texttt{RotCurves} against \texttt{dysmalpy} for variations in key model parameters. Each panel shows the residual rotation velocity (top figure) or the residual velocity dispersion (bottom figure), defined as $\Delta \equiv \left( \texttt{RotCurves} - \texttt{dysmalpy} \right) / \texttt{dysmalpy}$, at notable apertures: $r = (0.5, 1, 2)\times R_{\rm eff}$ given by the dashed, solid, and dash-dotted lines respectively (in addition the the center $r=0$ for the dispersion, shown as a dotted.). The fiducial value given by the red dashed line. From top left to bottom right, the parameters are: Disk mass $M_{\rm disk}$, bulge mass $M_{\rm bulge}$, dark matter fractions at the effective radius $f_{\rm DM} (R_{\rm eff})$, v-over-sigma ratio $V/\sigma_0$, FWHM of the circular PSF in units of the effective radius, $PSF_{\rm FWHM} / R_{\rm eff}$, and the inclination angle $i$.}
    \label{fig:fiducial_paramstudy_pcnt}
\end{figure*}

\subsection{Parameter study} \label{sec:fiducial_parameter_study}
To examine \texttt{RotCurves} over a wide range of model parameters, we examine how relative residuals react when varying a single parameter while keeping the rest fixed (as described in Table~\ref{tab:fiducial_parameters}). The residuals are
\begin{align*}
    \Delta V_{\mathrm{rot}} (r_{\mathrm{ap}}) &= \frac{V_{\mathrm{rot}}^{Rotcurve} (r_{\mathrm{ap}}) - V_{\mathrm{rot}}^{dysmalpy} (r_{\mathrm{ap}}) }{ V_{\mathrm{rot}}^{dysmalpy} (r_{\mathrm{ap}}) } \\
    \Delta \sigma (r_{\mathrm{ap}}) &= \frac{\sigma^{Rotcurve} (r_{\mathrm{ap}}) - \sigma^{dysmalpy} (r_{\mathrm{ap}}) }{ \sigma^{dysmalpy} (r_{\mathrm{ap}}) } \ \ \ .
\end{align*}
Figure~\ref{fig:fiducial_paramstudy_pcnt} shows the velocity and velocity dispersion residuals at three locations representing the inner and outer parts of the galaxy, $(0.5, 1, 2) \times R_{\rm eff}$, as a function of one parameter of the model. We focused on parameters that either have a physical importance (e.g., mass) or a strong beam-smearing effect (e.g., PSF size). Non-normalized versions of these plots can be found in Appendix~\ref{appendix:residuals_physical}.

The velocity residuals at all apertures are within $\lesssim 5\%$ with almost no dependence on the value of the parameter, except for PSF size. It is not surprising that the beam size has a strong effect, as it directly determines the mixing within the beam. Moreover, the inner aperture ($0.5\times R_{\rm eff}$) is more susceptible to changes in the beam size, either when it is smaller than $\lesssim 0.2\times R_{\rm eff}$ for which the velocity is overestimated, or when $\rm PSF_{FWHM} \gtrsim R_{\rm eff}$ for which the inner velocity is underestimated. This can be easily understood by considering the area the beam covers. When the PSF FWHM is larger than the distance to the center, the beam collects velocities with opposite signs (i.e., the ``blue" or ``red" sides), and by going off the midplane it reduces the radial distance of these particles along the line of sight. For this reason, $\Delta V_{\rm rot}$ decreases with the beam size, and changes sign roughly when the $PSF_{\rm FWHM}$ matches the distance of the aperture from the center.

The velocity dispersions show a very similar behavior, with residuals of $\lesssim 5\%$ at all apertures for almost the entire range of parameters considered, and even $\lesssim 2\%$ for large parts (see Figure~\ref{fig:fiducial_paramstudy_pcnt}). In general, \texttt{RotCurves} will tend to underestimate the velocity dispersion by design, as it considers only the midplane and mixes fewer regions in its beam. This is shown by the strong trends in the bottom-middle panel of Figure~\ref{fig:fiducial_paramstudy_pcnt}. Importantly, there are two locations where the velocity dispersion has great significance: at the outskirts, where the velocity dispersion is less susceptible to beam-smearing, and at the center, where the dispersion is a sensitive probe of the central mass concentration. The residuals at the outer parts are very small, $\left| \Delta \sigma (2 R_{\rm eff}) \right| < 3\%$ (dot-dashed line in Figure~\ref{fig:fiducial_paramstudy_pcnt}), yet in the center the velocity dispersion is underestimated on average by $\Delta \sigma (r=0) \approx -10\%$) and up to $20\%$ in extreme cases (dotted line). As discussed later, when fitting a model for given data points, \texttt{RotCurves} will typically find higher bulge masses than \texttt{dysmalpy} to compensate for the difference, as long as the uncertainties on the central data points are smaller than the residuals discussed here.

\subsection{Ftting a rotation curve}\label{sec:benchmark - mcmc fitter}
To test how well \texttt{RotCurves} can recover the intrinsic model parameters, in the following section we construct a rotation curve using \texttt{dysmalpy} that we use as a ``data" input for \texttt{RotCurves} with the fiducial model as our baseline. We sample the rotation curve out to $3 R_{\rm eff,disk}$ at each half-width half-maximum (HWHM) of the PSF, and assume a flat uncertainty of $5\%$. Our model consists of five free physical parameters ($M_{\rm baryon}, B/T, R_{\rm eff,disk}, \sigma_0$ and $M_{\rm vir}$), with the rest fixed to their true values. While the inclination is often a source of known degeneracies with other parameters, such as the mass of the galaxy, we assume a fixed value in this analysis as it can typically be recovered from photometric isophote fitting. However, in the lowest resolution regime isophote fitting may become unreliable, especially when the beam is comparable to, or larger, than the minor axis. In such cases, it is best to treat the inclination as a free model parameter in \texttt{RotCurves} and constrain it using the prior distribution, e.g., as a truncated Gaussian, to correctly propagate the uncertainties. In our fiducial model we fix the inclination to its true value and use very relaxed prior probabilities for the free parameters, assuming uniform distribution for all parameters and draw randomized initial steps for each fitting procedure. We use an MCMC fitting procedure (see \S~\ref{sec:mcmc_fitting}) with 100 walkers and 3000 steps (of which 500 are discarded as burn-ins), suitable for autocorrelation times of $\sim$70 steps. The free parameters are initialized randomly centered around their true values and are free to vary with a uniform prior probability distribution. To test the robustness of the MCMC fitter, we perform $n=10$ separate fitting procedures with their own initial states, finding the statistical uncertainty $\Delta_{\rm stat} =  \sigma / \sqrt{n}$ of the best fit values are an order-of-magnitude smaller than the uncertainties of the posterior itself. 

\begin{table}
    \centering
    \begin{tabular}{c|c|c|c}
         Parameter& Prior & Model& \texttt{RotCurves}\\ \hline \hline
         $M_{\rm baryon}$& Uniform & $10.6$&  $10.75 \pm 0.13$\\
         $B/T$& Uniform & $0.25$&  $0.23 \pm 0.06$\\
         $R_{\rm eff,disk}$& Uniform & $4$&  $5.4 \pm 0.8$\\
         $\sigma_0$& Uniform & $56$&  $49.0 \pm 1.1$\\
         $M_{\rm vir}$& Uniform & $11.8$&  $11.1 \pm 0.6$\\
         \end{tabular}
    \caption{Best fit values for the MCMC fitting using \texttt{RotCurves} for the fiducial model generated with \texttt{dysmalpy}. The value shown is the mean of the best fit values from $n=10$ separate fitting procedures, with the mean 1$\sigma$ uncertainty reported after the $\pm$ sign. The statistical error is $<0.01 \%$ the 1$\sigma$ for all parameters.}
    \label{tab:fiducial_bestfit_mcmc}
\end{table}

\begin{figure*}
    \centering
    \includegraphics[width=0.6\linewidth]{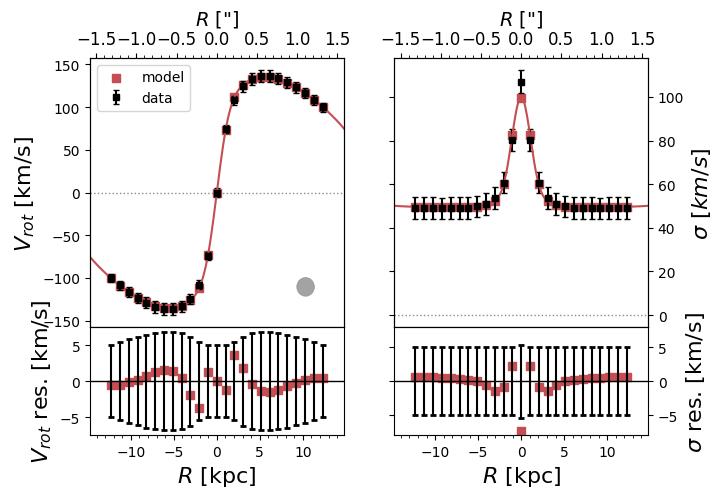}
    \includegraphics[width=0.75\linewidth]{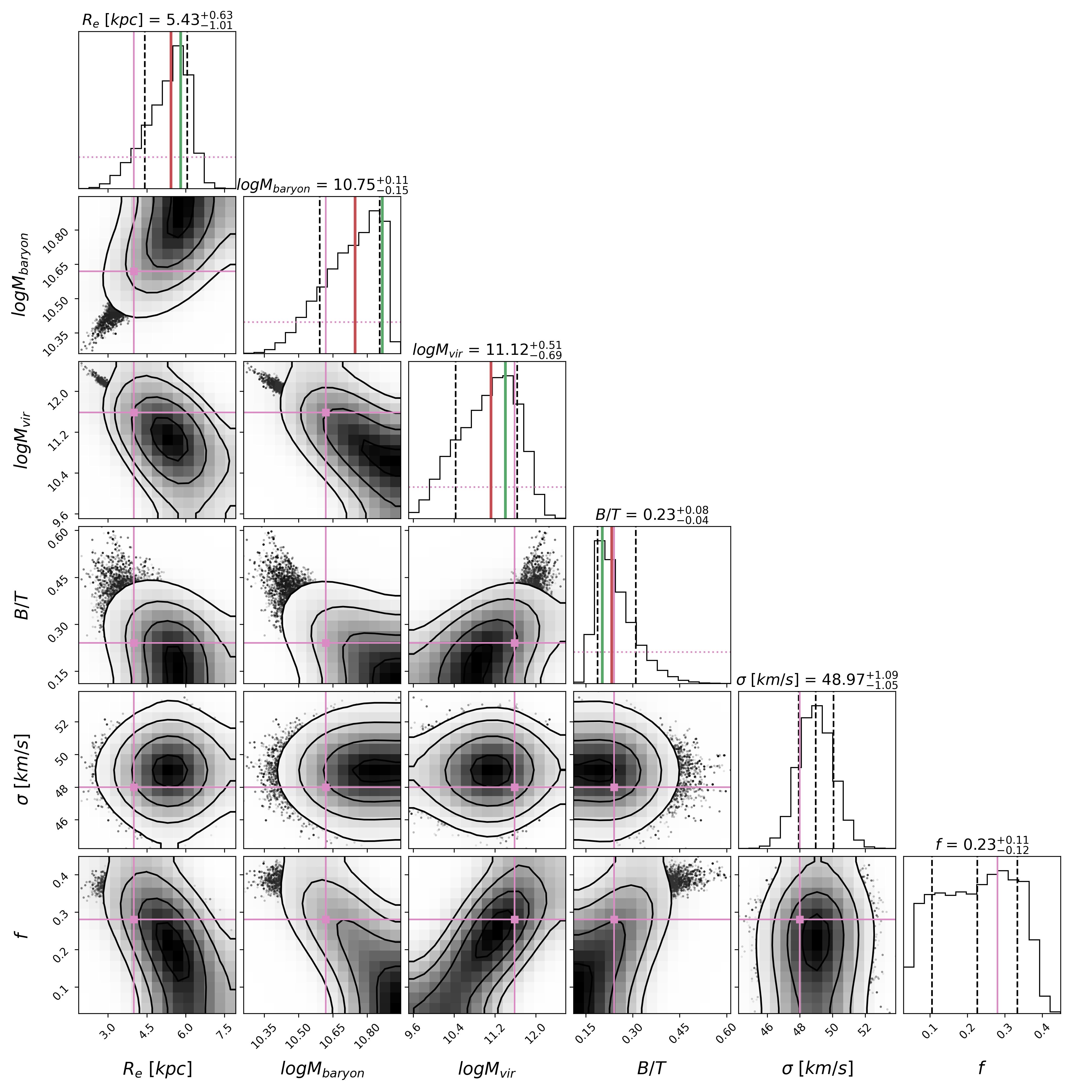}
    \caption{The best fit results for the fiducial model from the MCMC fitting procedure, fitted with \texttt{RotCurves} for a mock model created with \texttt{dysmalpy} up to 3 times the disk effective radius. Top figure: rotation curve (left) and velocity dispersion (right) with the \texttt{dysmalpy} mock model in black squares extracted in circular apertures given a PSF HWHM of $0.25"$, and the \texttt{RotCurves} best fit is shown in the red solid line. The residuals are shown in the bottom panels for a flat 5\% uncertainty on the \texttt{dysmalpy} points for the purpose of the fit. Bottom figure: corresponding corner plot for the 5 fitted parameters: $R_\mathrm{eff}, M_\mathrm{baryon}, M_\mathrm{vir}, \mathrm{B/T}, \sigma_0$ and the inferred dark matter fractions $f_\mathrm{DM}(R_\mathrm{eff})$. The median and MAP of the posterior is shown in the red and green lines, respectively, and the true value of the model in a pink line. the dashed lines show the 16$^{\rm th}$ and 84$^{\rm th}$ percentiles. The covariance panels show some degeneracies between the model parameters, namely between the halo mass and both baryon mass and disk size, or the halo mass and the B/T ratio. }
    \label{fig:fiducial_model_fitting}
\end{figure*}

The best fit values are given in Table~\ref{tab:fiducial_bestfit_mcmc} for both the median and maximum (MAP) of the posterior, and the rotation curves for the median values are shown in Figure~\ref{fig:fiducial_model_fitting}, along with the MCMC corner plot. The resulting rotation and dispersion curves are an excellent fit and \texttt{RotCurves} recovers the model parameters well. The differences in the intrinsic rotation curve transform into slight systematics in the results of the fit: the effective radius of the disk is overestimated by almost a factor of $1.4$, the baryon mass is slightly overestimated and the halo mass is lower than its true value. These are the result of \texttt{RotCurves} finding the combination of parameters that effectively minimize the differences of the two beam smearing procedures (see bottom of Figure~\ref{fig:fiducial_rc_comparison}). As such, there are some notable degeneracies between the parameters of the model. The bottom of Figure~\ref{fig:fiducial_model_fitting} shows the posterior distribution and covariances, from which we can see a degeneracy between the halo mass $M_\mathrm{vir}$ and both $M_\mathrm{baryon}$ and $R_\mathrm{eff}$. This is understood when we consider how they affect one another - larger disk sizes result in RCs that fall further out, requiring less dark matter to reach similar velocities. The same goes with the baryon mass, as it simply scales the velocities higher.

\begin{figure*}[t]
    \centering
    \includegraphics[width=0.85\linewidth]{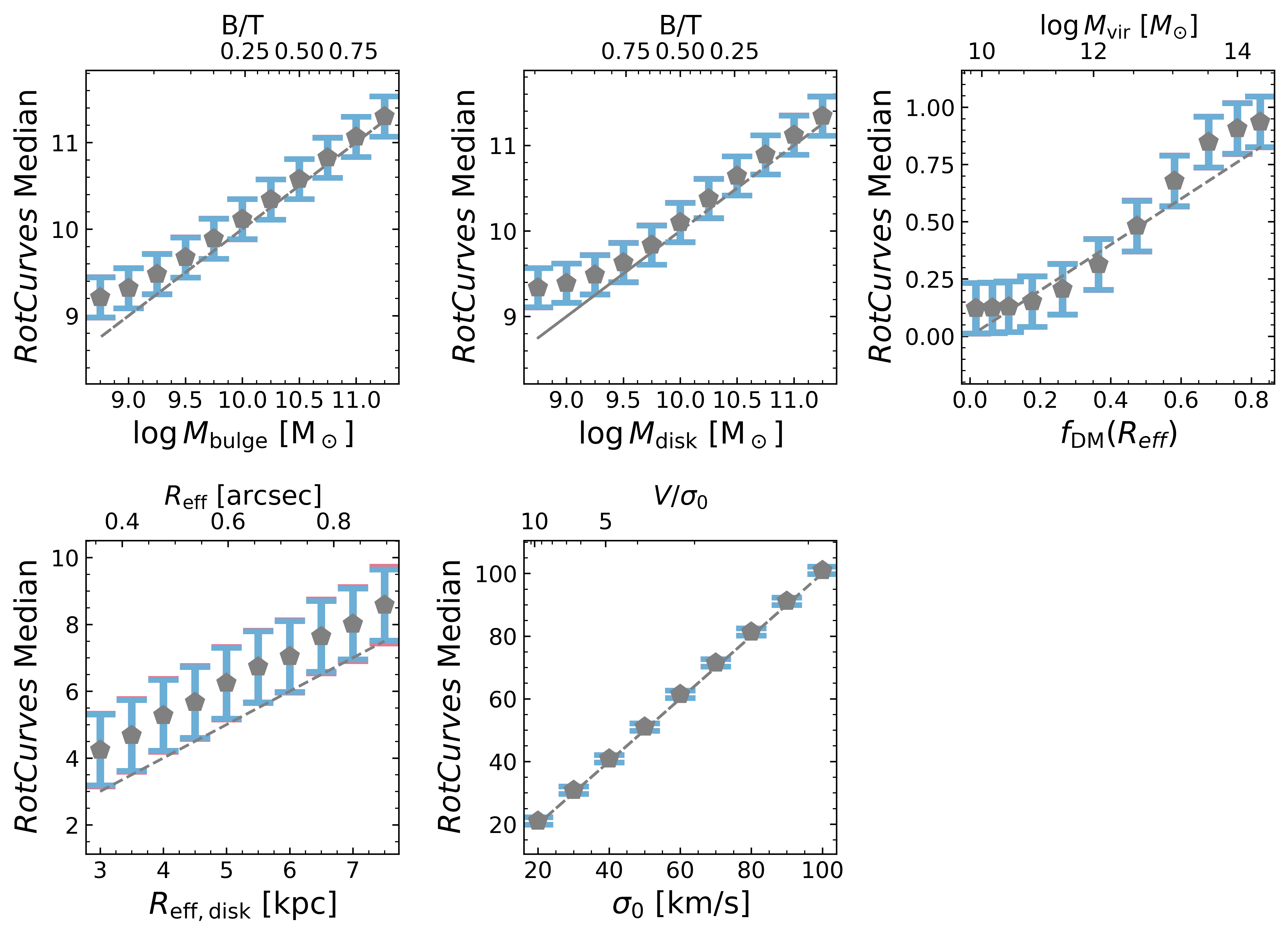}
    \caption{Model parameters inferred by \texttt{RotCurves} when fitting for a rotation curve created with \texttt{dysmalpy}. At each value the fit was run $n_{\rm fits}=10$ separate times, and the pentagons show the mean of best fit value of these runs. The blue error bar marks the average MCMC confidence interval shown, and the dashed line is a one-to-one ratio. When significant, the statistical error $\Delta _{stat} = \sigma_{\rm std}/\sqrt{n_{\rm fits}}$ is shown in an additional red error bar.}
    \label{fig:param_variation_medians}
\end{figure*}

Next, we vary each of the five main parameters of the model independently ($\log {M_{\rm disk} }, \log {M_{\rm bulge} }, B/T, R_{\rm eff,disk}, \sigma_0, M_{\rm vir}$) from the initial fiducial values, keeping the rest of the parameters fixed, and fit the data generate by \texttt{dysmalpy} as described previously. At each step we perform $n_{\rm fits}=10$ independent MCMC fits and take the mean of the best fit median and best fit MAP, verifying the statistical uncertainties are negligible. Figure~\ref{fig:param_variation_medians} shows the comparison for the \texttt{RotCurves} best fit median estimator. We can see that for a wide variety of parameters, \texttt{RotCurves} recovers the model parameters accurately, well within the confidence intervals of the MCMC fit. For brevity, we do not include the fitting results for all combinations, but some examples can be found in Appendix~\ref{appendix:param_variation_plots} However, there is a strong systematic error for recovering the disk effective radius $R_{\rm eff, disk}$, where \texttt{RotCurves} recovers higher values by a factor of $\sim 1.5$. This can be explained by looking at the velocity residuals in Figure~\ref{fig:fiducial_rc_comparison}, as differences in velocities are compensated by increasing the effective radius of the disk. For larger $R_{\rm eff,disk}$, the velocity peaks further out, so the values obtained at the actual $R_{\rm eff,disk}^{dysmalpy}$ will match better. Another, weaker systematic is when the bulge dominates over the disk, i.e. $M_{\rm disk} / M_{\rm bulge} \lesssim 0.3$, and the disk mass is poorly inferred in such that \texttt{RotCurves} can not accurately probe below $M_{\rm disk} / M_{\rm bulge} \approx 0.2$ (or $\mathrm{B/T} \gtrsim 0.8$).

\subsection{Effects of noise}\label{sec:mock_observation}
To simulate more realistic data, we would need to consider how noise affects the mock-observed rotation curve from a given model. Because \texttt{RotCurves} does not input a full data cube, we cannot add noise in the underlying \texttt{dysmalpy} model without relying too heavily on the line fitting methods employed there. Instead, we adopt a simpler approach to achieve behavior similar to observed data, in which: (i) for smooth distributions the S/N follows the flux, peaking at the center and dropping with radius, (ii) the noise level $\sigma_{\rm noise}$ is spatially uniform, and (iii) the noise is drawn from a Gaussian distribution. Given that, we can reconstruct the emission line at a range of apertures, $r_{\rm ap}$, separated by the size of the PSF FWHM, as pure Gaussian line profiles. The line centers, $V_{\rm rot}$, and standard deviation, $\sigma$, are extracted from the \texttt{dysmalpy} modelling. The line is initially constructed with a fine spectral width of 10 km/s, then convolved over a Gaussian LSF and binned to a degraded spectral resolution $\Delta v$, mimicking the instrumental channel width, with an additional random phase of $\Delta v / 6$. Next, as the noise-rms can vary slightly between spectral channels, we add a $10\%$ variation of the noise level $\sigma_{\rm noise}$ in each channel, before drawing a Gaussian noise for each channel. Finally, the line profile is fitted with a Gaussian to recover the centroid velocity $V_{\rm dys,nosie}$ and velocity dispersion $\sigma_{\rm dys,noise}$, at each aperture.

We set the S/N level as the ratio between the line peak and the noise standard deviation $\sigma_{\rm noise}$ at the effective radius $R_{\rm eff, disk}$. The radial flux profile $I_0(r_{\rm ap})$ traces the mass of the disk (assuming $\mathrm{M/L} = 1$ in the fiducial model), and so the line peak will drop exponentially with radius, leading to higher S/N in the center and lower S/N as one goes outwards. If we normalize to the central flux, we simply get an exponential profile $I_0(r) = e^{-r/R_{\rm d}}$. In that case, the S/N is:
\begin{equation}
    \mathrm{S/N} = \frac{I_0(R_{\rm eff,disk})}{\sigma_{\rm noise}} \approx \frac{0.186}{\sigma_{\rm noise}} 
\end{equation}
where we used $R_{\rm eff,disk} / R_{\rm d} = 1.68$ for an exponential disk. 
We define the limit of the observation as the point where the S/N drops below 3, which can be written as:
\begin{align*}
    r_{\rm edge} &= R_{\rm eff, disk} \left[ 1 + \frac{1}{1.68} \ln { \left( \frac{\rm S/N}{3} \right)} \right] \\
    & \approx R_{\rm eff, disk} \left[ 2 + \ln{ \left( \rm \frac{S/N}{16} \right) } \right] \ \ \ .
\end{align*} 

\begin{figure}
    \centering
    \includegraphics[width=\columnwidth]{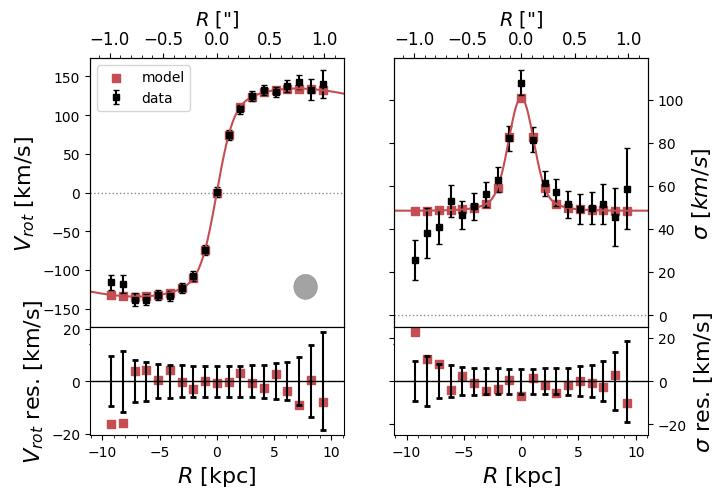}
    \caption{Results of the MCMC fit using \texttt{RotCurves} (red line, labeled ``model") to for data generated with \texttt{dysmalpy} included added noise with $\mathrm{S/N}=25$, $\Delta v = 35 \kms$ and $\sigma_\mathrm{LSF}=30 \kms$ (black squares, labeled ``data"), for our fiducial model. The deviation and uncertainty caused by the additional noise increase outwards, until they fall below $\mathrm{S/N} < 3$. }
    \label{fig:mock_mcmc_rc_SN_25}
\end{figure}
\begin{figure}
    \centering
    \includegraphics[width=\columnwidth]{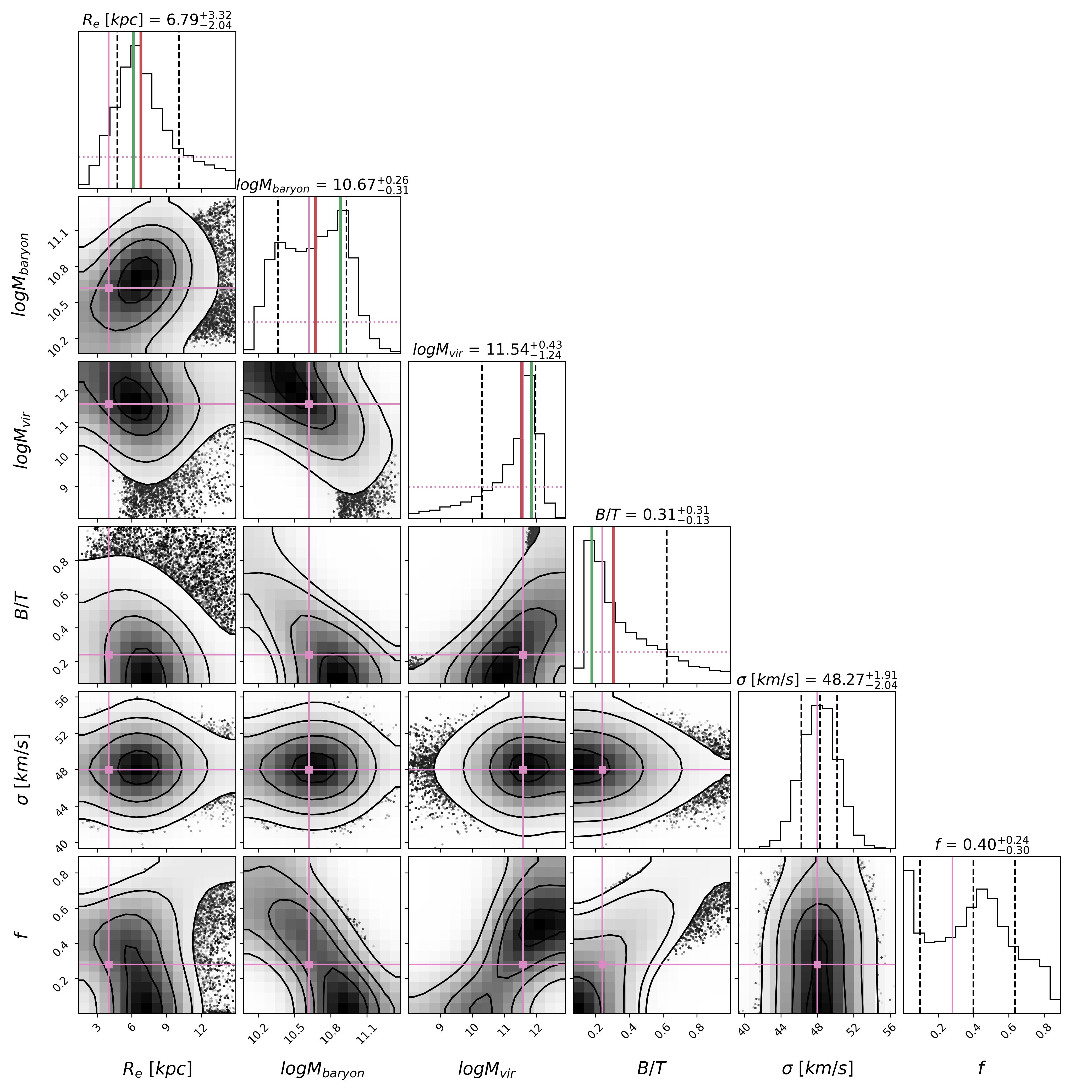}
    \caption{Corner of the MCMC fitting results for the fit shown in Figure~\ref{fig:mock_mcmc_rc_SN_25}, for $\mathrm{S/N}=25$, $\Delta v=35 \kms$, $\sigma_\mathrm{LSF}=30 \kms$. The recovered parameters are recovered well given the posterior distribution, with $\mathrm{B/T}$ and $f_\mathrm{DM}$ showing a bias for overestimating the true value.}
    \label{fig:mock_mcmc_cornerplot_SN_25}
\end{figure}
At each S/N level we generate 10 individual mock-data, and fit each one using \texttt{RotCurves}, so we can average the random effects of the noise addition process. 

Figures~\ref{fig:mock_mcmc_rc_SN_25}-\ref{fig:mock_mcmc_cornerplot_SN_25} show the results of an MCMC fitting for $\mathrm{S/N}=25$ (and $\Delta v=35 \kms$, $\sigma_\mathrm{LSF}=30 \kms$). The effect of the noise is clear on the mock data (black squared). At the outer RC, where the S/N is smaller, the line fitting becomes less secure and the uncertainties increase. The scatter in the mock data points to the centroid shifting around its ``true" value, increasing as the S/N drops outwards. In the velocity dispersion, the width of the fitted line shifts more dramatically in the outer regions, even creating a completely artificial drop signature in one end. The best fit model generated with \texttt{RotCurves} is drawn in a red line, and the residuals show no clear systematic bias. The physical parameters are recovered well, albeit with large posterior uncertainties especially for $M_\mathrm{baryon}$, $\mathrm{B/T}$ and $f_\mathrm{DM}$. This result is encouraging, as we have assumed no prior knowledge at all for the free parameters and let them vary freely in the parameter space. In real observations, some parameters can usually be constrained (e.g. baryonic mass from $M_\star$, the effective radius from imaging of the stellar continuum, or B/T ratios from aperture photometry) such that only smaller regions are explored and the posterior distribution becomes narrower.

\begin{figure*}
    \centering
    \includegraphics[width=0.70\linewidth]{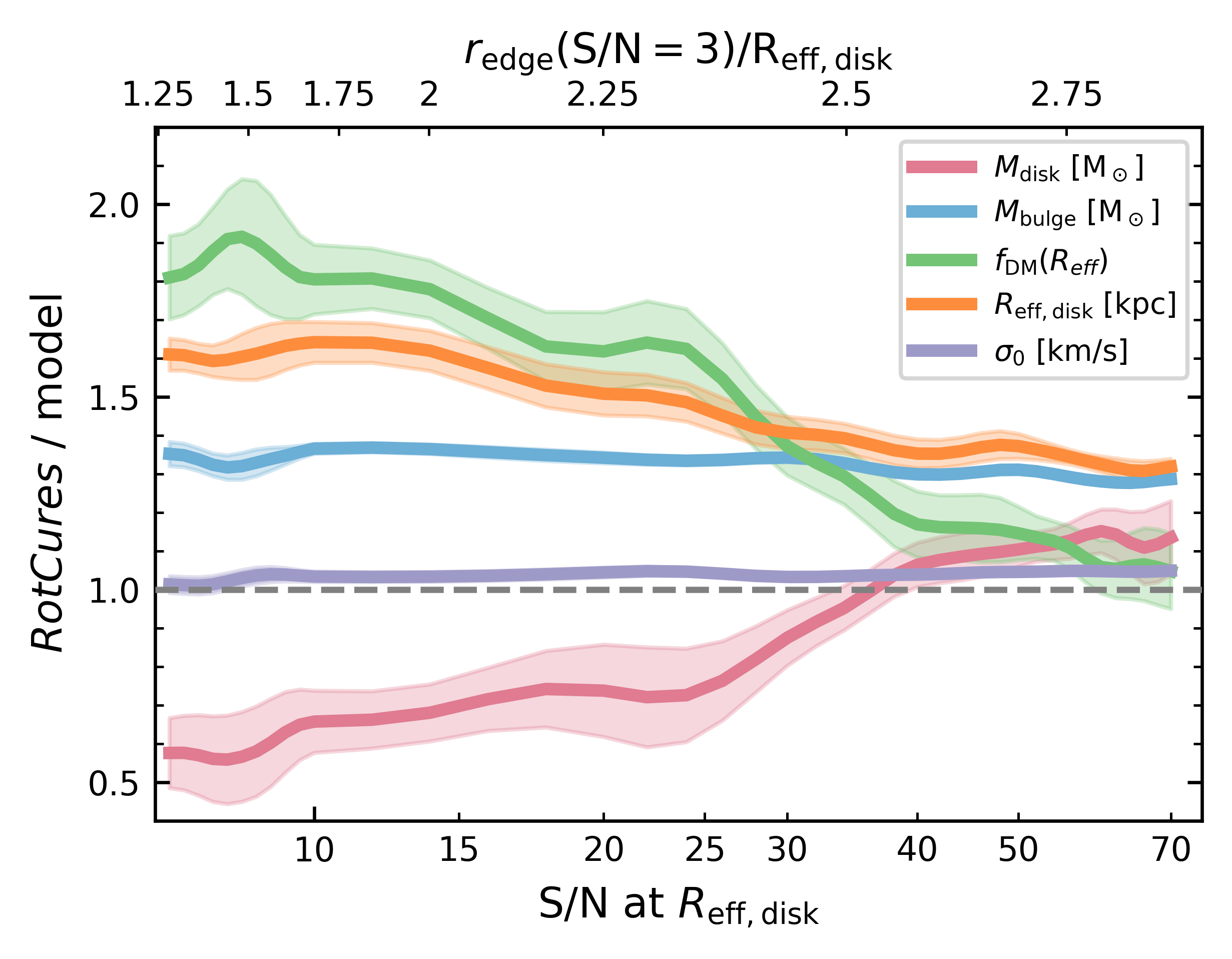}
    \caption{Best fit values obtained while fitting a mock-observed rotation curve created with \texttt{dysmalpy}, and fitted using the MCMC fitter with \texttt{RotCurves}. The y-axis shows the difference in each parameter $X$ as $\Delta X = X^{fit} / X^{model}$. The S/N is defined by the ratio of the line peak to the noise level at the disk effective radius, and the edge of the mock-observation is where the S/N drops below 3. At higher S/N, the difference decreases as the mock RC probes the outer parts of the RC, better constraining the model parameters.}
    \label{fig:mock_observation_fit_median}
\end{figure*}

We explore the trend on the parameter recovery based on S/N level in Figure~\ref{fig:mock_observation_fit_median}. The line shows a running median and the shaded region shows the statistical scatter over the 10 realizations at each S/N level. \texttt{Rotcurves} improves with increasing S/N for all parameters, as more kinematic information is included as $r_{\rm edge}$ increases, with significant improvement at $\mathrm{S/N} > 25$, until it converges at the intrinsic biases discussed in previous sections. Interestingly, the recovery of the bulge mass $M_{\rm bulge}$ and velocity dispersion $\sigma_0$ are not sensitive to changes in S/N, as they can be well determined by the inner RC alone. Yet, as we already discussed earlier, the systematic overestimation of the bulge is evident and does not improve with higher S/N.

The parameters most affected are the disk mass $M_{\rm disk}$ and dark matter fractions $f_{\rm DM} (R_{\rm eff})$, with great improvements as S/N$\gtrsim 25$ (corresponding to $r_{\rm edge} \gtrsim 2.3$. At high S/N $\gtrsim 40$, or $r_{\rm edge} \gtrsim 2.6$, both of them are recovered very well, with no apparent systematic biases. Reaching far out into the RC is crucial for a reliable interpretation of the parameters, as in the low S/N regime the disk is underestimated by a factor $\sim \times 0.5-0.7$ and the dark matter fraction (a proxy for the halo mass) is overestimated by $\sim \times 1.5-2$. Finally, the effective radius has only a slight improvement with S/N, with a clear systematic bias of larger disk sizes, by a factor of $\sim \times 1.5$ (similar to the values discussed in \S~\ref{sec:mcmc_fitting}). 

\section{Conclusions}\label{sec:discussion}
In this paper we presented \texttt{RotCurves}\footnote{https://github.com/nestoramit/RotCurves} as a tool to carry out rotation curve modelling, allowing for fast exploration of various assumptions on mass models. The primary goal of this work has been to test the performance of our beam smearing correction, performed by projecting the beam in the plane of the disk (see \S~\ref{sec:sky_projection}), against the full 3D treatment performed by the code \texttt{dysmalpy}. We use \texttt{dysmalpy} as the choice for comparison as it is also a forward mass modelling tool, assuming analytical function for the mass components. Such benchmarking exercises are critical for ensuring reliability and reproducibility in kinematic studies, particularly as the field moves toward larger data sets from future IFU surveys. \texttt{RotCurves} created a beam-smeared rotation curve in a few $10$ms, which is about $200-300$ faster than \texttt{dysmalpy}. By systematically comparing the two codes under controlled conditions, we find small differences in the mock observed rotation curves, and show that \texttt{RotCurves} can recover well the intrinsic model parameters, even when including a more realistic noise signal. Performance is impacted negatively mostly for low S/N and large beam sizes.

We first compare the beam smearing treatment between \texttt{RotCurves} and \texttt{dysmalpy} for an identical model galaxy (the ``fiducial model"), with typical values for a main sequence SFG at $z=2$, and the galaxy is well resolved with a PSF FWHM of half an effective radius. We find that despite the methodological differences, both mock observed rotation curves agree to within $5\%$ ($\approx 7 \kms$) across all radii. As expected, discrepancies are largest in the inner regions where the velocity gradients are steepest due to the presence of the bulge, and become smaller beyond the effective radius. The same is true for the velocity dispersion, where both approaches agree to within $5\%$ ($\approx 5 \kms$), with the largest discrepancy in the center. Due to the 2D approach of \texttt{RotCurves}, the velocity dispersion is always under-estimated compared to \texttt{dysmalpy}, whereas the velocity difference varies. The PSF size has the greatest effect on the rotation curves, and as the system becomes less resolved differences are much larger, up to $\approx 15 \%$. This is most notable in marginally resolved systems, when the PSF FWHM becomes larger than one effective radius. Other parameters, such as changing the mass of the disk or the bulge, the DM fractions, intrinsic velocity dispersion or inclination have little effect on the residual values.

\texttt{RotCurve} recovers the intrinsic model parameters well for a mock-observed rotation curve generated with \texttt{dysmalpy}. We use an MCMC approach to recover the posterior distribution simultaneously for five model parameters: the total baryon mass $M_{\rm baryon}$, the disk effective radius $R_{\rm eff}$, the bulge-to-total ratio $B/T$, the dark matter fraction $f_{\rm DM} (R_{\rm eff})$, and the intrinsic velocity dispersion $\sigma_0$. We vary each parameter over a wide range of values (see figure~\ref{fig:param_variation_medians}), performing 10 individual fits for each value for a total of $\sim$550 individual runs over the entire parameter space. Each fitting procedure has 100 walkers and 3000 steps (of which 500 are burn-in steps), which is $\approx 40-50$ times the autocorrelation time, and assuming uniform priors and randomized initial phases at each run. The fit procedure is robust, as the best fit values have very little scatter and negligible statistical errors. The posterior distribution uncertainty is significant, but the recovered median (and maximum a-posteriori, MAP) values are in very good agreement with the intrinsic values. However, the disk effective radius was systematically inferred to be larger ($\approx 50\%$), and well outside the 1$\sigma$ uncertainty for $R_{\rm eff} > 6\, \rm kpc$. Similarly, the fit increasingly misses for disk masses below $M_{\rm disk} < 10^{9.5} \Msun$, overestimating the disk mass by $\approx 10\%$. Yet, overall, \texttt{RotCurves} can accurately recover the intrinsic parameters for a wide range in mass models.

Finally, we test the performance of \texttt{RotCurves} under realistic observational conditions by including a line spread function and adding Gaussian noise to the \texttt{dysmalpy}-generated velocity maps at levels typical of current IFU surveys (see figure~\ref{fig:mock_observation_fit_median}). We explore signal-to-noise ratios ranging from S/N$5-70$, that we define by the peak flux over a beam element centered at the effective radius. At each S/N level we generate a mock-data from \texttt{dysmalpy} which we fit using \texttt{RotCurves}. At high S/N the recovered values are in excellent agreement with the intrinsic parameters, with the same intrinsic  biases discussed earlier, and as the S/N drop we start seeing larger discrepancies. The bulge mass $M_{\rm bulge}$ and velocity dispersion $\sigma_0$ have almost no dependency on S/N, whereas the disk mass $M_{\rm disk}$, DM fractions $f_{\rm DM} (R_{\rm eff})$ and effective radius $R_{\rm eff}$ start to deviate as S/N$\lesssim 25$. However, even at very low S/N we recover reasonable values that can be used as an estimate. These results demonstrate that \texttt{RotCurves} can perform reliably for the majority of observations, but we advise to be cautious when interpreting results from poorly resolved, low S/N systems - a regime that is unfortunately common in high-redshift studies.

Looking forward, \texttt{RotCurves} is designed to serve multiple roles in the analysis of galaxy kinematics. As an exploratory tool, it enables rapid testing of different mass model assumptions—such as varying dark matter profiles, baryon distributions, or including/excluding bulge components—without the computational overhead of full 3D modeling. This makes it particularly useful in the early stages of analysis when the appropriate model complexity is not yet clear, or when investigating systematic effects from different modeling choices. For real observational data, \texttt{RotCurves} can provide quick parameter estimates and identify potential issues (e.g., warps, asymmetries, or kinematic substructure) before committing to more computationally expensive fitting procedures.
Perhaps most critically, as we enter the era of large ground-based IFU surveys (e.g., VLT/MAVIS) or space telescopes (e.g. JWST/NIRSpec), and next generation ground-based instruments (e.g., ELT/MOSAIC, GMT IFS), computational efficiency will become essential to effectively analyze the data. The ability to process large samples with consistent methodology, while maintaining the flexibility to adapt models to different mass regimes and redshifts, positions \texttt{RotCurves} as a valuable tool for future statistical studies of galaxy dynamics and evolution. The code is publicly available on \href{https://github.com/nestoramit/RotCurves}{github}.

\section*{Acknowledgements}
We thank the referee for their insightful comments and helpful suggestions.
ANS and AS are supported by the Center for Computational Astrophysics (CCA) of the Flatiron Institute, and the Mathematics and Science Division of the Simons Foundation, USA. We thank the German Science Foundation (DFG) for support via German-Israel Project (DIP) grant STE/1869-2 GE 625/17-1.
H{\"U} acknowledges funding by the European Union (ERC APEX, 101164796).
NMFS, CB, GT, JMES and JC acknowledge funding by the European Union (ERC Advanced Grant GALPHYS, 101055023). Views and opinions expressed are, however, those of the authors only and do not necessarily reflect those of the European Union or the European Research Council. Neither the European Union nor the granting authority can be held responsible for them.

\section*{Data availability}
No new data were generated or analysed in support of this research. The software described in this paper is publicly available at https://github.com/nestoramit/RotCurves.

\section*{Software}
This paper makes use of the publicly available \texttt{PYTHON} packages: \texttt{scipy} \citep{scipy2020}, \texttt{numpy} \cite{numpy2011}, \texttt{matplotlib} \citep{matplotlib2007}, \texttt{seaborn} \citep{seaborn}, \texttt{emcee} \citep{emcee}, \texttt{corner} \citep{corner} \texttt{pandas} \citep{pandas2022}, \texttt{Astropy} \citep{astropy2013, astropy2018, astropy2022} and \texttt{dysmalpy} \citep{Davies2004_dysmalpy, Davies2004b_dysmalpy, Davies2011, Cresci2009_dysmalpy, Wuyts2016_KMOS3d, Lang2017_stacking_rc, Price2021_rc41, Lee2025_dysmalpy}.



\clearpage

\appendix
\section{residuals for beam smearing comparison in physical units} \label{appendix:residuals_physical}
\begin{figure*}
    \centering
    \includegraphics[width=0.7\linewidth]{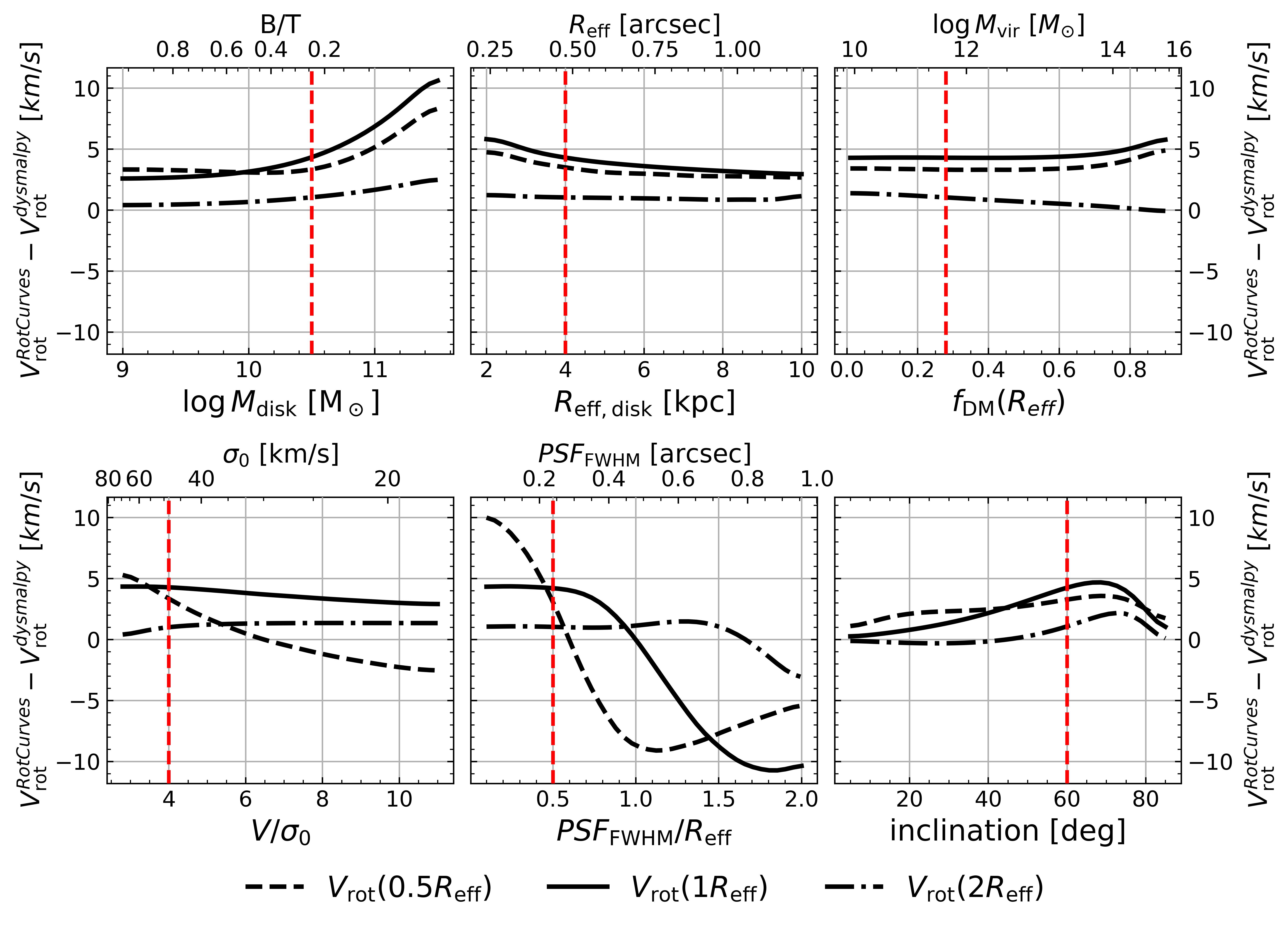}
    \includegraphics[width=0.7\linewidth]{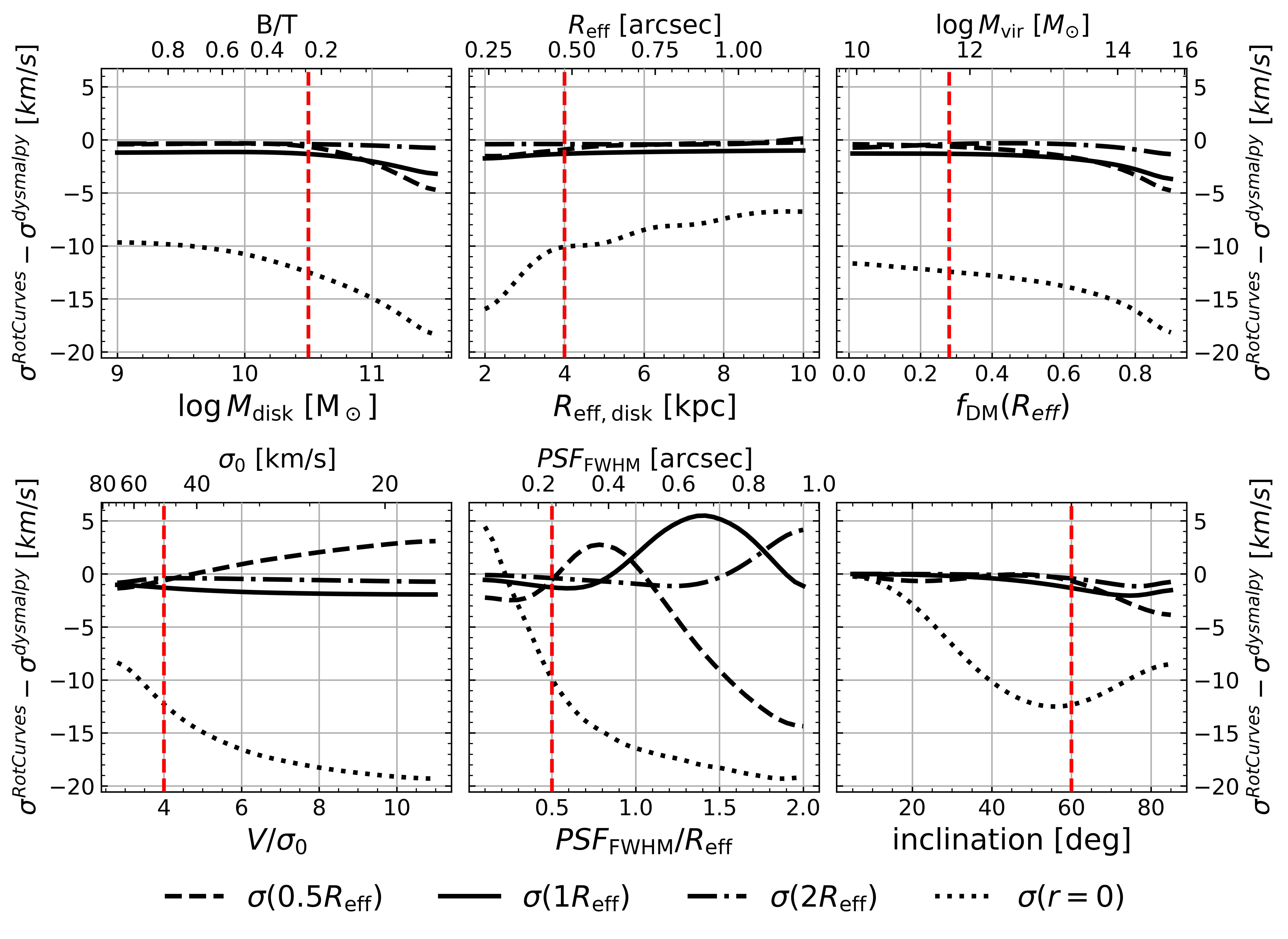}
    \caption{Same as Figure~\ref{fig:fiducial_paramstudy_pcnt} but with residuals in physical units of km/s.}
    \label{fig:fiducial_paramstudy_num}
\end{figure*}

Following the discussion in section \S~\ref{sec:benchmark - beam smearing}, Figure~\ref{fig:fiducial_paramstudy_num} presents the residuals between the beam-smeared rotation curves as a function of model parameters in absolute terms, without normalizing by the \texttt{dysmalpy} values (as is done in Figure~\ref{fig:fiducial_paramstudy_pcnt}).

\clearpage
\section{Fit results for various parameter combinations} \label{appendix:param_variation_plots}
\begin{figure*}
\centering
\medskip
\includegraphics[width=\columnwidth]{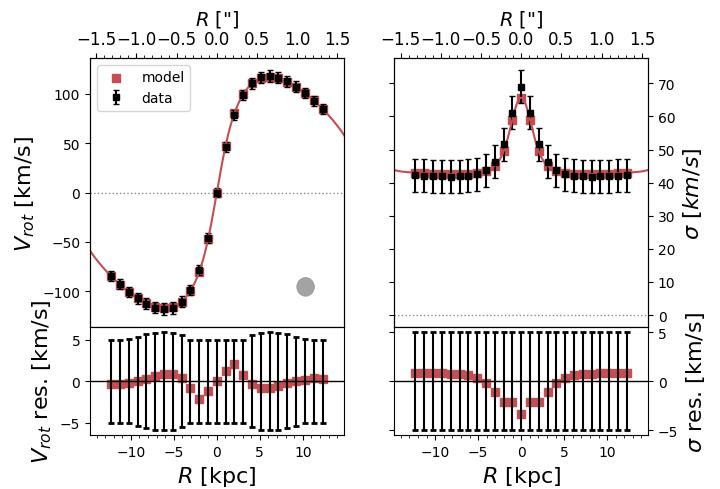}\hfill
\includegraphics[width=\columnwidth]{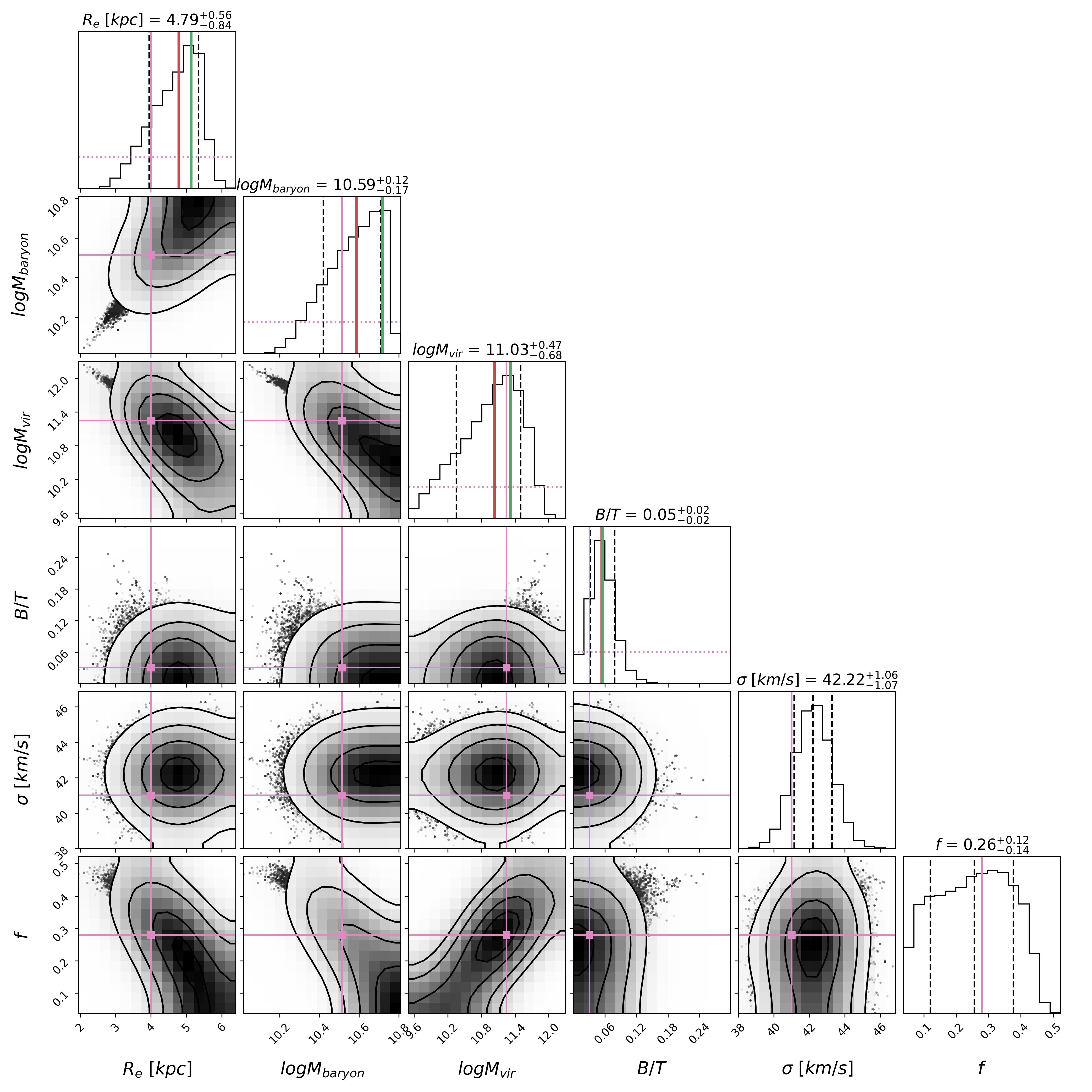}
\smallskip
\makebox[\textwidth][c]{$\log M_\mathrm{bulge} = 9$}
\medskip
\includegraphics[width=\columnwidth]{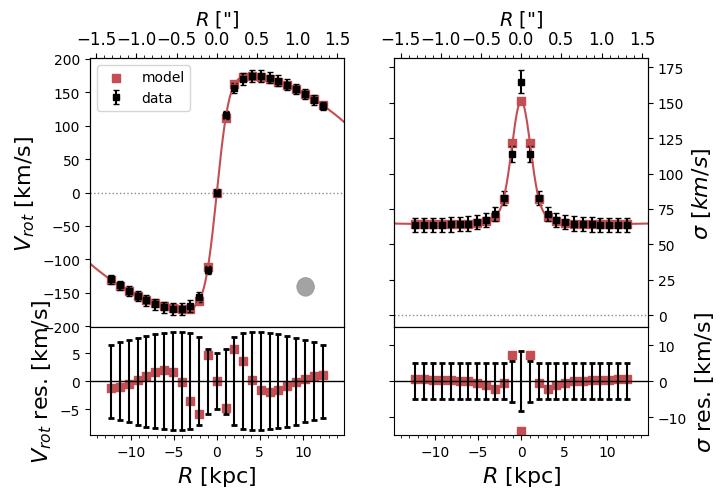}\hfill
\includegraphics[width=\columnwidth]{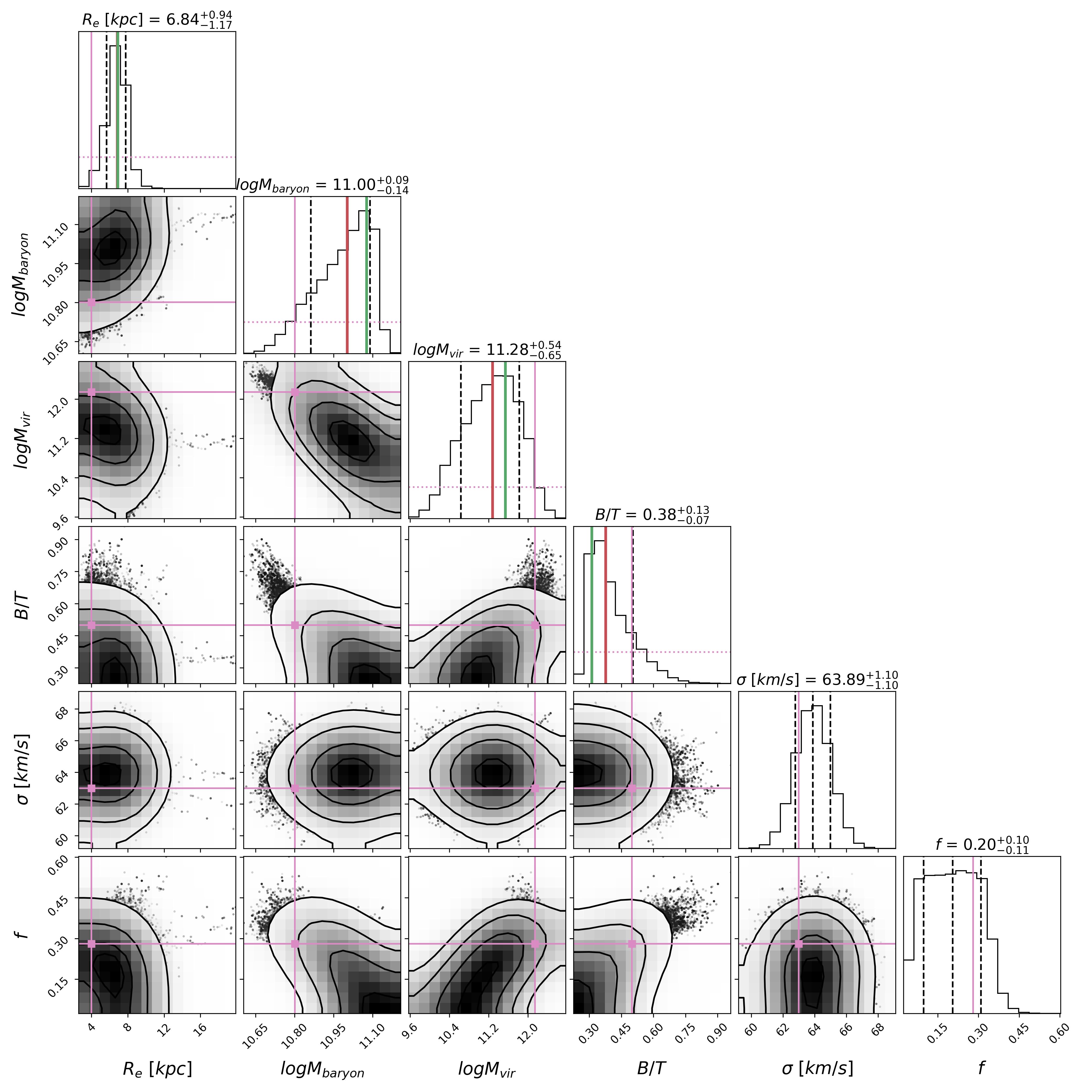}
\smallskip
\makebox[\textwidth][c]{$\log M_\mathrm{bulge} = 10.5$}
\caption{Two variation in bulge mass for the MCMC fit results with \texttt{RotCurves} on a model generated by \texttt{dysmalpy}.}
\label{fig:app_paramvariation_mbulge}
\end{figure*}

\begin{figure*}
\centering
\medskip
\includegraphics[width=\columnwidth]{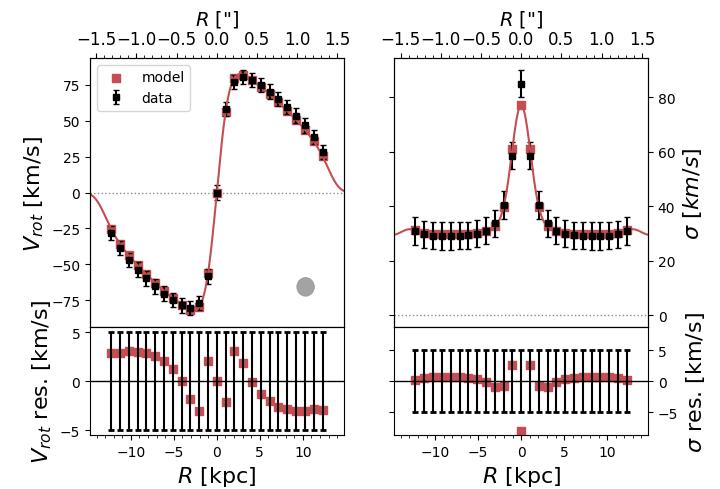}\hfill
\includegraphics[width=\columnwidth]{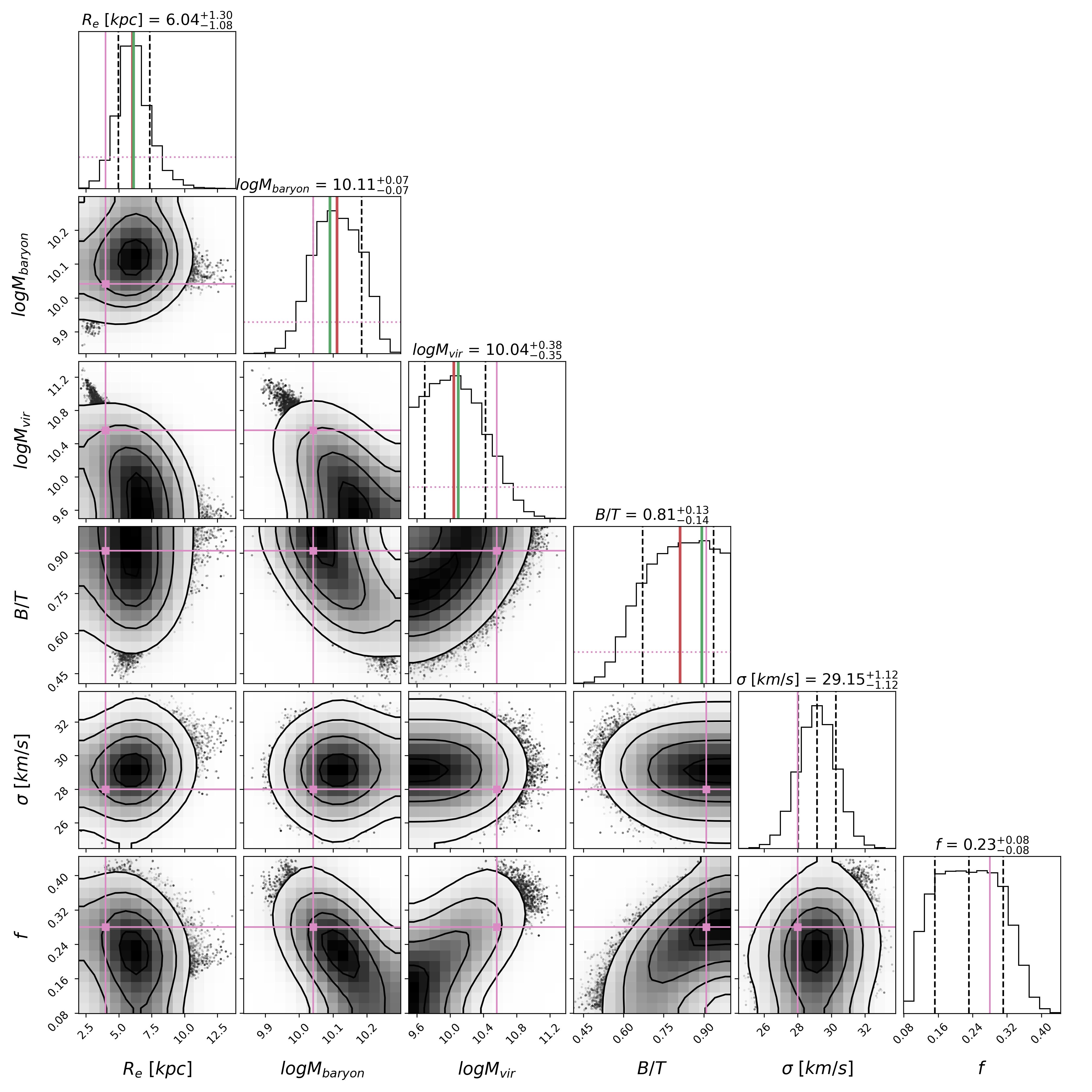}
\smallskip
\makebox[\textwidth][c]{$\log M_\mathrm{disk} = 9$}
\medskip
\includegraphics[width=\columnwidth]{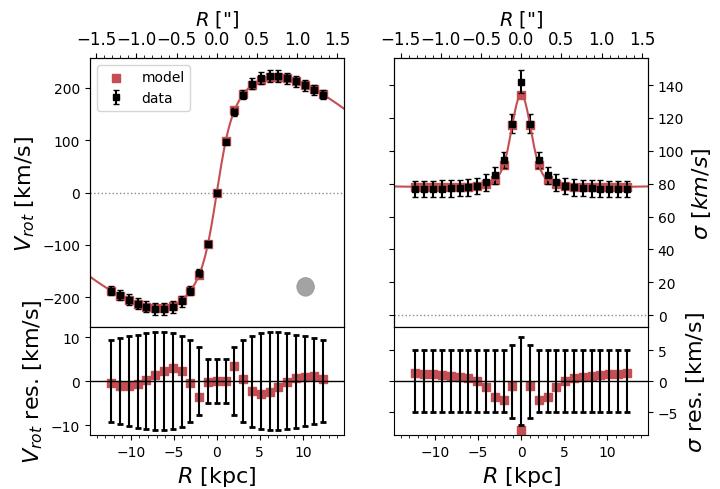}\hfill
\includegraphics[width=\columnwidth]{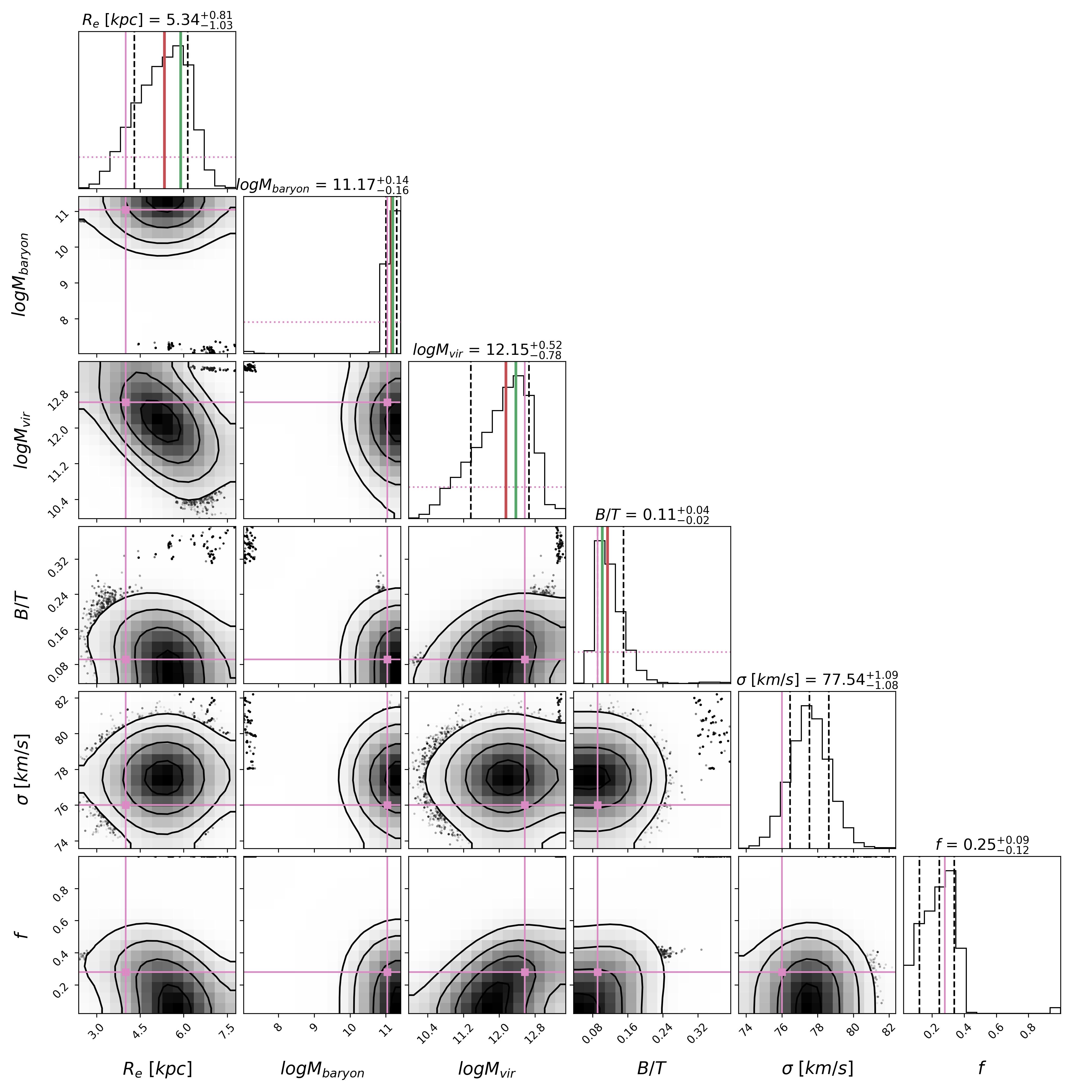}
\smallskip
\makebox[\textwidth][c]{$\log M_\mathrm{disk} = 11$}
\caption{Two variation in disk mass for the MCMC fit results with \texttt{RotCurves} on a model generated by \texttt{dysmalpy}.}
\label{fig:app_paramvariation_mdisk}
\end{figure*}

\begin{figure*}
\centering
\medskip
\includegraphics[width=\columnwidth]{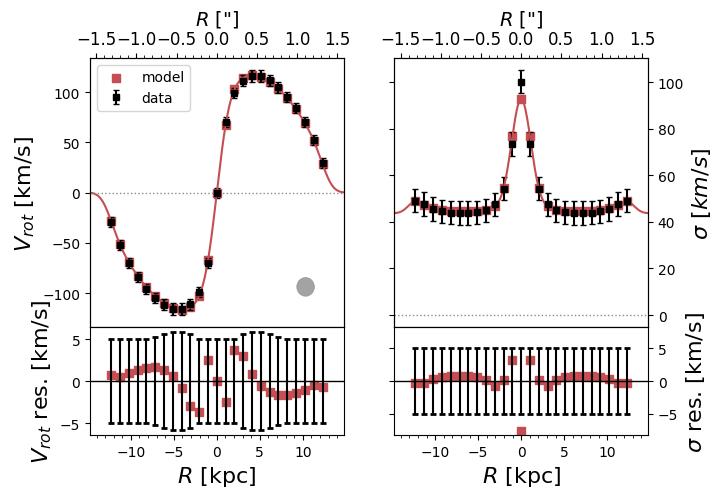}\hfill
\includegraphics[width=\columnwidth]{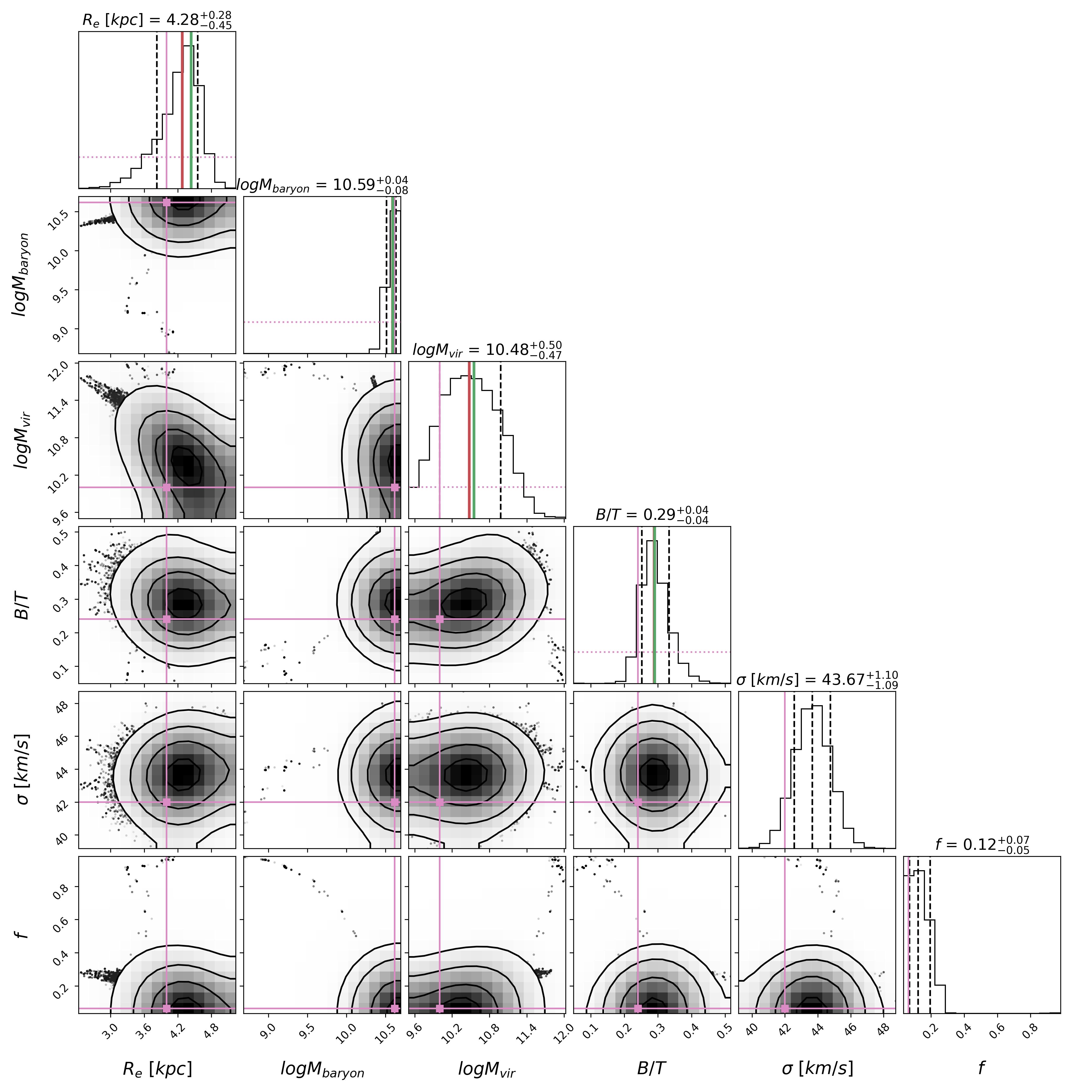}
\smallskip
\makebox[\textwidth][c]{$\log M_\mathrm{vir} = 10$, $f_\mathrm{DM}(R_\mathrm{eff}) = 0.06$}
\medskip
\includegraphics[width=\columnwidth]{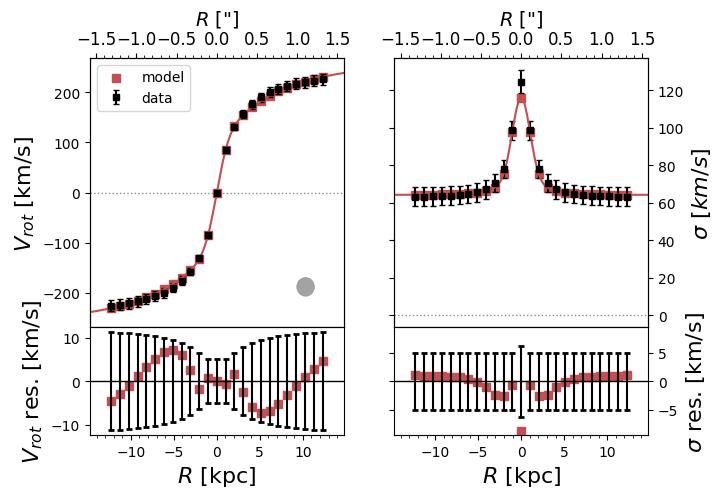}\hfill
\includegraphics[width=\columnwidth]{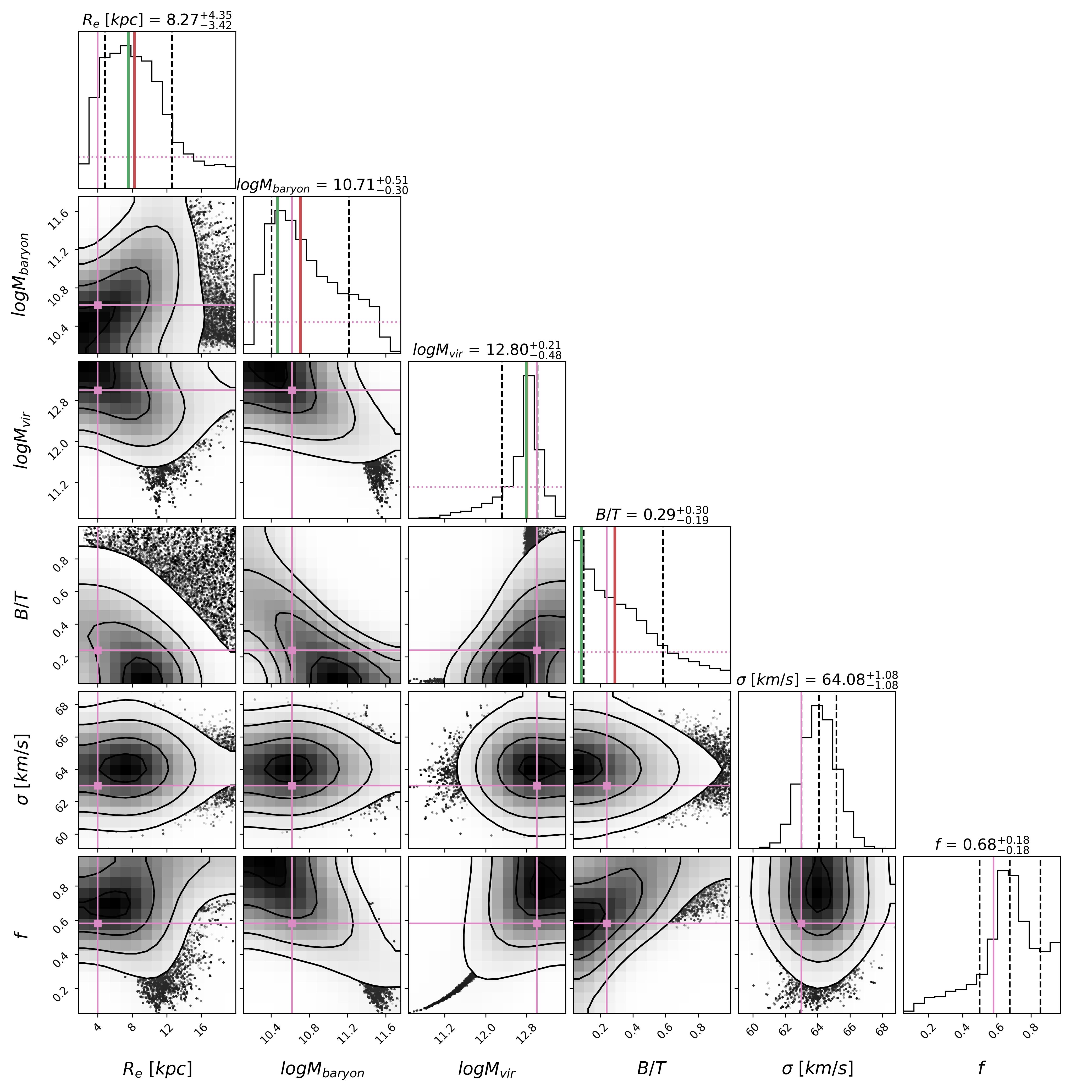}
\smallskip
\makebox[\textwidth][c]{$\log M_\mathrm{vir} = 13$, $f_\mathrm{DM}(R_\mathrm{eff}) = 0.58$}
\caption{Two variation in halo mass for the MCMC fit results with \texttt{RotCurves} on a model generated by \texttt{dysmalpy}.}
\label{fig:app_paramvariation_mvir}
\end{figure*}

\begin{figure*}
\centering
\medskip
\includegraphics[width=\columnwidth]{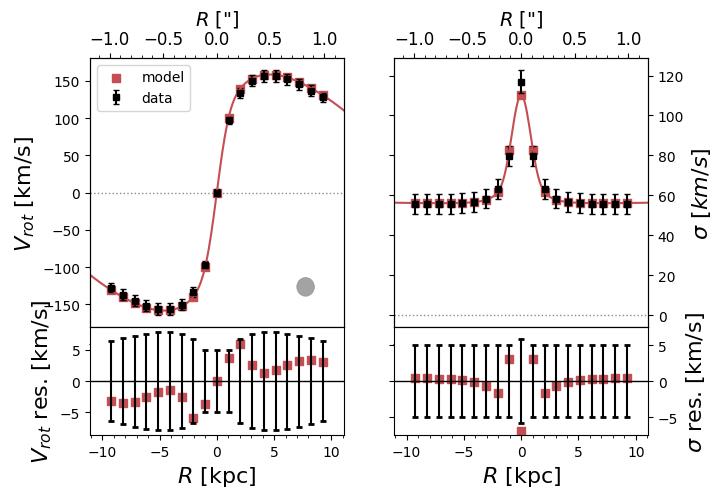}\hfill
\includegraphics[width=\columnwidth]{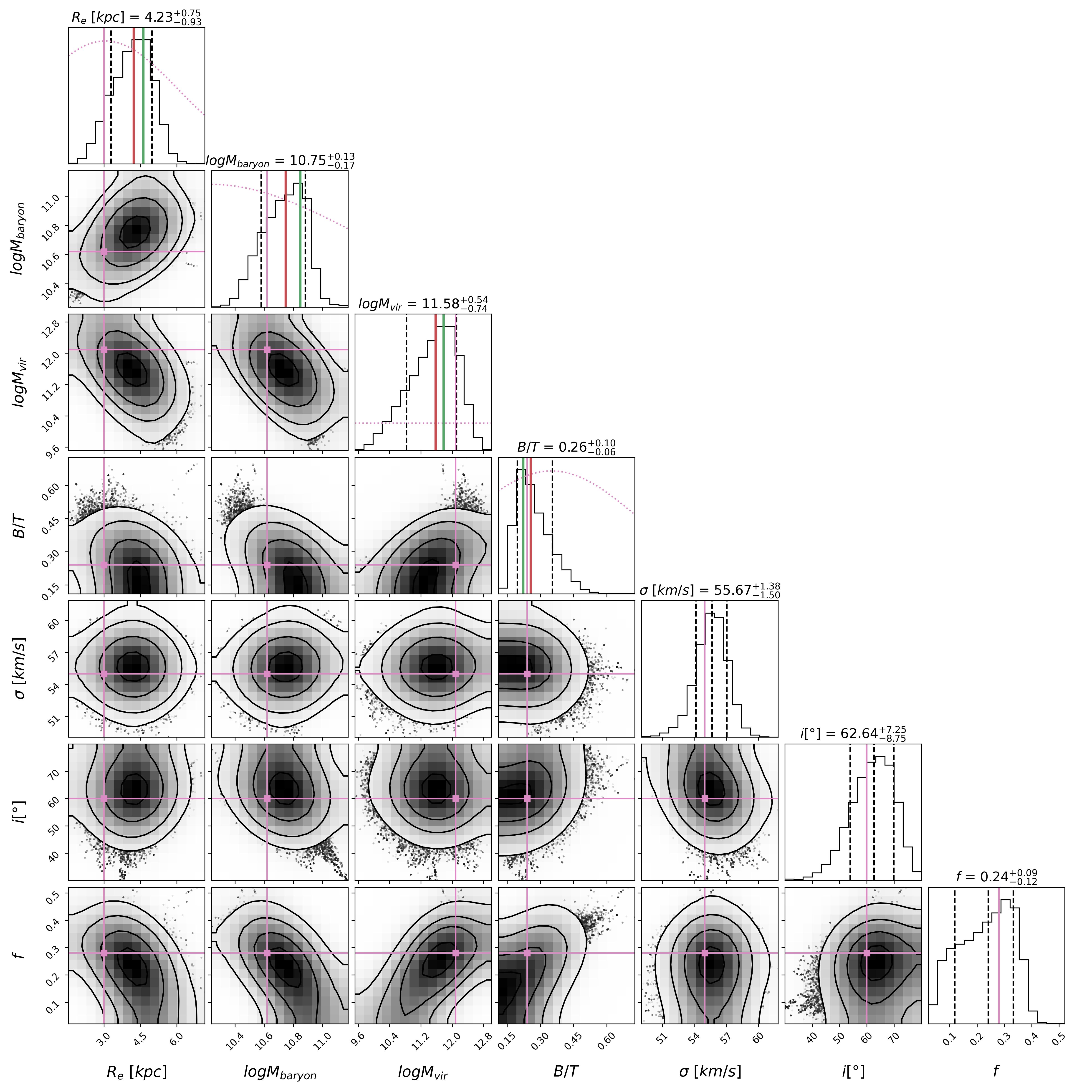}
\smallskip
\makebox[\textwidth][c]{$R_\mathrm{eff} = 3$ kpc}
\medskip
\includegraphics[width=\columnwidth]{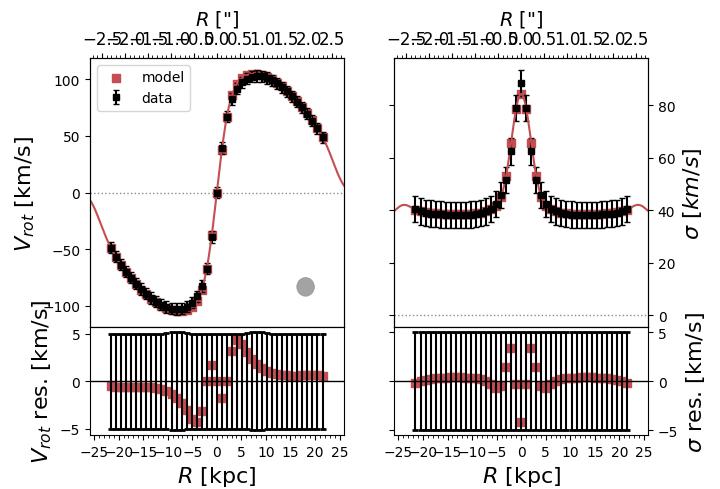}\hfill
\includegraphics[width=\columnwidth]{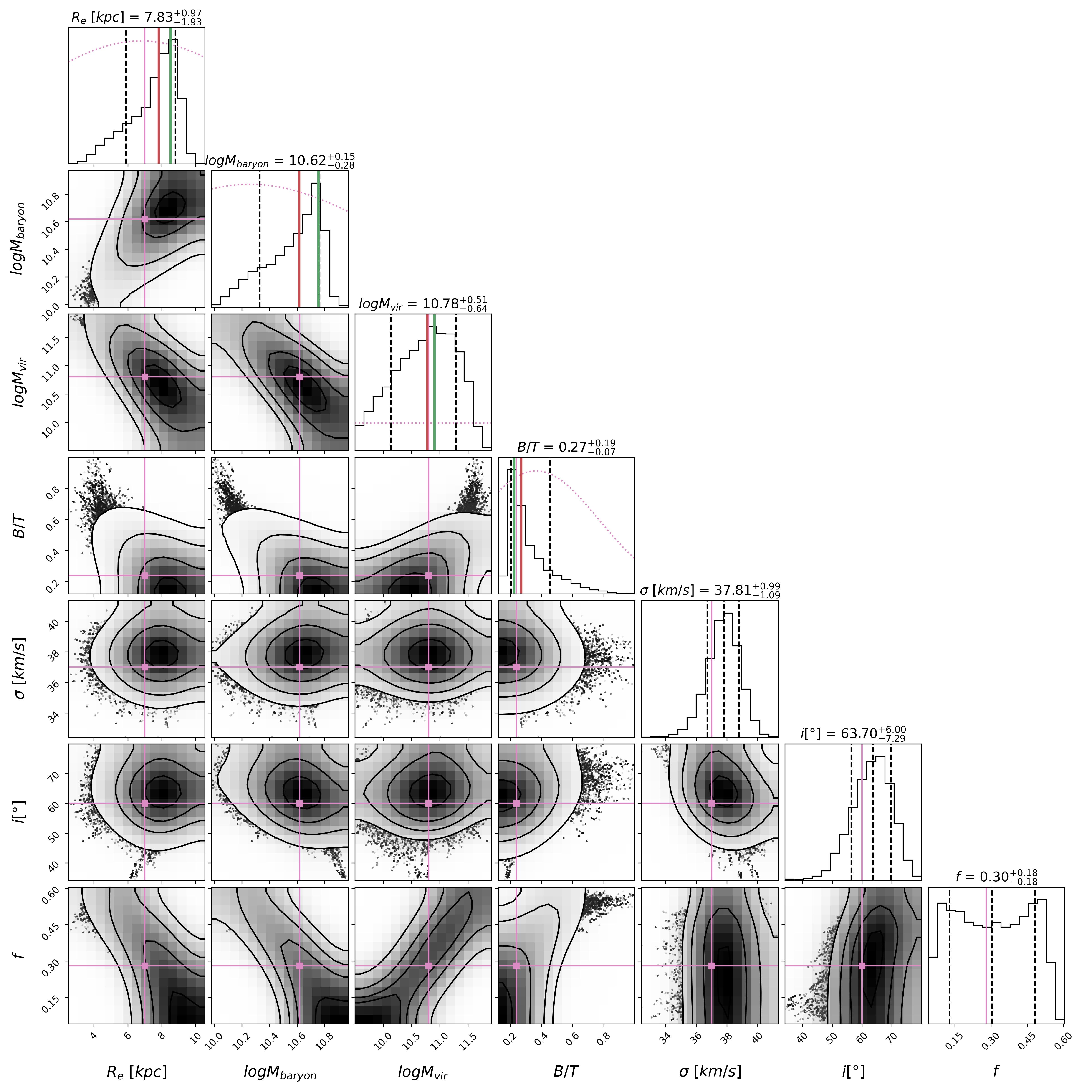}
\smallskip
\makebox[\textwidth][c]{$R_\mathrm{eff} = 7$ kpc}
\caption{Two variation in the disk effective radius for the MCMC fit results with \texttt{RotCurves} on a model generated by \texttt{dysmalpy}.}
\label{fig:app_paramvariation_reff}
\end{figure*}

\begin{figure*}
\centering
\medskip
\includegraphics[width=0.45\textwidth]{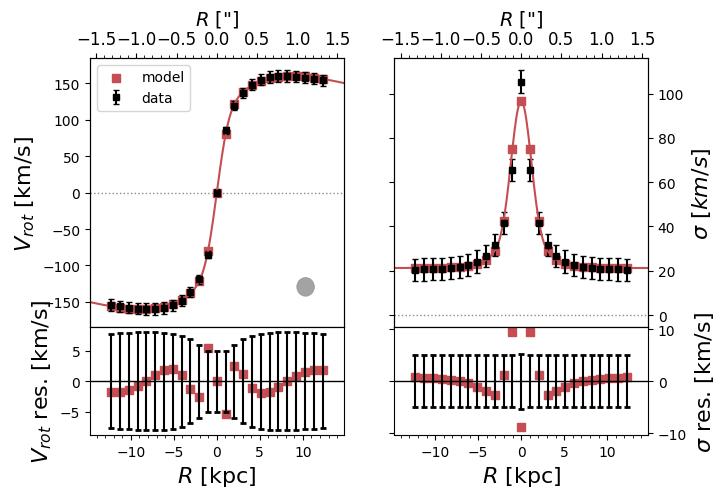}\hfill
\includegraphics[width=0.45\textwidth]{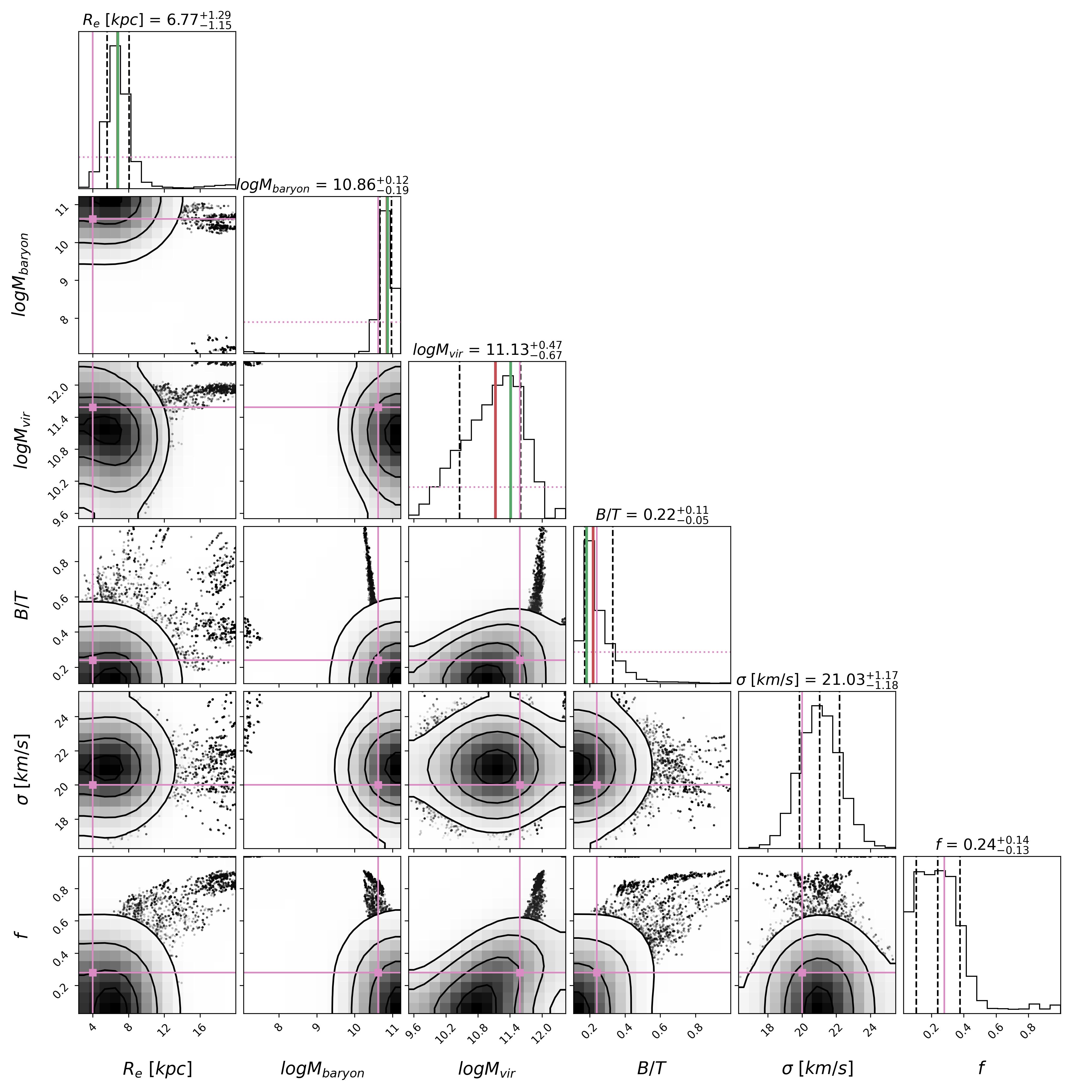}
\smallskip
\makebox[\textwidth][c]{$\sigma_0 = 20 \kms$, $V/\sigma_0 = 10$}
\medskip
\includegraphics[width=0.45\textwidth]{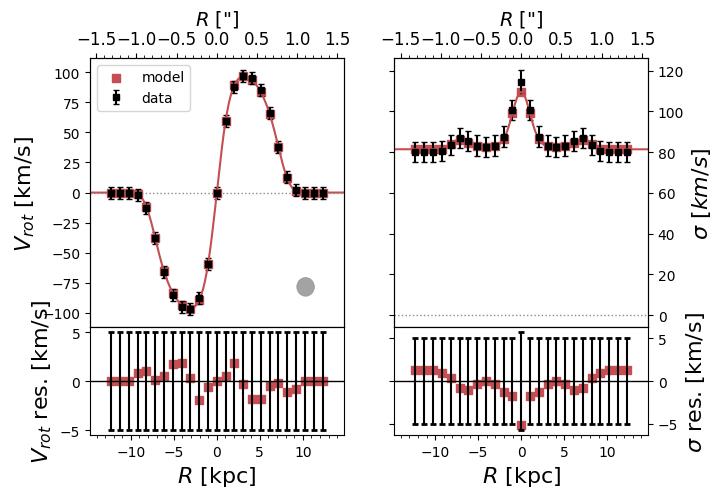}\hfill
\includegraphics[width=0.45\textwidth]{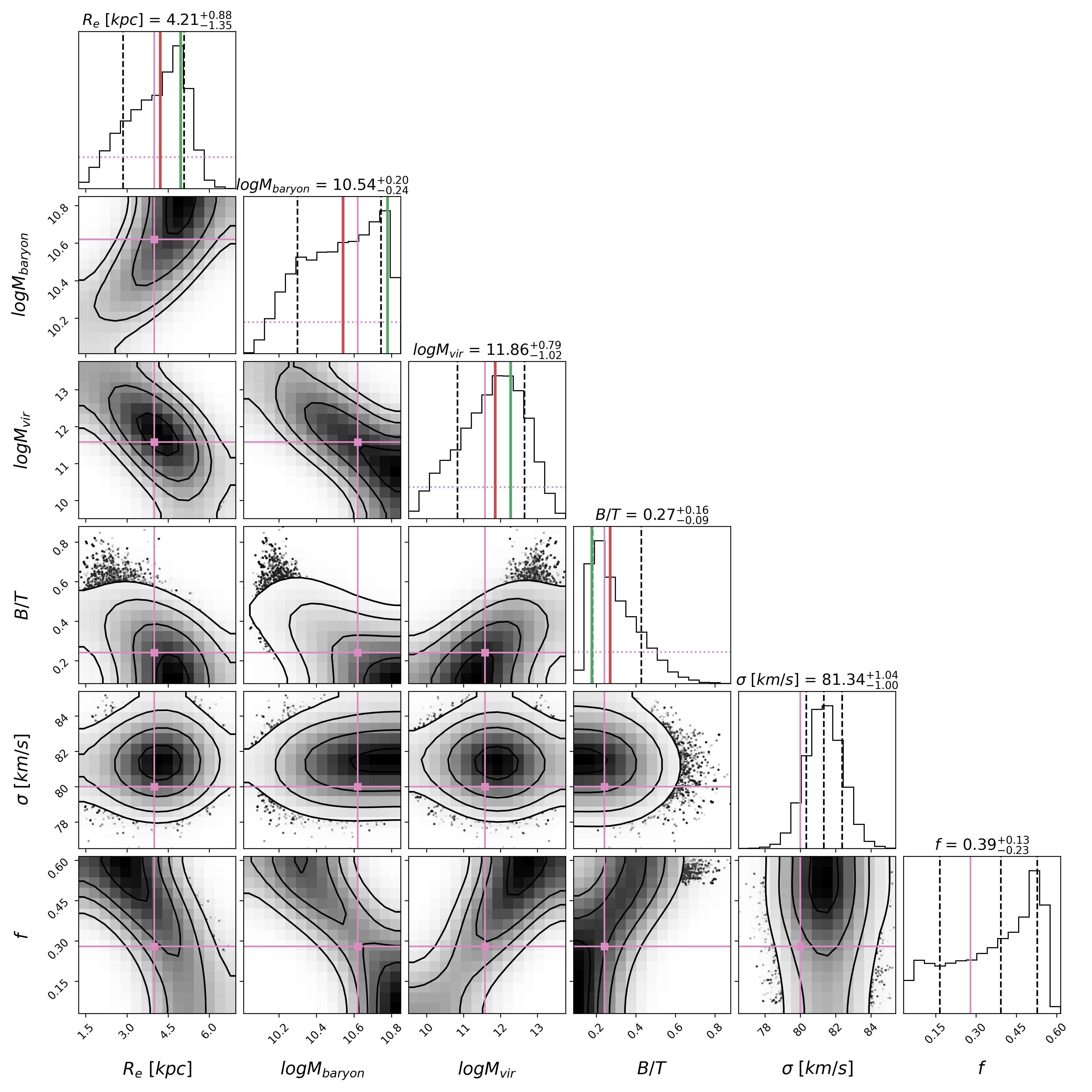}
\smallskip
\makebox[\textwidth][c]{$\sigma_0 = 80 \kms$, $V/\sigma_0 = 2.5$}
\caption{Two variation in the intrinsic velocity dispresion for the MCMC fit results with \texttt{RotCurves} on a model generated by \texttt{dysmalpy}. At very high $\sigma_0$, the rotation curve of our idealized axisymmetric model falls extremely rapidly and loses the information contained in the outer RC about the dark matter content, resulting in a close to flat posterior on $f_\mathrm{DM}$. }
\label{fig:app_paramvariation_sigma0}
\end{figure*}

Following section~\ref{sec:fiducial_parameter_study}, we performed variations in the model parameters and examined the results of the MCMC fit of \texttt{RotCurves}. Figure~\ref{fig:param_variation_medians} shows the comparison between the best fit and the true values. Each combination was fit with 10 MCMC fitting procedures, resulting in a large number of figures. In the following appendix, we present some representative fitting results allowing to examine the degeneracies and residuals for some of these cases in Figures~\ref{fig:app_paramvariation_mbulge}-\ref{fig:app_paramvariation_sigma0}. In all cases, the fiducial parameters describe the true values of the model, except for the parameter explicitly written in the figure caption, noting its value.

\clearpage
\section{Mock observation and noise}
\begin{figure*}
\centering
\medskip
\includegraphics[width=0.95\textwidth]{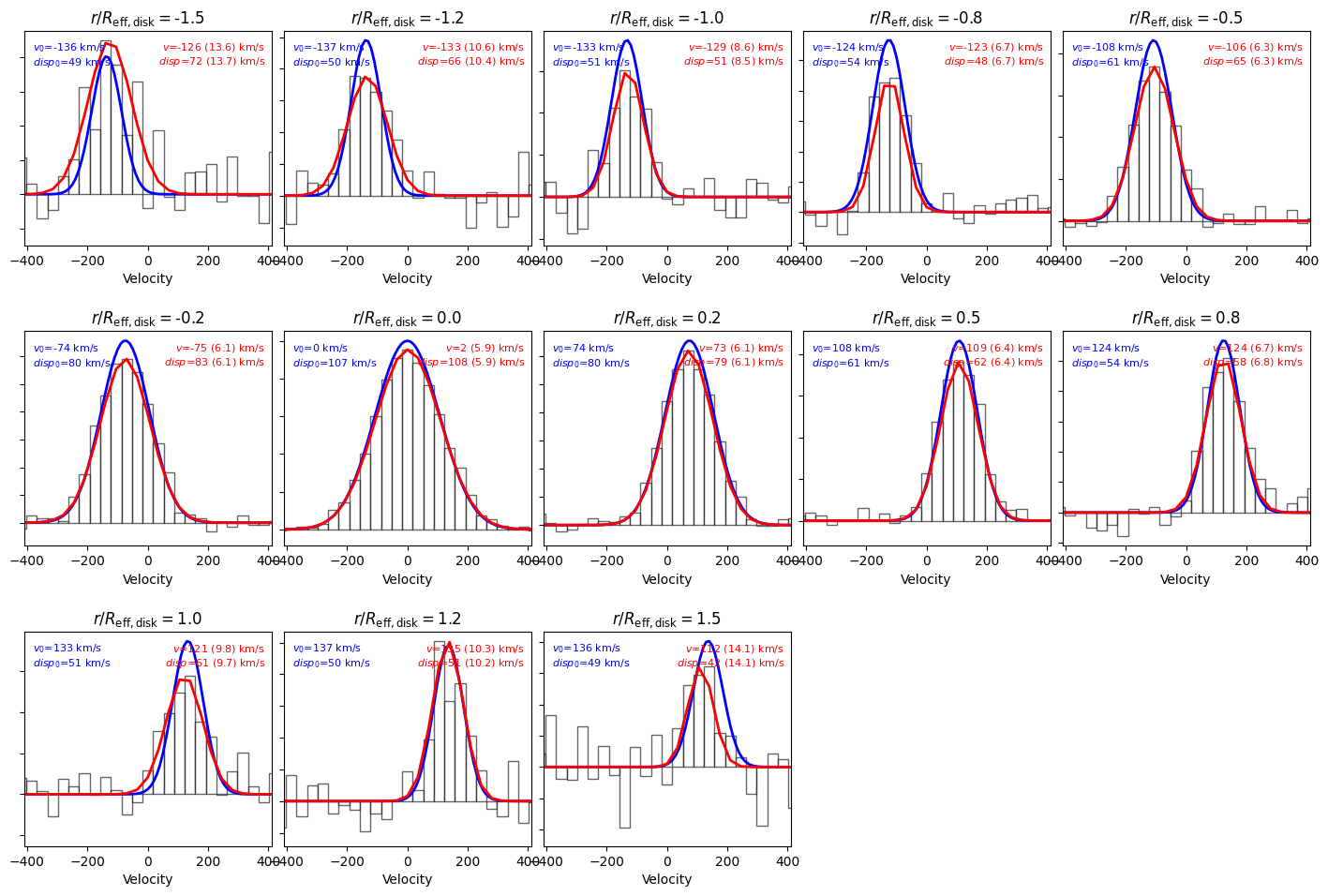}\hfill
\includegraphics[width=0.48\textwidth]{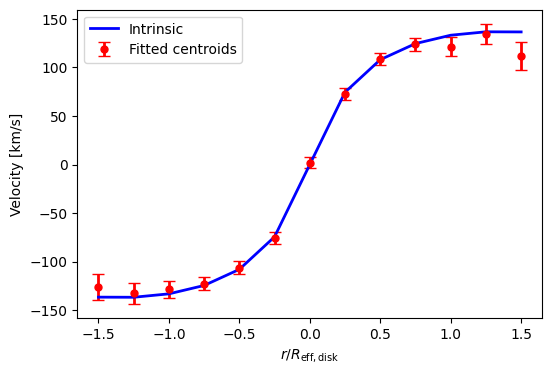}\hfill
\includegraphics[width=0.48\textwidth]{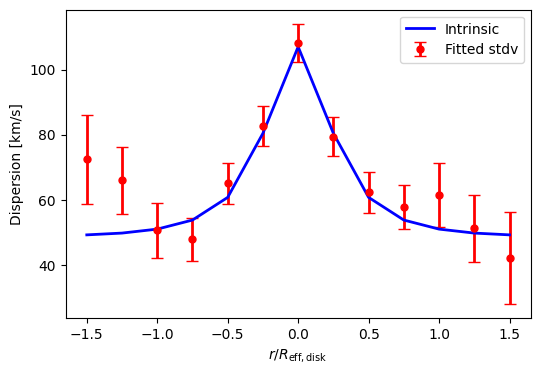}
\smallskip
\caption{Mock observation process for a S/N of 10 at $R_{\rm eff}$, a channel width of $\Delta v=35\,  \rm km/s$ and with a Gaussian LSF with $\sigma _{|rm LSF}=30\, \rm km/s$. The emission lines are constructed from a pure Gaussian with ($V_{\rm rot}, \sigma$, in blue) taken from \texttt{dysmalpy}, with a M/L following the exponential disk. The line is convolved and binned over the LSF, then added a noise level drawn from a Gaussian distribution. The noisy emission line is fitted with a Gaussian (red line), which is used as the mock data points in the rotation curve (middle panel) and dispersion curve (bottom panel) for the MCMC fitting with \texttt{RotCurves}.}
\label{fig:mock_SN10_appendix}
\end{figure*}

\begin{figure*}
\centering
\medskip
\includegraphics[width=0.70\textwidth]{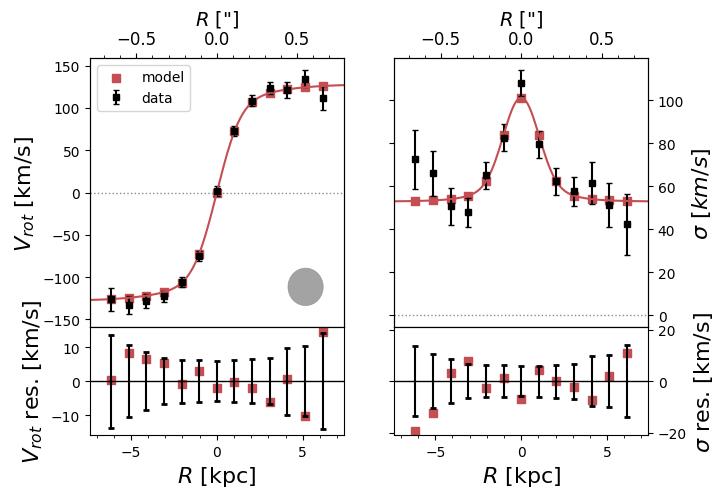}\hfill
\includegraphics[width=0.80\textwidth]{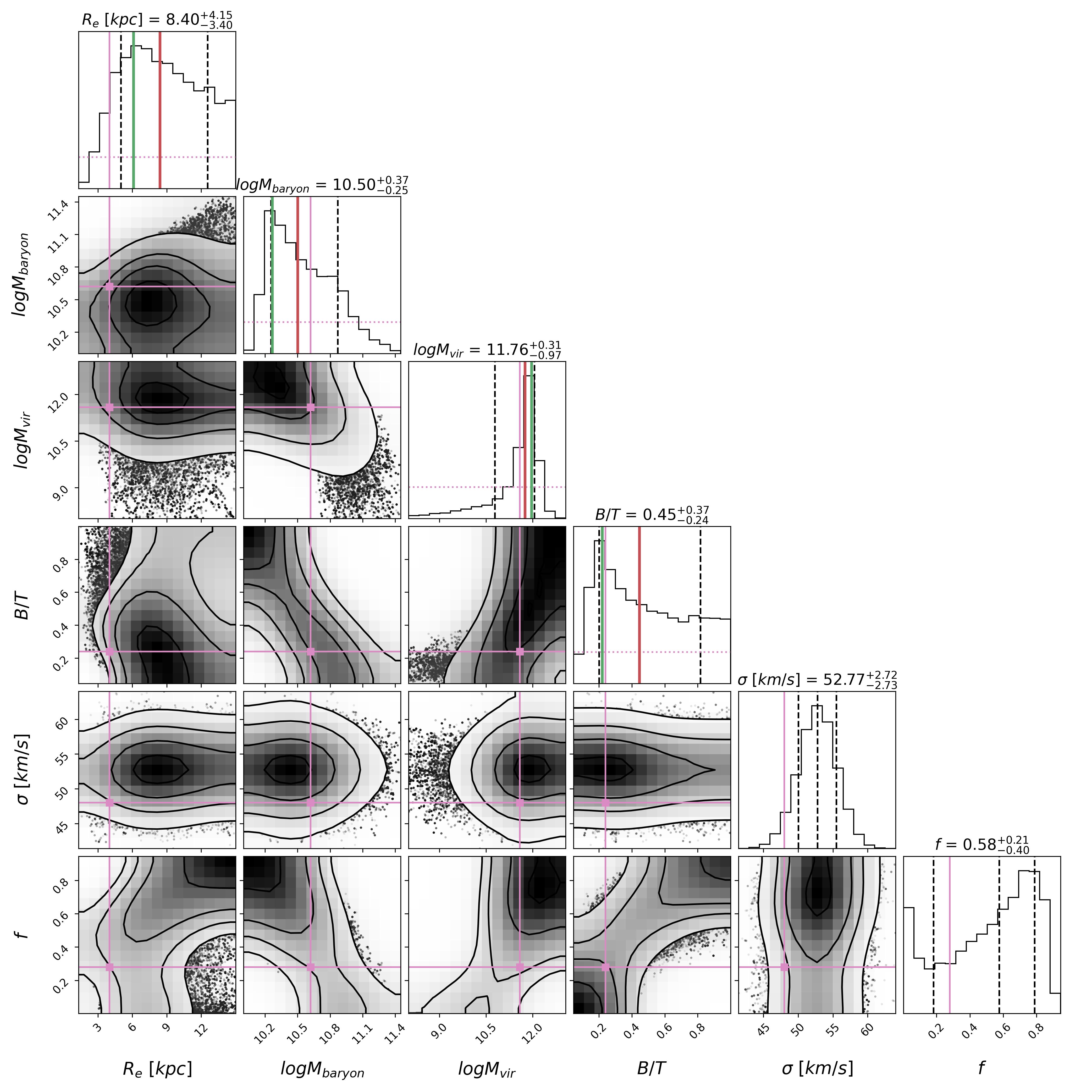}
\smallskip
\caption{Results of the MCMC fit for a mock observation with a S/N of 10 at $R_{\rm eff}$, given in Figure~\ref{fig:mock_SN10_appendix}.}
\label{fig:mock_SN10_fit_appendix}
\end{figure*}

\begin{figure*}
\centering
\medskip
\includegraphics[width=0.7\textwidth]{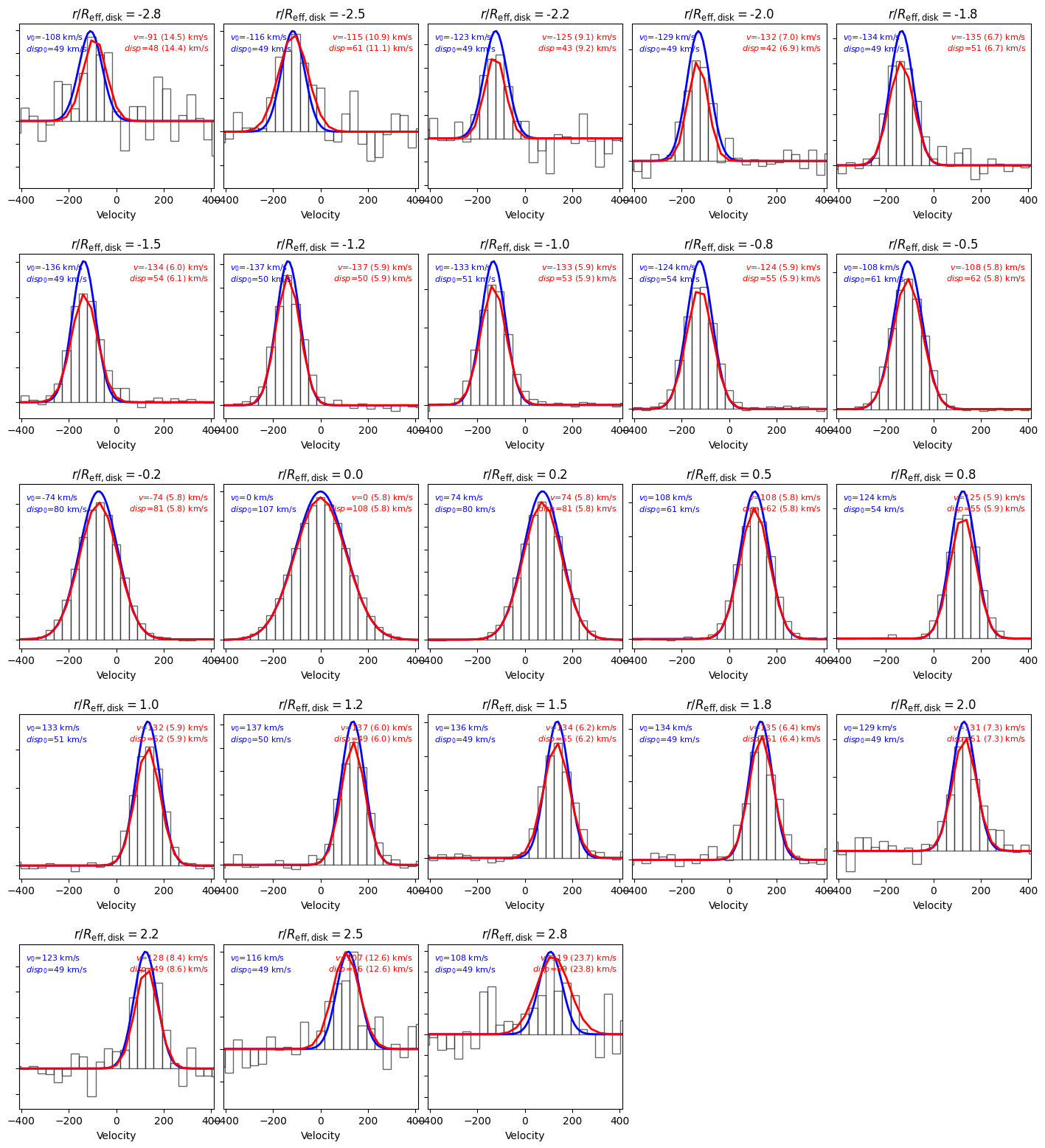}\hfill
\includegraphics[width=0.48\textwidth]{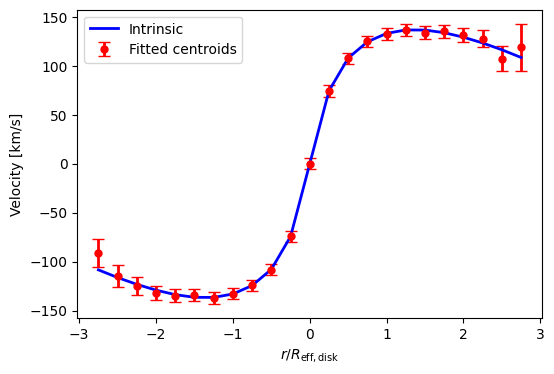}\hfill
\includegraphics[width=0.48\textwidth]{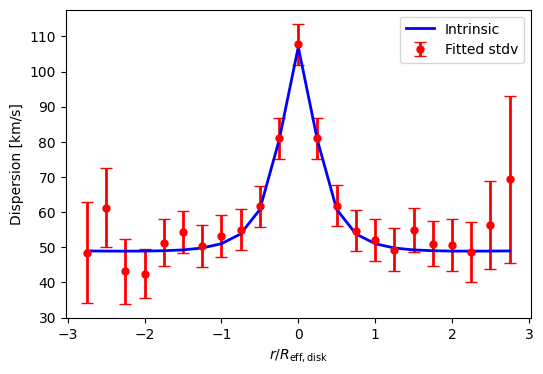}
\smallskip
\caption{Same as \ref{fig:mock_SN10_appendix}, but for a S/N of 70 at $R_{\rm eff}$.}
    \label{fig:mock_SN70_appendix}
\end{figure*}

\begin{figure*}
\centering
\medskip
\includegraphics[width=0.70\textwidth]{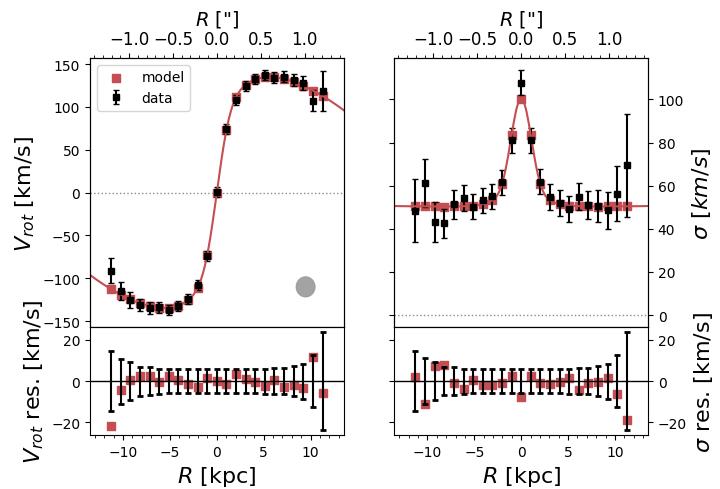}\hfill
\includegraphics[width=0.80\textwidth]{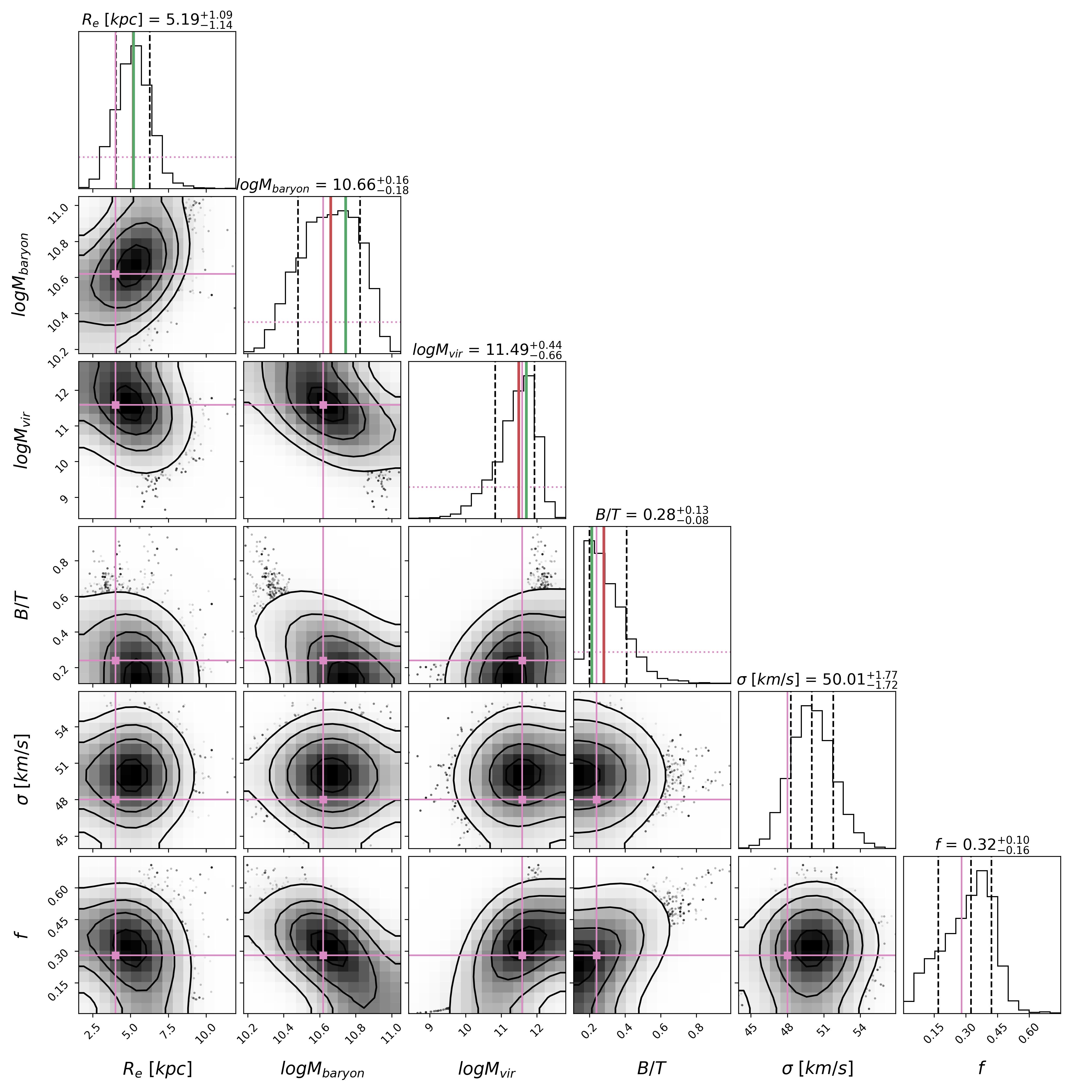}
\smallskip
\caption{Results of the MCMC fit for a mock observation with a S/N of 70 at $R_{\rm eff}$, given in Figure~\ref{fig:mock_SN70_appendix}.}
\label{fig:mock_SN70_fit_appendix}
\end{figure*}

In section~\ref{sec:mock_observation} we discuss the procedure for adding noise to a rotation curve calculated with \texttt{RotCurves}, defined by the S/N ratio of the line peak at the effective radius $R_{\rm eff}$, with a given spectral list-spread-function (LSF) and a channel resolution of $\Delta v$. Figure~\ref{fig:mock_observation_fit_median} show the how the recovery of \texttt{RotCurves} improves as a function of the S/N, given the systematics discussed in the paper. We show two cases of the process of generating the noise and the mock data points, and the consequent results of the MCMC fitting with \texttt{RotCurves}. Figure~\ref{fig:mock_SN10_appendix}-\ref{fig:mock_SN10_fit_appendix} show the results for a low S/N of 10, and Figure~\ref{fig:mock_SN70_appendix}-\ref{fig:mock_SN70_fit_appendix} for a high S/N of 70. As the edge of the mock data points depends on the S/N, as well as the lower uncertainties, it is clear to see the recovery of the physical parameters is much better in the high S/N regime.

\clearpage

\bibliographystyle{mnras}
\bibliography{bibliography}

@ARTICLE{Yttergren2025_code_comparison,
       author = {{Yttergren}, M. and {Knudsen}, K.~K. and {Molina}, J. and {Jones}, G.~C. and {Kade}, K. and {Scholtz}, J. and {Bewketu Belete}, A.},
        title = "{Kinematics of synthetically observed high-z rotating discs: reliability and biases of 3D fitting tools}",
      journal = {\mnras},
         year = 2025,
        month = nov,
       volume = {543},
       number = {4},
        pages = {3103-3122},
          doi = {10.1093/mnras/staf1489},
archivePrefix = {arXiv},
       eprint = {2509.18328},
 primaryClass = {astro-ph.GA},
       adsurl = {https://ui.adsabs.harvard.edu/abs/2025MNRAS.543.3103Y}
}

@ARTICLE{Bisaria2022_diskfit,
       author = {{Bisaria}, Dhruv and {Spekkens}, Kristine and {Huang}, Shan and {Hallenbeck}, Gregory and {Haynes}, Martha P.},
        title = "{Non-circular flows in HIghMass galaxies in a test of the late accretion hypothesis}",
      journal = {\mnras},
         year = 2022,
        month = jan,
       volume = {509},
       number = {1},
        pages = {100-113},
          doi = {10.1093/mnras/stab2851},
archivePrefix = {arXiv},
       eprint = {2111.01806},
 primaryClass = {astro-ph.GA},
       adsurl = {https://ui.adsabs.harvard.edu/abs/2022MNRAS.509..100B}
}

@ARTICLE{Belete2021_diskfit,
       author = {{Bewketu Belete}, A. and {Andreani}, P. and {Fern{\'a}ndez-Ontiveros}, J.~A. and {Hatziminaoglou}, E. and {Combes}, F. and {Sirressi}, M. and {Slater}, R. and {Ricci}, C. and {Dasyra}, K. and {Cicone}, C. and {Aalto}, S. and {Spinoglio}, L. and {Imanishi}, M. and {De Medeiros}, J.~R.},
        title = "{Molecular gas kinematics in the nuclear region of nearby Seyfert galaxies with ALMA}",
      journal = {\aap},
         year = 2021,
        month = oct,
       volume = {654},
          eid = {A24},
        pages = {A24},
          doi = {10.1051/0004-6361/202140492},
archivePrefix = {arXiv},
       eprint = {2105.06867},
 primaryClass = {astro-ph.GA},
       adsurl = {https://ui.adsabs.harvard.edu/abs/2021A&A...654A..24B}
}

@ARTICLE{diFolco2020_diskfit,
       author = {{Di Folco}, E. and {P{\'e}ricaud}, J. and {Dutrey}, A. and {Augereau}, J.-C. and {Chapillon}, E. and {Guilloteau}, S. and {Pi{\'e}tu}, V. and {Boccaletti}, A.},
        title = "{An ALMA/NOEMA study of gas dissipation and dust evolution in the 5 Myr-old HD 141569A hybrid disc}",
      journal = {\aap},
         year = 2020,
        month = mar,
       volume = {635},
          eid = {A94},
        pages = {A94},
          doi = {10.1051/0004-6361/201732243},
       adsurl = {https://ui.adsabs.harvard.edu/abs/2020A&A...635A..94D}
}

@ARTICLE{Liu2025_diskfit,
       author = {{Liu}, Jie and {Li}, Zhi and {Shen}, Juntai},
        title = "{The Impact of Bar-induced Noncircular Motions on the Measurement of Galactic Rotation Curves}",
      journal = {\apj},
         year = 2025,
        month = feb,
       volume = {980},
       number = {1},
          eid = {146},
        pages = {146},
          doi = {10.3847/1538-4357/adabe0},
archivePrefix = {arXiv},
       eprint = {2501.12760},
 primaryClass = {astro-ph.GA},
       adsurl = {https://ui.adsabs.harvard.edu/abs/2025ApJ...980..146L}
}

@ARTICLE{Sellwood2015_diskfit,
       author = {{Sellwood}, J.~A. and {Spekkens}, Kristine},
        title = "{DiskFit: a code to fit simple non-axisymmetric galaxy models either to photometric images or to kinematic maps}",
      journal = {arXiv e-prints},
         year = 2015,
        month = sep,
          eid = {arXiv:1509.07120},
        pages = {arXiv:1509.07120},
          doi = {10.48550/arXiv.1509.07120},
archivePrefix = {arXiv},
       eprint = {1509.07120},
 primaryClass = {astro-ph.GA},
       adsurl = {https://ui.adsabs.harvard.edu/abs/2015arXiv150907120S}
}

@software{deNaray2012_diskfit_soft,
       author = {{Kuzio de Naray}, Rachel and {Arsenault}, Cameron A. and {Spekkens}, Kristine and {Sellwood}, J.~A. and {McDonald}, Michael and {Simon}, Joshua D. and {Teuben}, Peter},
        title = "{DiskFit: Modeling Asymmetries in Disk Galaxies}",
 howpublished = {Astrophysics Source Code Library, record ascl:1209.011},
         year = 2012,
        month = sep,
          eid = {ascl:1209.011},
archivePrefix = {ascl},
       eprint = {1209.011},
       adsurl = {https://ui.adsabs.harvard.edu/abs/2012ascl.soft09011K}
}

@software{Oh2020_2dbat_soft,
       author = {{Oh}, Se-Heon and {Staveley-Smith}, Lister and {Spekkens}, Kristine and {Kamphuis}, Peter and {Koribalski}, B{\"a}rbel S.},
        title = "{2DBAT: 2D Bayesian Automated Tilted-ring fitter}",
 howpublished = {Astrophysics Source Code Library, record ascl:2005.012},
         year = 2020,
        month = may,
          eid = {ascl:2005.012},
archivePrefix = {ascl},
       eprint = {2005.012},
       adsurl = {https://ui.adsabs.harvard.edu/abs/2020ascl.soft05012O}
}

@ARTICLE{Kim2020_2dbat,
       author = {{Kim}, Shinna and {Oh}, Se-Heon and {For}, Bi-Qing and {Sheen}, Yun-Kyeong},
        title = "{Mass models of the Large Magellanic Cloud: HI gas kinematics}",
      journal = {The Bulletin of The Korean Astronomical Society},
         year = 2020,
        month = jan,
       volume = {45},
       number = {1},
        pages = {60.3-61},
       adsurl = {https://ui.adsabs.harvard.edu/abs/2020BKAS...45R..60K}
}

@ARTICLE{Oh2025_2dbat_b,
       author = {{Oh}, Se-Heon and {Kim}, Shinna and {Kim}, Shin-Jeong and {Koribalski}, B{\"a}rbel S.},
        title = "{Rotation curve analysis of dwarf galaxies from LVHIS}",
      journal = {\apss},
         year = 2025,
        month = nov,
       volume = {370},
       number = {11},
          eid = {129},
        pages = {129},
          doi = {10.1007/s10509-025-04520-w},
       adsurl = {https://ui.adsabs.harvard.edu/abs/2025Ap&SS.370..129O}
}

@ARTICLE{Elagali2018_2dbat,
       author = {{Elagali}, Ahmed and {Wong}, O. Ivy and {Oh}, Se-Heon and {Staveley-Smith}, Lister and {Koribalski}, B{\"a}rbel S. and {Bekki}, Kenji and {Zwaan}, Martin},
        title = "{An H I study of the collisional ring galaxy NGC 922}",
      journal = {\mnras},
         year = 2018,
        month = jun,
       volume = {476},
       number = {4},
        pages = {5681-5691},
          doi = {10.1093/mnras/sty741},
archivePrefix = {arXiv},
       eprint = {1804.07037},
 primaryClass = {astro-ph.GA},
       adsurl = {https://ui.adsabs.harvard.edu/abs/2018MNRAS.476.5681E}
}

@ARTICLE{Oh2025_2dbat_a,
       author = {{Oh}, Se-Heon and {Wang}, Jing},
        title = "{Extraction of H I gas with bulk motions in the disc of galaxies}",
      journal = {\mnras},
         year = 2025,
        month = apr,
       volume = {538},
       number = {3},
        pages = {1816-1829},
          doi = {10.1093/mnras/staf298},
archivePrefix = {arXiv},
       eprint = {2503.11173},
 primaryClass = {astro-ph.GA},
       adsurl = {https://ui.adsabs.harvard.edu/abs/2025MNRAS.538.1816O}
}

@ARTICLE{Oh2022_2dbat,
       author = {{Oh}, Se-Heon and {Kim}, Shinna and {For}, Bi-Qing and {Staveley-Smith}, Lister},
        title = "{Kinematic Decomposition of the H I Gaseous Component in the Large Magellanic Cloud}",
      journal = {\apj},
         year = 2022,
        month = apr,
       volume = {928},
       number = {2},
          eid = {177},
        pages = {177},
          doi = {10.3847/1538-4357/ac5905},
       adsurl = {https://ui.adsabs.harvard.edu/abs/2022ApJ...928..177O}
}

@ARTICLE{Wong2021_2dbat,
       author = {{Wong}, O.~I. and {Stevens}, A.~R.~H. and {For}, B.-Q. and {Westmeier}, T. and {Dixon}, M. and {Oh}, S.-H. and {J{\'o}zsa}, G.~I.~G. and {Reynolds}, T.~N. and {Lee-Waddell}, K. and {Rom{\'a}n}, J. and {Verdes-Montenegro}, L. and {Courtois}, H.~M. and {Pomar{\`e}de}, D. and {Murugeshan}, C. and {Whiting}, M.~T. and {Bekki}, K. and {Bigiel}, F. and {Bosma}, A. and {Catinella}, B. and {D{\'e}nes}, H. and {Elagali}, A. and {Holwerda}, B.~W. and {Kamphuis}, P. and {Kilborn}, V.~A. and {Kleiner}, D. and {Koribalski}, B.~S. and {Lelli}, F. and {Madrid}, J.~P. and {McQuinn}, K.~B.~W. and {Popping}, A. and {Rhee}, J. and {Roychowdhury}, S. and {Scott}, T.~C. and {Sengupta}, C. and {Spekkens}, K. and {Staveley-Smith}, L. and {Wakker}, B.~P.},
        title = "{WALLABY pre-pilot survey: two dark clouds in the vicinity of NGC 1395}",
      journal = {\mnras},
         year = 2021,
        month = oct,
       volume = {507},
       number = {2},
        pages = {2905-2921},
          doi = {10.1093/mnras/stab2262},
archivePrefix = {arXiv},
       eprint = {2108.04412},
 primaryClass = {astro-ph.GA},
       adsurl = {https://ui.adsabs.harvard.edu/abs/2021MNRAS.507.2905W}
}

@ARTICLE{Zhang2020_2dbat,
       author = {{Zhang}, Hong-Xin and {Smith}, Rory and {Oh}, Se-Heon and {Paudel}, Sanjaya and {Duc}, Pierre-Alain and {Boselli}, Alessandro and {C{\^o}t{\'e}}, Patrick and {Ferrarese}, Laura and {Gao}, Yu and {Hunter}, Deidre A. and {Puzia}, Thomas H. and {Peng}, Eric W. and {Rong}, Yu and {Shin}, Jihye and {Zhao}, Yinghe},
        title = "{The Blue Compact Dwarf Galaxy VCC 848 Formed by Dwarf-Dwarf Merging: H I Gas, Star Formation, and Numerical Simulations}",
      journal = {\apj},
         year = 2020,
        month = sep,
       volume = {900},
       number = {2},
          eid = {152},
        pages = {152},
          doi = {10.3847/1538-4357/abab96},
archivePrefix = {arXiv},
       eprint = {2007.15667},
 primaryClass = {astro-ph.GA},
       adsurl = {https://ui.adsabs.harvard.edu/abs/2020ApJ...900..152Z}
}

@ARTICLE{Oh2018_2DBAT,
       author = {{Oh}, Se-Heon and {Staveley-Smith}, Lister and {Spekkens}, Kristine and {Kamphuis}, Peter and {Koribalski}, B{\"a}rbel S.},
        title = "{2D Bayesian automated tilted-ring fitting of disc galaxies in large H I galaxy surveys: 2DBAT}",
      journal = {\mnras},
         year = 2018,
        month = jan,
       volume = {473},
       number = {3},
        pages = {3256-3298},
          doi = {10.1093/mnras/stx2304},
archivePrefix = {arXiv},
       eprint = {1709.02049},
 primaryClass = {astro-ph.GA},
       adsurl = {https://ui.adsabs.harvard.edu/abs/2018MNRAS.473.3256O}
}

@ARTICLE{NestorShachar2025,
       author = {{Nestor Shachar}, A. and {Sternberg}, A. and {Genzel}, R. and {Liu}, D. and {Price}, S.~H. and {Pulsoni}, C. and {Tacconi}, L.~J. and {Herrera-Camus}, R. and {F{\"o}rster Schreiber}, N.~M. and {Burkert}, A. and {Jolly}, J.~B. and {Lutz}, D. and {Wuyts}, S. and {Barfety}, C. and {Cao}, Y. and {Chen}, J. and {Davies}, R. and {Eisenhauer}, F. and {Espejo Salcedo}, J.~M. and {Lee}, L.~L. and {Lee}, M. and {Naab}, T. and {Pastras}, S. and {Shimizu}, T.~T. and {Sturm}, E. and {Tozzi}, G. and {{\"U}bler}, H.},
        title = "{A Large-scale Ring Galaxy at z = 2.2 Revealed by JWST/NIRCam: Kinematic Observations and Analytical Modelling}",
      journal = {\apj},
         year = 2025,
        month = aug,
       volume = {988},
       number = {2},
          eid = {182},
        pages = {182},
          doi = {10.3847/1538-4357/ade2d0},
archivePrefix = {arXiv},
       eprint = {2503.00839},
 primaryClass = {astro-ph.GA},
       adsurl = {https://ui.adsabs.harvard.edu/abs/2025ApJ...988..182N}
}

@ARTICLE{Jones2025,
       author = {{Jones}, Gareth C. and {Maiolino}, Roberto and {D'Eugenio}, Francesco and {Arribas}, Santiago and {Bunker}, Andrew J. and {Charlot}, Stephane and {Perna}, Michele and {Rodriguez del Pino}, Bruno and {{\"U}bler}, Hannah and {B{\"o}ker}, Torsten and {Cresci}, Giovanni and {Lamperti}, Isabella and {Parlanti}, Eleonora and {Pascalau}, Robert and {Scholtz}, Jan and {Zamora}, Sandra},
        title = "{GA-NIFS: A smouldering disk galaxy undergoing ordered rotation at z=4.26}",
      journal = {arXiv e-prints},
         year = 2025,
        month = dec,
          eid = {arXiv:2512.05213},
        pages = {arXiv:2512.05213},
          doi = {10.48550/arXiv.2512.05213},
archivePrefix = {arXiv},
       eprint = {2512.05213},
 primaryClass = {astro-ph.GA},
       adsurl = {https://ui.adsabs.harvard.edu/abs/2025arXiv251205213J}
}

@ARTICLE{HerreraCamus2025_cristal,
       author = {{Herrera-Camus}, R. and {Gonz{\'a}lez-L{\'o}pez}, J. and {F{\"o}rster Schreiber}, N. and {Aravena}, M. and {de Looze}, I. and {Spilker}, J. and {Tadaki}, K. and {Barcos-Mu{\~n}oz}, L. and {Assef}, R.~J. and {Birkin}, J.~E. and {Bolatto}, A.~D. and {Bouwens}, R. and {Bovino}, S. and {Bowler}, R.~A.~A. and {Calistro Rivera}, G. and {da Cunha}, E. and {Davies}, R.~I. and {Davies}, R.~L. and {D{\'\i}az-Santos}, T. and {Ferrara}, A. and {Fisher}, D. and {Genzel}, R. and {Hodge}, J. and {Ikeda}, R. and {Killi}, M. and {Lee}, L. and {Li}, Y. and {Li}, J. and {Liu}, D. and {Lutz}, D. and {Mitsuhashi}, I. and {Narayanan}, D. and {Naab}, T. and {Palla}, M. and {Price}, S.~H. and {Posses}, A. and {Rela{\~n}o}, M. and {Smit}, R. and {Solimano}, M. and {Sternberg}, A. and {Tacconi}, L. and {Telikova}, K. and {{\"U}bler}, H. and {van der Giessen}, S.~A. and {Veilleux}, S. and {Villanueva}, V. and {Baeza-Garay}, M.},
        title = "{The ALMA-CRISTAL survey: Gas, dust, and stars in star-forming galaxies when the Universe was {\ensuremath{\sim}}1 Gyr old: I. Survey overview and case studies}",
      journal = {\aap},
         year = 2025,
        month = jul,
       volume = {699},
          eid = {A80},
        pages = {A80},
          doi = {10.1051/0004-6361/202553896},
archivePrefix = {arXiv},
       eprint = {2505.06340},
 primaryClass = {astro-ph.GA},
       adsurl = {https://ui.adsabs.harvard.edu/abs/2025A&A...699A..80H}
}

@ARTICLE{Wisnioski2025_dispersion,
       author = {{Wisnioski}, E. and {Mendel}, J.~T. and {Leaman}, R. and {Tsukui}, T. and {{\"U}bler}, H. and {Schreiber}, N.~M. F{\"o}rster},
        title = "{Evolution of Gas Velocity Dispersion in Discs from z \raisebox{-0.5ex}\textasciitilde 8 to z \raisebox{-0.5ex}\textasciitilde 0.5}",
      journal = {\mnras},
         year = 2025,
        month = sep,
          doi = {10.1093/mnras/staf1606},
archivePrefix = {arXiv},
       eprint = {2505.24129},
 primaryClass = {astro-ph.GA},
       adsurl = {https://ui.adsabs.harvard.edu/abs/2025MNRAS.tmp.1526W}
}

@ARTICLE{Barisic2025jwst_msa3d,
       author = {{Bari{\v{s}}i{\'c}}, Ivana and {Jones}, Tucker and {Mortensen}, Kris and {Nanayakkara}, Themiya and {Chen}, Yuguang and {Sanders}, Ryan and {Bullock}, James S. and {Bundy}, Kevin and {Faucher-Gigu{\`e}re}, Claude-Andr{\'e} and {Glazebrook}, Karl and {Henry}, Alaina and {Ju}, Mengting and {Malkan}, Matthew and {Morishita}, Takahiro and {Obreschkow}, Danail and {Roy}, Namrata and {Espejo Salcedo}, Juan M. and {Shapley}, Alice E. and {Treu}, Tommaso and {Wang}, Xin and {Westfall}, Kyle B.},
        title = "{MSA-3D: Dissecting Galaxies at z {\ensuremath{\sim}} 1 with High Spatial and Spectral Resolution}",
      journal = {\apj},
         year = 2025,
        month = apr,
       volume = {983},
       number = {2},
          eid = {139},
        pages = {139},
          doi = {10.3847/1538-4357/ada617},
archivePrefix = {arXiv},
       eprint = {2408.08350},
 primaryClass = {astro-ph.GA},
       adsurl = {https://ui.adsabs.harvard.edu/abs/2025ApJ...983..139B}
}

@ARTICLE{Marconcini2024_jwst_ganifs,
       author = {{Marconcini}, C. and {D'Eugenio}, F. and {Maiolino}, R. and {Arribas}, S. and {Bunker}, A. and {Carniani}, S. and {Charlot}, S. and {Perna}, M. and {Rodr{\'\i}guez Del Pino}, B. and {{\"U}bler}, H. and {Willott}, C.~J. and {B{\"o}ker}, T. and {Cresci}, G. and {Curti}, M. and {Jones}, G.~C. and {Lamperti}, I. and {Parlanti}, E. and {Venturi}, G.},
        title = "{GA-NIFS: the interplay between merger, star formation, and chemical enrichment in MACS1149-JD1 at z = 9.11 with JWST/NIRSpec}",
      journal = {\mnras},
         year = 2024,
        month = sep,
       volume = {533},
       number = {2},
        pages = {2488-2501},
          doi = {10.1093/mnras/stae1971},
archivePrefix = {arXiv},
       eprint = {2407.08616},
 primaryClass = {astro-ph.GA},
       adsurl = {https://ui.adsabs.harvard.edu/abs/2024MNRAS.533.2488M}
}

@ARTICLE{Bertola2025_jwst_ganifs,
       author = {{Bertola}, E. and {Cresci}, G. and {Venturi}, G. and {Perna}, M. and {Circosta}, C. and {Tozzi}, G. and {Lamperti}, I. and {Vignali}, C. and {Arribas}, S. and {Bunker}, A.~J. and {Charlot}, S. and {Carniani}, S. and {Maiolino}, R. and {Rodr{\'\i}guez Del Pino}, B. and {{\"U}bler}, H. and {Willott}, C.~J. and {B{\"o}ker}, T. and {Marshall}, M.~A. and {Parlanti}, E. and {Scholtz}, J.},
        title = "{GA-NIFS: Mapping z ≃ 3.5 AGN-driven ionized outflows in the COSMOS field}",
      journal = {\aap},
         year = 2025,
        month = jul,
       volume = {699},
          eid = {A220},
        pages = {A220},
          doi = {10.1051/0004-6361/202554281},
archivePrefix = {arXiv},
       eprint = {2505.08867},
 primaryClass = {astro-ph.GA},
       adsurl = {https://ui.adsabs.harvard.edu/abs/2025A&A...699A.220B}
}

@ARTICLE{Smit2018_alma_cii,
       author = {{Smit}, Renske and {Bouwens}, Rychard J. and {Carniani}, Stefano and {Oesch}, Pascal A. and {Labb{\'e}}, Ivo and {Illingworth}, Garth D. and {van der Werf}, Paul and {Bradley}, Larry D. and {Gonzalez}, Valentino and {Hodge}, Jacqueline A. and {Holwerda}, Benne W. and {Maiolino}, Roberto and {Zheng}, Wei},
        title = "{Rotation in [C II]-emitting gas in two galaxies at a redshift of 6.8}",
      journal = {\nat},
         year = 2018,
        month = jan,
       volume = {553},
       number = {7687},
        pages = {178-181},
          doi = {10.1038/nature24631},
archivePrefix = {arXiv},
       eprint = {1706.04614},
 primaryClass = {astro-ph.GA},
       adsurl = {https://ui.adsabs.harvard.edu/abs/2018Natur.553..178S}
}

@ARTICLE{Chen2017_alma,
       author = {{Chen}, Chian-Chou and {Hodge}, J.~A. and {Smail}, Ian and {Swinbank}, A.~M. and {Walter}, Fabian and {Simpson}, J.~M. and {Calistro Rivera}, Gabriela and {Bertoldi}, F. and {Brandt}, W.~N. and {Chapman}, S.~C. and {da Cunha}, Elisabete and {Dannerbauer}, H. and {De Breuck}, C. and {Harrison}, C.~M. and {Ivison}, R.~J. and {Karim}, A. and {Knudsen}, K.~K. and {Wardlow}, J.~L. and {Wei{\ss}}, A. and {van der Werf}, P.~P.},
        title = "{A Spatially Resolved Study of Cold Dust, Molecular Gas, H II Regions, and Stars in the z = 2.12 Submillimeter Galaxy ALESS67.1}",
      journal = {\apj},
         year = 2017,
        month = sep,
       volume = {846},
       number = {2},
          eid = {108},
        pages = {108},
          doi = {10.3847/1538-4357/aa863a},
archivePrefix = {arXiv},
       eprint = {1708.08937},
 primaryClass = {astro-ph.GA},
       adsurl = {https://ui.adsabs.harvard.edu/abs/2017ApJ...846..108C}
}

@ARTICLE{Tadaki2018_alma_co,
       author = {{Tadaki}, K. and {Iono}, D. and {Yun}, M.~S. and {Aretxaga}, I. and {Hatsukade}, B. and {Hughes}, D.~H. and {Ikarashi}, S. and {Izumi}, T. and {Kawabe}, R. and {Kohno}, K. and {Lee}, M. and {Matsuda}, Y. and {Nakanishi}, K. and {Saito}, T. and {Tamura}, Y. and {Ueda}, J. and {Umehata}, H. and {Wilson}, G.~W. and {Michiyama}, T. and {Ando}, M. and {Kamieneski}, P.},
        title = "{The gravitationally unstable gas disk of a starburst galaxy 12 billion years ago}",
      journal = {\nat},
         year = 2018,
        month = aug,
       volume = {560},
       number = {7720},
        pages = {613-616},
          doi = {10.1038/s41586-018-0443-1},
archivePrefix = {arXiv},
       eprint = {1808.09592},
 primaryClass = {astro-ph.GA},
       adsurl = {https://ui.adsabs.harvard.edu/abs/2018Natur.560..613T}
}

@ARTICLE{Molina2019_co,
       author = {{Molina}, J. and {Ibar}, Edo and {Smail}, I. and {Swinbank}, A.~M. and {Villard}, E. and {Escala}, A. and {Sobral}, D. and {Hughes}, T.~M.},
        title = "{The kiloparsec-scale gas kinematics in two star-forming galaxies at z {\ensuremath{\sim}} 1.47 seen with ALMA and VLT-SINFONI}",
      journal = {\mnras},
         year = 2019,
        month = aug,
       volume = {487},
       number = {4},
        pages = {4856-4869},
          doi = {10.1093/mnras/stz1643},
archivePrefix = {arXiv},
       eprint = {1906.06245},
 primaryClass = {astro-ph.GA},
       adsurl = {https://ui.adsabs.harvard.edu/abs/2019MNRAS.487.4856M}
}

@ARTICLE{Genzel2013_phibss,
       author = {{Genzel}, R. and {Tacconi}, L.~J. and {Kurk}, J. and {Wuyts}, S. and {Combes}, F. and {Freundlich}, J. and {Bolatto}, A. and {Cooper}, M.~C. and {Neri}, R. and {Nordon}, R. and {Bournaud}, F. and {Burkert}, A. and {Comerford}, J. and {Cox}, P. and {Davis}, M. and {F{\"o}rster Schreiber}, N.~M. and {Garc{\'\i}a-Burillo}, S. and {Gracia-Carpio}, J. and {Lutz}, D. and {Naab}, T. and {Newman}, S. and {Saintonge}, A. and {Shapiro Griffin}, K. and {Shapley}, A. and {Sternberg}, A. and {Weiner}, B.},
        title = "{Phibss: Molecular Gas, Extinction, Star Formation, and Kinematics in the z = 1.5 Star-forming Galaxy EGS13011166}",
      journal = {\apj},
         year = 2013,
        month = aug,
       volume = {773},
       number = {1},
          eid = {68},
        pages = {68},
          doi = {10.1088/0004-637X/773/1/68},
archivePrefix = {arXiv},
       eprint = {1304.0668},
 primaryClass = {astro-ph.CO},
       adsurl = {https://ui.adsabs.harvard.edu/abs/2013ApJ...773...68G}
}

@ARTICLE{Tacconi2013_phibss,
       author = {{Tacconi}, L.~J. and {Neri}, R. and {Genzel}, R. and {Combes}, F. and {Bolatto}, A. and {Cooper}, M.~C. and {Wuyts}, S. and {Bournaud}, F. and {Burkert}, A. and {Comerford}, J. and {Cox}, P. and {Davis}, M. and {F{\"o}rster Schreiber}, N.~M. and {Garc{\'\i}a-Burillo}, S. and {Gracia-Carpio}, J. and {Lutz}, D. and {Naab}, T. and {Newman}, S. and {Omont}, A. and {Saintonge}, A. and {Shapiro Griffin}, K. and {Shapley}, A. and {Sternberg}, A. and {Weiner}, B.},
        title = "{Phibss: Molecular Gas Content and Scaling Relations in z \raisebox{-0.5ex}\textasciitilde 1-3 Massive, Main-sequence Star-forming Galaxies}",
      journal = {\apj},
         year = 2013,
        month = may,
       volume = {768},
       number = {1},
          eid = {74},
        pages = {74},
          doi = {10.1088/0004-637X/768/1/74},
archivePrefix = {arXiv},
       eprint = {1211.5743},
 primaryClass = {astro-ph.CO},
       adsurl = {https://ui.adsabs.harvard.edu/abs/2013ApJ...768...74T}
}

@ARTICLE{Tacconi2018_phibss,
       author = {{Tacconi}, L.~J. and {Genzel}, R. and {Saintonge}, A. and {Combes}, F. and {Garc{\'\i}a-Burillo}, S. and {Neri}, R. and {Bolatto}, A. and {Contini}, T. and {F{\"o}rster Schreiber}, N.~M. and {Lilly}, S. and {Lutz}, D. and {Wuyts}, S. and {Accurso}, G. and {Boissier}, J. and {Boone}, F. and {Bouch{\'e}}, N. and {Bournaud}, F. and {Burkert}, A. and {Carollo}, M. and {Cooper}, M. and {Cox}, P. and {Feruglio}, C. and {Freundlich}, J. and {Herrera-Camus}, R. and {Juneau}, S. and {Lippa}, M. and {Naab}, T. and {Renzini}, A. and {Salome}, P. and {Sternberg}, A. and {Tadaki}, K. and {{\"U}bler}, H. and {Walter}, F. and {Weiner}, B. and {Weiss}, A.},
        title = "{PHIBSS: Unified Scaling Relations of Gas Depletion Time and Molecular Gas Fractions}",
      journal = {\apj},
         year = 2018,
        month = feb,
       volume = {853},
       number = {2},
          eid = {179},
        pages = {179},
          doi = {10.3847/1538-4357/aaa4b4},
archivePrefix = {arXiv},
       eprint = {1702.01140},
 primaryClass = {astro-ph.GA},
       adsurl = {https://ui.adsabs.harvard.edu/abs/2018ApJ...853..179T}
}

@ARTICLE{Contini2016_muse,
       author = {{Contini}, T. and {Epinat}, B. and {Bouch{\'e}}, N. and {Brinchmann}, J. and {Boogaard}, L.~A. and {Ventou}, E. and {Bacon}, R. and {Richard}, J. and {Weilbacher}, P.~M. and {Wisotzki}, L. and {Krajnovi{\'c}}, D. and {Vielfaure}, J.-B. and {Emsellem}, E. and {Finley}, H. and {Inami}, H. and {Schaye}, J. and {Swinbank}, M. and {Gu{\'e}rou}, A. and {Martinsson}, T. and {Michel-Dansac}, L. and {Schroetter}, I. and {Shirazi}, M. and {Soucail}, G.},
        title = "{Deep MUSE observations in the HDFS. Morpho-kinematics of distant star-forming galaxies down to {}10$^{8}$M$_{{\ensuremath{\odot}}}$}",
      journal = {\aap},
         year = 2016,
        month = jun,
       volume = {591},
          eid = {A49},
        pages = {A49},
          doi = {10.1051/0004-6361/201527866},
archivePrefix = {arXiv},
       eprint = {1512.00246},
 primaryClass = {astro-ph.GA},
       adsurl = {https://ui.adsabs.harvard.edu/abs/2016A&A...591A..49C}
}

@ARTICLE{Guerou2017_muse,
       author = {{Gu{\'e}rou}, Adrien and {Krajnovi{\'c}}, Davor and {Epinat}, Benoit and {Contini}, Thierry and {Emsellem}, Eric and {Bouch{\'e}}, Nicolas and {Bacon}, Roland and {Michel-Dansac}, Leo and {Richard}, Johan and {Weilbacher}, Peter M. and {Schaye}, Joop and {Marino}, Raffaella Anna and {den Brok}, Mark and {Erroz-Ferrer}, Santiago},
        title = "{The MUSE Hubble Ultra Deep Field Survey. V. Spatially resolved stellar kinematics of galaxies at redshift 0.2 {\ensuremath{\lesssim}} z {\ensuremath{\lesssim}} 0.8}",
      journal = {\aap},
         year = 2017,
        month = nov,
       volume = {608},
          eid = {A5},
        pages = {A5},
          doi = {10.1051/0004-6361/201730905},
archivePrefix = {arXiv},
       eprint = {1710.07694},
 primaryClass = {astro-ph.GA},
       adsurl = {https://ui.adsabs.harvard.edu/abs/2017A&A...608A...5G}
}

@ARTICLE{Leethochawalit2016_lensed,
       author = {{Leethochawalit}, Nicha and {Jones}, Tucker A. and {Ellis}, Richard S. and {Stark}, Daniel P. and {Richard}, Johan and {Zitrin}, Adi and {Auger}, Matthew},
        title = "{A Keck Adaptive Optics Survey of a Representative Sample of Gravitationally Lensed Star-forming Galaxies: High Spatial Resolution Studies of Kinematics and Metallicity Gradients}",
      journal = {\apj},
         year = 2016,
        month = apr,
       volume = {820},
       number = {2},
          eid = {84},
        pages = {84},
          doi = {10.3847/0004-637X/820/2/84},
archivePrefix = {arXiv},
       eprint = {1509.01279},
 primaryClass = {astro-ph.GA},
       adsurl = {https://ui.adsabs.harvard.edu/abs/2016ApJ...820...84L}
}

@ARTICLE{Liu2023_lensed,
       author = {{Liu}, Daizhong and {F{\"o}rster Schreiber}, N.~M. and {Genzel}, R. and {Lutz}, D. and {Price}, S.~H. and {Lee}, L.~L. and {Baker}, Andrew J. and {Burkert}, A. and {Coogan}, R.~T. and {Davies}, R.~I. and {Davies}, R.~L. and {Herrera-Camus}, R. and {Kodama}, Tadayuki and {Lee}, Minju M. and {Nestor}, A. and {Pulsoni}, C. and {Renzini}, A. and {Sharon}, Chelsea E. and {Shimizu}, T.~T. and {Tacconi}, L.~J. and {Tadaki}, Ken-ichi and {{\"U}bler}, H.},
        title = "{An  600 pc View of the Strongly Lensed, Massive Main-sequence Galaxy J0901: A Baryon-dominated, Thick Turbulent Rotating Disk with a Clumpy Cold Gas Ring at z = 2.259}",
      journal = {\apj},
         year = 2023,
        month = jan,
       volume = {942},
       number = {2},
          eid = {98},
        pages = {98},
          doi = {10.3847/1538-4357/aca46b},
archivePrefix = {arXiv},
       eprint = {2211.08488},
 primaryClass = {astro-ph.GA},
       adsurl = {https://ui.adsabs.harvard.edu/abs/2023ApJ...942...98L}
}

@ARTICLE{Livermore2015_lensed,
       author = {{Livermore}, R.~C. and {Jones}, T.~A. and {Richard}, J. and {Bower}, R.~G. and {Swinbank}, A.~M. and {Yuan}, T.-T. and {Edge}, A.~C. and {Ellis}, R.~S. and {Kewley}, L.~J. and {Smail}, Ian and {Coppin}, K.~E.~K. and {Ebeling}, H.},
        title = "{Resolved spectroscopy of gravitationally lensed galaxies: global dynamics and star-forming clumps on {\ensuremath{\sim}}100 pc scales at 1 < z < 4}",
      journal = {\mnras},
         year = 2015,
        month = jun,
       volume = {450},
       number = {2},
        pages = {1812-1835},
          doi = {10.1093/mnras/stv686},
archivePrefix = {arXiv},
       eprint = {1503.07873},
 primaryClass = {astro-ph.GA},
       adsurl = {https://ui.adsabs.harvard.edu/abs/2015MNRAS.450.1812L}
}

@ARTICLE{Law2009_keck,
       author = {{Law}, David R. and {Steidel}, Charles C. and {Erb}, Dawn K. and {Larkin}, James E. and {Pettini}, Max and {Shapley}, Alice E. and {Wright}, Shelley A.},
        title = "{The Kiloparsec-scale Kinematics of High-redshift Star-forming Galaxies}",
      journal = {\apj},
         year = 2009,
        month = jun,
       volume = {697},
       number = {2},
        pages = {2057-2082},
          doi = {10.1088/0004-637X/697/2/2057},
archivePrefix = {arXiv},
       eprint = {0901.2930},
 primaryClass = {astro-ph.GA},
       adsurl = {https://ui.adsabs.harvard.edu/abs/2009ApJ...697.2057L}
}

@ARTICLE{Gnerucci2011_sinfoni,
       author = {{Gnerucci}, A. and {Marconi}, A. and {Cresci}, G. and {Maiolino}, R. and {Mannucci}, F. and {Calura}, F. and {Cimatti}, A. and {Cocchia}, F. and {Grazian}, A. and {Matteucci}, F. and {Nagao}, T. and {Pozzetti}, L. and {Troncoso}, P.},
        title = "{Dynamical properties of AMAZE and LSD galaxies from gas kinematics and the Tully-Fisher relation at z \raisebox{-0.5ex}\textasciitilde 3}",
      journal = {\aap},
         year = 2011,
        month = apr,
       volume = {528},
          eid = {A88},
        pages = {A88},
          doi = {10.1051/0004-6361/201015465},
archivePrefix = {arXiv},
       eprint = {1007.4180},
 primaryClass = {astro-ph.CO},
       adsurl = {https://ui.adsabs.harvard.edu/abs/2011A&A...528A..88G}
}

@ARTICLE{Epinat2009_sinfoni,
       author = {{Epinat}, B. and {Contini}, T. and {Le F{\`e}vre}, O. and {Vergani}, D. and {Garilli}, B. and {Amram}, P. and {Queyrel}, J. and {Tasca}, L. and {Tresse}, L.},
        title = "{Integral field spectroscopy with SINFONI of VVDS galaxies. I. Galaxy dynamics and mass assembly at 1.2 < z < 1.6}",
      journal = {\aap},
         year = 2009,
        month = sep,
       volume = {504},
       number = {3},
        pages = {789-805},
          doi = {10.1051/0004-6361/200911995},
archivePrefix = {arXiv},
       eprint = {0903.1216},
 primaryClass = {astro-ph.CO},
       adsurl = {https://ui.adsabs.harvard.edu/abs/2009A&A...504..789E}
}

@ARTICLE{mancini2011_zcosmos,
       author = {{Mancini}, C. and {F{\"o}rster Schreiber}, N.~M. and {Renzini}, A. and {Cresci}, G. and {Hicks}, E.~K.~S. and {Peng}, Y. and {Vergani}, D. and {Lilly}, S. and {Carollo}, M. and {Pozzetti}, L. and {Zamorani}, G. and {Daddi}, E. and {Genzel}, R. and {Maraston}, C. and {McCracken}, H.~J. and {Tacconi}, L. and {Bouch{\'e}}, N. and {Davies}, R. and {Oesch}, P. and {Shapiro}, K. and {Mainieri}, V. and {Lutz}, D. and {Mignoli}, M. and {Sternberg}, A.},
        title = "{The zCOSMOS-SINFONI Project. I. Sample Selection and Natural-seeing Observations}",
      journal = {\apj},
         year = 2011,
        month = dec,
       volume = {743},
       number = {1},
          eid = {86},
        pages = {86},
          doi = {10.1088/0004-637X/743/1/86},
archivePrefix = {arXiv},
       eprint = {1109.5952},
 primaryClass = {astro-ph.CO},
       adsurl = {https://ui.adsabs.harvard.edu/abs/2011ApJ...743...86M}
}

@ARTICLE{Sanchez2023_califa,
       author = {{S{\'a}nchez}, S.~F. and {Galbany}, L. and {Walcher}, C.~J. and {Garc{\'\i}a-Benito}, R. and {Barrera-Ballesteros}, J.~K.},
        title = "{The Calar Alto Legacy Integral Field Area survey: extended and remastered data release}",
      journal = {\mnras},
         year = 2023,
        month = dec,
       volume = {526},
       number = {4},
        pages = {5555-5589},
          doi = {10.1093/mnras/stad3119},
archivePrefix = {arXiv},
       eprint = {2304.13022},
 primaryClass = {astro-ph.GA},
       adsurl = {https://ui.adsabs.harvard.edu/abs/2023MNRAS.526.5555S}
}

@ARTICLE{Sanchez2016_califa,
       author = {{S{\'a}nchez}, S.~F. and {Garc{\'\i}a-Benito}, R. and {Zibetti}, S. and {Walcher}, C.~J. and {Husemann}, B. and {Mendoza}, M.~A. and {Galbany}, L. and {Falc{\'o}n-Barroso}, J. and {Mast}, D. and {Aceituno}, J. and {Aguerri}, J.~A.~L. and {Alves}, J. and {Amorim}, A.~L. and {Ascasibar}, Y. and {Barrado-Navascues}, D. and {Barrera-Ballesteros}, J. and {Bekerait{\`e}}, S. and {Bland-Hawthorn}, J. and {Cano D{\'\i}az}, M. and {Cid Fernandes}, R. and {Cavichia}, O. and {Cortijo}, C. and {Dannerbauer}, H. and {Demleitner}, M. and {D{\'\i}az}, A. and {Dettmar}, R.~J. and {de Lorenzo-C{\'a}ceres}, A. and {del Olmo}, A. and {Galazzi}, A. and {Garc{\'\i}a-Lorenzo}, B. and {Gil de Paz}, A. and {Gonz{\'a}lez Delgado}, R. and {Holmes}, L. and {Igl{\'e}sias-P{\'a}ramo}, J. and {Kehrig}, C. and {Kelz}, A. and {Kennicutt}, R.~C. and {Kleemann}, B. and {Lacerda}, E.~A.~D. and {L{\'o}pez Fern{\'a}ndez}, R. and {L{\'o}pez S{\'a}nchez}, A.~R. and {Lyubenova}, M. and {Marino}, R. and {M{\'a}rquez}, I. and {Mendez-Abreu}, J. and {Moll{\'a}}, M. and {Monreal-Ibero}, A. and {Ortega Minakata}, R. and {Torres-Papaqui}, J.~P. and {P{\'e}rez}, E. and {Rosales-Ortega}, F.~F. and {Roth}, M.~M. and {S{\'a}nchez-Bl{\'a}zquez}, P. and {Schilling}, U. and {Spekkens}, K. and {Vale Asari}, N. and {van den Bosch}, R.~C.~E. and {van de Ven}, G. and {Vilchez}, J.~M. and {Wild}, V. and {Wisotzki}, L. and {Y{\i}ld{\i}r{\i}m}, A. and {Ziegler}, B.},
        title = "{CALIFA, the Calar Alto Legacy Integral Field Area survey. IV. Third public data release}",
      journal = {\aap},
         year = 2016,
        month = oct,
       volume = {594},
          eid = {A36},
        pages = {A36},
          doi = {10.1051/0004-6361/201628661},
archivePrefix = {arXiv},
       eprint = {1604.02289},
 primaryClass = {astro-ph.GA},
       adsurl = {https://ui.adsabs.harvard.edu/abs/2016A&A...594A..36S}
}

@ARTICLE{Croom2021_sami_survey,
       author = {{Croom}, Scott M. and {Owers}, Matt S. and {Scott}, Nicholas and {Poetrodjojo}, Henry and {Groves}, Brent and {van de Sande}, Jesse and {Barone}, Tania M. and {Cortese}, Luca and {D'Eugenio}, Francesco and {Bland-Hawthorn}, Joss and {Bryant}, Julia and {Oh}, Sree and {Brough}, Sarah and {Agostino}, James and {Casura}, Sarah and {Catinella}, Barbara and {Colless}, Matthew and {Cecil}, Gerald and {Davies}, Roger L. and {Drinkwater}, Michael J. and {Driver}, Simon P. and {Ferreras}, Ignacio and {Foster}, Caroline and {Fraser-McKelvie}, Amelia and {Lawrence}, Jon and {Leslie}, Sarah K. and {Liske}, Jochen and {L{\'o}pez-S{\'a}nchez}, {\'A}ngel R. and {Lorente}, Nuria P.~F. and {McElroy}, Rebecca and {Medling}, Anne M. and {Obreschkow}, Danail and {Richards}, Samuel N. and {Sharp}, Rob and {Sweet}, Sarah M. and {Taranu}, Dan S. and {Taylor}, Edward N. and {Tescari}, Edoardo and {Thomas}, Adam D. and {Tocknell}, James and {Vaughan}, Sam P.},
        title = "{The SAMI Galaxy Survey: the third and final data release}",
      journal = {\mnras},
         year = 2021,
        month = jul,
       volume = {505},
       number = {1},
        pages = {991-1016},
          doi = {10.1093/mnras/stab229},
archivePrefix = {arXiv},
       eprint = {2101.12224},
 primaryClass = {astro-ph.GA},
       adsurl = {https://ui.adsabs.harvard.edu/abs/2021MNRAS.505..991C}
}

@ARTICLE{Courteau2015_rc,
       author = {{Courteau}, St{\'e}phane and {Dutton}, Aaron A.},
        title = "{On the Global Mass Distribution in Disk Galaxies}",
      journal = {\apjl},
         year = 2015,
        month = mar,
       volume = {801},
       number = {2},
          eid = {L20},
        pages = {L20},
          doi = {10.1088/2041-8205/801/2/L20},
archivePrefix = {arXiv},
       eprint = {1502.04709},
 primaryClass = {astro-ph.GA},
       adsurl = {https://ui.adsabs.harvard.edu/abs/2015ApJ...801L..20C}
}

@ARTICLE{Courteau1997_rc,
       author = {{Courteau}, Stephane},
        title = "{Optical Rotation Curves and Linewidths for Tully-Fisher Applications}",
      journal = {\aj},
         year = 1997,
        month = dec,
       volume = {114},
        pages = {2402},
          doi = {10.1086/118656},
archivePrefix = {arXiv},
       eprint = {astro-ph/9709201},
 primaryClass = {astro-ph},
       adsurl = {https://ui.adsabs.harvard.edu/abs/1997AJ....114.2402C}
}

@ARTICLE{Tiley2019_kges,
       author = {{Tiley}, Alfred L. and {Swinbank}, A.~M. and {Harrison}, C.~M. and {Smail}, Ian and {Turner}, O.~J. and {Schaller}, M. and {Stott}, J.~P. and {Sobral}, D. and {Theuns}, T. and {Sharples}, R.~M. and {Gillman}, S. and {Bower}, R.~G. and {Bunker}, A.~J. and {Best}, P. and {Richard}, J. and {Bacon}, Roland and {Bureau}, M. and {Cirasuolo}, M. and {Magdis}, G.},
        title = "{The shapes of the rotation curves of star-forming galaxies over the last {\ensuremath{\approx}}10 Gyr}",
      journal = {\mnras},
         year = 2019,
        month = may,
       volume = {485},
       number = {1},
        pages = {934-960},
          doi = {10.1093/mnras/stz428},
archivePrefix = {arXiv},
       eprint = {1811.05982},
 primaryClass = {astro-ph.GA},
       adsurl = {https://ui.adsabs.harvard.edu/abs/2019MNRAS.485..934T}
}

@ARTICLE{Harrison2017_kross,
       author = {{Harrison}, C.~M. and {Johnson}, H.~L. and {Swinbank}, A.~M. and {Stott}, J.~P. and {Bower}, R.~G. and {Smail}, Ian and {Tiley}, A.~L. and {Bunker}, A.~J. and {Cirasuolo}, M. and {Sobral}, D. and {Sharples}, R.~M. and {Best}, P. and {Bureau}, M. and {Jarvis}, M.~J. and {Magdis}, G.},
        title = "{The KMOS Redshift One Spectroscopic Survey (KROSS): rotational velocities and angular momentum of z {\ensuremath{\approx}} 0.9 galaxies★}",
      journal = {\mnras},
         year = 2017,
        month = may,
       volume = {467},
       number = {2},
        pages = {1965-1983},
          doi = {10.1093/mnras/stx217},
archivePrefix = {arXiv},
       eprint = {1701.05561},
 primaryClass = {astro-ph.GA},
       adsurl = {https://ui.adsabs.harvard.edu/abs/2017MNRAS.467.1965H}
}

@ARTICLE{Deg2022_barolo,
       author = {{Deg}, N. and {Spekkens}, K. and {Westmeier}, T. and {Reynolds}, T.~N. and {Venkataraman}, P. and {Goliath}, S. and {Shen}, A.~X. and {Halloran}, R. and {Bosma}, A. and {Catinella}, B. and {de Blok}, W.~J.~G. and {D{\'e}nes}, H. and {Di Teodoro}, E.~M. and {Elagali}, A. and {For}, B.-Q. and {Howlett}, C. and {J{\'o}zsa}, G.~I.~G. and {Kamphuis}, P. and {Kleiner}, D. and {Koribalski}, B. and {Lee-Waddell}, K. and {Lelli}, F. and {Lin}, X. and {Murugeshan}, C. and {Oh}, S. and {Rhee}, J. and {Scott}, T.~C. and {Staveley-Smith}, L. and {van der Hulst}, J.~M. and {Verdes-Montenegro}, L. and {Wang}, J. and {Wong}, O.~I.},
        title = "{WALLABY Pilot Survey: Public release of HI kinematic models for more than 100 galaxies from phase 1 of ASKAP pilot observations}",
      journal = {\pasa},
         year = 2022,
        month = nov,
       volume = {39},
          eid = {e059},
        pages = {e059},
          doi = {10.1017/pasa.2022.43},
archivePrefix = {arXiv},
       eprint = {2211.07333},
 primaryClass = {astro-ph.GA},
       adsurl = {https://ui.adsabs.harvard.edu/abs/2022PASA...39...59D}
}

@ARTICLE{Westmeier2022_barolo,
       author = {{Westmeier}, T. and {Deg}, N. and {Spekkens}, K. and {Reynolds}, T.~N. and {Shen}, A.~X. and {Gaudet}, S. and {Goliath}, S. and {Huynh}, M.~T. and {Venkataraman}, P. and {Lin}, X. and {O'Beirne}, T. and {Catinella}, B. and {Cortese}, L. and {D{\'e}nes}, H. and {Elagali}, A. and {For}, B.-Q. and {J{\'o}zsa}, G.~I.~G. and {Howlett}, C. and {van der Hulst}, J.~M. and {Jurek}, R.~J. and {Kamphuis}, P. and {Kilborn}, V.~A. and {Kleiner}, D. and {Koribalski}, B.~S. and {Lee-Waddell}, K. and {Murugeshan}, C. and {Rhee}, J. and {Serra}, P. and {Shao}, L. and {Staveley-Smith}, L. and {Wang}, J. and {Wong}, O.~I. and {Zwaan}, M.~A. and {Allison}, J.~R. and {Anderson}, C.~S. and {Ball}, Lewis and {Bock}, D.~C.-J. and {Brodrick}, D. and {Bunton}, J.~D. and {Cooray}, F.~R. and {Gupta}, N. and {Hayman}, D.~B. and {Mahony}, E.~K. and {Moss}, V.~A. and {Ng}, A. and {Pearce}, S.~E. and {Raja}, W. and {Roxby}, D.~N. and {Voronkov}, M.~A. and {Warhurst}, K.~A. and {Courtois}, H.~M. and {Said}, K.},
        title = "{WALLABY pilot survey: Public release of H I data for almost 600 galaxies from phase 1 of ASKAP pilot observations}",
      journal = {\pasa},
         year = 2022,
        month = nov,
       volume = {39},
          eid = {e058},
        pages = {e058},
          doi = {10.1017/pasa.2022.50},
archivePrefix = {arXiv},
       eprint = {2211.07094},
 primaryClass = {astro-ph.GA},
       adsurl = {https://ui.adsabs.harvard.edu/abs/2022PASA...39...58W}
}

@ARTICLE{Iorio2017_galpak,
       author = {{Iorio}, G. and {Fraternali}, F. and {Nipoti}, C. and {Di Teodoro}, E. and {Read}, J.~I. and {Battaglia}, G.},
        title = "{LITTLE THINGS in 3D: robust determination of the circular velocity of dwarf irregular galaxies}",
      journal = {\mnras},
         year = 2017,
        month = apr,
       volume = {466},
       number = {4},
        pages = {4159-4192},
          doi = {10.1093/mnras/stw3285},
archivePrefix = {arXiv},
       eprint = {1611.03865},
 primaryClass = {astro-ph.GA},
       adsurl = {https://ui.adsabs.harvard.edu/abs/2017MNRAS.466.4159I}
}

@ARTICLE{DiTeodoro2016_barolo,
       author = {{Di Teodoro}, E.~M. and {Fraternali}, F. and {Miller}, S.~H.},
        title = "{Flat rotation curves and low velocity dispersions in KMOS star-forming galaxies at z \raisebox{-0.5ex}\textasciitilde 1}",
      journal = {\aap},
         year = 2016,
        month = oct,
       volume = {594},
          eid = {A77},
        pages = {A77},
          doi = {10.1051/0004-6361/201628315},
archivePrefix = {arXiv},
       eprint = {1602.04942},
 primaryClass = {astro-ph.GA},
       adsurl = {https://ui.adsabs.harvard.edu/abs/2016A&A...594A..77D}
}

@ARTICLE{Mason2017galpak,
       author = {{Mason}, Charlotte A. and {Treu}, Tommaso and {Fontana}, Adriano and {Jones}, Tucker and {Morishita}, Takahiro and {Amorin}, Ricardo and {Brada{\v{c}}}, Maru{\v{s}}a and {Quinn Finney}, Emily and {Grillo}, Claudio and {Henry}, Alaina and {Hoag}, Austin and {Huang}, Kuang-Han and {Schmidt}, Kasper B. and {Trenti}, Michele and {Vulcani}, Benedetta},
        title = "{First Results from the KMOS Lens-Amplified Spectroscopic Survey (KLASS): Kinematics of Lensed Galaxies at Cosmic Noon}",
      journal = {\apj},
         year = 2017,
        month = mar,
       volume = {838},
       number = {1},
          eid = {14},
        pages = {14},
          doi = {10.3847/1538-4357/aa60c4},
archivePrefix = {arXiv},
       eprint = {1610.03075},
 primaryClass = {astro-ph.GA},
       adsurl = {https://ui.adsabs.harvard.edu/abs/2017ApJ...838...14M}
}

@ARTICLE{Szakacs2021galpak,
       author = {{Szakacs}, Roland and {P{\'e}roux}, C{\'e}line and {Zwaan}, Martin and {Hamanowicz}, Aleksandra and {Klitsch}, Anne and {Fresco}, Alejandra Y. and {Augustin}, Ramona and {Biggs}, Andrew and {Kulkarni}, Varsha and {Rahmani}, Hadi},
        title = "{MUSE-ALMA haloes VI: coupling atomic, ionized, and molecular gas kinematics of galaxies}",
      journal = {\mnras},
         year = 2021,
        month = aug,
       volume = {505},
       number = {4},
        pages = {4746-4761},
          doi = {10.1093/mnras/stab1434},
archivePrefix = {arXiv},
       eprint = {2105.07280},
 primaryClass = {astro-ph.GA},
       adsurl = {https://ui.adsabs.harvard.edu/abs/2021MNRAS.505.4746S}
}

@ARTICLE{Ciocan2022_galpak,
       author = {{Ciocan}, B.~I. and {Ziegler}, B.~L. and {B{\"o}hm}, A. and {Verdugo}, M. and {Maier}, C.},
        title = "{Morpho-kinematics of MACS J0416.1-2403 low-mass galaxies}",
      journal = {\aap},
         year = 2022,
        month = nov,
       volume = {667},
          eid = {A61},
        pages = {A61},
          doi = {10.1051/0004-6361/202244131},
archivePrefix = {arXiv},
       eprint = {2208.10388},
 primaryClass = {astro-ph.GA},
       adsurl = {https://ui.adsabs.harvard.edu/abs/2022A&A...667A..61C}
}

@ARTICLE{Bacon2015_galpak,
       author = {{Bacon}, R. and {Brinchmann}, J. and {Richard}, J. and {Contini}, T. and {Drake}, A. and {Franx}, M. and {Tacchella}, S. and {Vernet}, J. and {Wisotzki}, L. and {Blaizot}, J. and {Bouch{\'e}}, N. and {Bouwens}, R. and {Cantalupo}, S. and {Carollo}, C.~M. and {Carton}, D. and {Caruana}, J. and {Cl{\'e}ment}, B. and {Dreizler}, S. and {Epinat}, B. and {Guiderdoni}, B. and {Herenz}, C. and {Husser}, T.-O. and {Kamann}, S. and {Kerutt}, J. and {Kollatschny}, W. and {Krajnovic}, D. and {Lilly}, S. and {Martinsson}, T. and {Michel-Dansac}, L. and {Patricio}, V. and {Schaye}, J. and {Shirazi}, M. and {Soto}, K. and {Soucail}, G. and {Steinmetz}, M. and {Urrutia}, T. and {Weilbacher}, P. and {de Zeeuw}, T.},
        title = "{The MUSE 3D view of the Hubble Deep Field South}",
      journal = {\aap},
         year = 2015,
        month = mar,
       volume = {575},
          eid = {A75},
        pages = {A75},
          doi = {10.1051/0004-6361/201425419},
archivePrefix = {arXiv},
       eprint = {1411.7667},
 primaryClass = {astro-ph.GA},
       adsurl = {https://ui.adsabs.harvard.edu/abs/2015A&A...575A..75B}
}

@ARTICLE{Bouche2021_galpak,
       author = {{Bouch{\'e}}, Nicolas F. and {Genel}, Shy and {Pellissier}, Alisson and {Dubois}, C{\'e}dric and {Contini}, Thierry and {Epinat}, Beno{\^\i}t and {Pillepich}, Annalisa and {Krajnovi{\'c}}, Davor and {Nelson}, Dylan and {Abril-Melgarejo}, Valentina and {Richard}, Johan and {Boogaard}, Leindert and {Maseda}, Michael and {Mercier}, Wilfried and {Bacon}, Roland and {Steinmetz}, Matthias and {Vogelsberger}, Mark},
        title = "{The MUSE Hubble Ultra Deep Field Survey. XVI. The angular momentum of low-mass star-forming galaxies: A cautionary tale and insights from TNG50}",
      journal = {\aap},
         year = 2021,
        month = oct,
       volume = {654},
          eid = {A49},
        pages = {A49},
          doi = {10.1051/0004-6361/202040225},
archivePrefix = {arXiv},
       eprint = {2101.12250},
 primaryClass = {astro-ph.GA},
       adsurl = {https://ui.adsabs.harvard.edu/abs/2021A&A...654A..49B}
}

@ARTICLE{EspejoSalcedo2025_ang,
       author = {{Espejo Salcedo}, Juan M. and {Glazebrook}, Karl and {Fisher}, Deanne B. and {Sweet}, Sarah M. and {Obreschkow}, Danail and {Schreiber}, N.~M. F{\"o}rster},
        title = "{A shallow slope for the stellar mass-angular momentum relation of star-forming galaxies at 1.5 < z < 2.5}",
      journal = {\mnras},
         year = 2025,
        month = jan,
       volume = {536},
       number = {2},
        pages = {1188-1216},
          doi = {10.1093/mnras/stae2647},
archivePrefix = {arXiv},
       eprint = {2411.17312},
 primaryClass = {astro-ph.GA},
       adsurl = {https://ui.adsabs.harvard.edu/abs/2025MNRAS.536.1188E}
}

@ARTICLE{Rizzo2023_dispersion,
       author = {{Rizzo}, F. and {Roman-Oliveira}, F. and {Fraternali}, F. and {Frickmann}, D. and {Valentino}, F.~M. and {Brammer}, G. and {Zanella}, A. and {Kokorev}, V. and {Popping}, G. and {Whitaker}, K.~E. and {Kohandel}, M. and {Magdis}, G.~E. and {Di Mascolo}, L. and {Ikeda}, R. and {Jin}, S. and {Toft}, S.},
        title = "{The ALMA-ALPAKA survey. I. High-resolution CO and [CI] kinematics of star-forming galaxies at z = 0.5-3.5}",
      journal = {\aap},
         year = 2023,
        month = nov,
       volume = {679},
          eid = {A129},
        pages = {A129},
          doi = {10.1051/0004-6361/202346444},
archivePrefix = {arXiv},
       eprint = {2303.16227},
 primaryClass = {astro-ph.GA},
       adsurl = {https://ui.adsabs.harvard.edu/abs/2023A&A...679A.129R}
}

@ARTICLE{RomanOliveira2024_dispersion,
       author = {{Roman-Oliveira}, F. and {Rizzo}, F. and {Fraternali}, F.},
        title = "{Dynamical modelling and the origin of gas turbulence in z {\ensuremath{\sim}} 4.5 galaxies}",
      journal = {\aap},
         year = 2024,
        month = jul,
       volume = {687},
          eid = {A35},
        pages = {A35},
          doi = {10.1051/0004-6361/202348828},
archivePrefix = {arXiv},
       eprint = {2403.00904},
 primaryClass = {astro-ph.GA},
       adsurl = {https://ui.adsabs.harvard.edu/abs/2024A&A...687A..35R}
}

@ARTICLE{Danhaive2025_fdm,
       author = {{Danhaive}, A. Lola and {Tacchella}, Sandro and {Bunker}, Andrew J. and {Curtis-Lake}, Emma and {de Graaff}, Anna and {D'Eugenio}, Francesco and {Duan}, Qiao and {Egami}, Eiichi and {Eisenstein}, Daniel J. and {Johnson}, Benjamin D. and {Maiolino}, Roberto and {McClymont}, William and {Rieke}, Marcia and {Robertson}, Brant and {Sun}, Fengwu and {Willmer}, Christopher N.~A. and {Wu}, Zihao and {Zhu}, Yongda},
        title = "{The dark side of early galaxies: $\texttt{geko}$ uncovers dark-matter fractions at $z\sim4-6$}",
      journal = {arXiv e-prints},
         year = 2025,
        month = oct,
          eid = {arXiv:2510.14779},
        pages = {arXiv:2510.14779},
          doi = {10.48550/arXiv.2510.14779},
archivePrefix = {arXiv},
       eprint = {2510.14779},
 primaryClass = {astro-ph.GA},
       adsurl = {https://ui.adsabs.harvard.edu/abs/2025arXiv251014779D}
}

@ARTICLE{Danhaive2025_dispersion,
       author = {{Danhaive}, A. Lola and {Tacchella}, Sandro and {{\"U}bler}, Hannah and {de Graaff}, Anna and {Egami}, Eiichi and {Johnson}, Benjamin D. and {Sun}, Fengwu and {Arribas}, Santiago and {Bunker}, Andrew J. and {Carniani}, Stefano and {Jones}, Gareth C. and {Maiolino}, Roberto and {McClymont}, William and {Parlanti}, Eleonora and {Simmonds}, Charlotte and {Villanueva}, Natalia C. and {Baker}, William M. and {Jaffe}, Daniel T. and {Eisenstein}, Daniel and {Hainline}, Kevin and {Helton}, Jakob M. and {Ji}, Zhiyuan and {Lin}, Xiaojing and {Liu}, Yichen and {Pusk{\'a}s}, D{\'a}vid and {Rieke}, Marcia and {Rinaldi}, Pierluigi and {Robertson}, Brant and {Scholz}, Jan and {Williams}, Christina C. and {Willmer}, Christopher N.~A.},
        title = "{The dawn of discs: unveiling the turbulent ionized gas kinematics of the galaxy population at z {\ensuremath{\sim}} 4{\textendash}6 with JWST/NIRCam grism spectroscopy}",
      journal = {\mnras},
         year = 2025,
        month = nov,
       volume = {543},
       number = {4},
        pages = {3249-3302},
          doi = {10.1093/mnras/staf1540},
       adsurl = {https://ui.adsabs.harvard.edu/abs/2025MNRAS.543.3249D}
}

@ARTICLE{vanDerWel2025_disks,
       author = {{van der Wel}, Arjen and {Meidt}, Sharon E.},
        title = "{Transforming galaxies with EASE: widespread structural changes enabled by short-lived spirals}",
      journal = {arXiv e-prints},
         year = 2025,
        month = sep,
          eid = {arXiv:2509.02847},
        pages = {arXiv:2509.02847},
          doi = {10.48550/arXiv.2509.02847},
archivePrefix = {arXiv},
       eprint = {2509.02847},
 primaryClass = {astro-ph.GA},
       adsurl = {https://ui.adsabs.harvard.edu/abs/2025arXiv250902847V}
}

@article{Davies2011,
	title = {How well can we measure the intrinsic velocity dispersion of distant disk galaxies?},
	volume = {741},
	issn = {15384357},
	doi = {10.1088/0004-637X/741/2/69},
	number = {2},
	urldate = {2020-12-30},
	journal = {Astrophysical Journal},
	author = {Davies, R and Schreiber, N. M.Förster and Cresci, G and Genzel, R and Bouché, N and Burkert, A and Buschkamp, P and Genel, S and Hicks, E and Kurk, J and Lutz, D and Newman, S and Shapiro, K and Sternberg, A and Tacconi, L J and Wuyts, S},
	year = {2011},
	note = {arXiv: 1108.0285},
	pages = {69},
}

@ARTICLE{Cresci2009_dysmalpy,
       author = {{Cresci}, G. and {Hicks}, E.~K.~S. and {Genzel}, R. and {F{\"o}rster Schreiber}, N.~M. and {Davies}, R. and {Bouch{\'e}}, N. and {Buschkamp}, P. and {Genel}, S. and {Shapiro}, K. and {Tacconi}, L. and {Sommer-Larsen}, J. and {Burkert}, A. and {Eisenhauer}, F. and {Gerhard}, O. and {Lutz}, D. and {Naab}, T. and {Sternberg}, A. and {Cimatti}, A. and {Daddi}, E. and {Erb}, D.~K. and {Kurk}, J. and {Lilly}, S.~L. and {Renzini}, A. and {Shapley}, A. and {Steidel}, C.~C. and {Caputi}, K.},
        title = "{The SINS Survey: Modeling the Dynamics of z \raisebox{-0.5ex}\textasciitilde 2 Galaxies and the High-z Tully-Fisher Relation}",
      journal = {\apj},
         year = 2009,
        month = may,
       volume = {697},
       number = {1},
        pages = {115-132},
          doi = {10.1088/0004-637X/697/1/115},
archivePrefix = {arXiv},
       eprint = {0902.4701},
 primaryClass = {astro-ph.CO},
       adsurl = {https://ui.adsabs.harvard.edu/abs/2009ApJ...697..115C}
}

@ARTICLE{Davies2004b_dysmalpy,
       author = {{Davies}, R.~I. and {Tacconi}, L.~J. and {Genzel}, R.},
        title = "{The Nuclear Gas Dynamics and Star Formation of Markarian 231}",
      journal = {\apj},
         year = 2004,
        month = oct,
       volume = {613},
       number = {2},
        pages = {781-793},
          doi = {10.1086/423315},
archivePrefix = {arXiv},
       eprint = {astro-ph/0406342},
 primaryClass = {astro-ph},
       adsurl = {https://ui.adsabs.harvard.edu/abs/2004ApJ...613..781D}
}

@ARTICLE{Davies2004_dysmalpy,
       author = {{Davies}, R.~I. and {Tacconi}, L.~J. and {Genzel}, R.},
        title = "{The Nuclear Gasdynamics and Star Formation of NGC 7469}",
      journal = {\apj},
         year = 2004,
        month = feb,
       volume = {602},
       number = {1},
        pages = {148-161},
          doi = {10.1086/380995},
archivePrefix = {arXiv},
       eprint = {astro-ph/0310681},
 primaryClass = {astro-ph},
       adsurl = {https://ui.adsabs.harvard.edu/abs/2004ApJ...602..148D}
}

@ARTICLE{Genzel2011_clumps,
       author = {{Genzel}, R. and {Newman}, S. and {Jones}, T. and {F{\"o}rster Schreiber}, N.~M. and {Shapiro}, K. and {Genel}, S. and {Lilly}, S.~J. and {Renzini}, A. and {Tacconi}, L.~J. and {Bouch{\'e}}, N. and {Burkert}, A. and {Cresci}, G. and {Buschkamp}, P. and {Carollo}, C.~M. and {Ceverino}, D. and {Davies}, R. and {Dekel}, A. and {Eisenhauer}, F. and {Hicks}, E. and {Kurk}, J. and {Lutz}, D. and {Mancini}, C. and {Naab}, T. and {Peng}, Y. and {Sternberg}, A. and {Vergani}, D. and {Zamorani}, G.},
        title = "{The Sins Survey of z \raisebox{-0.5ex}\textasciitilde 2 Galaxy Kinematics: Properties of the Giant Star-forming Clumps}",
      journal = {\apj},
         year = 2011,
        month = jun,
       volume = {733},
       number = {2},
          eid = {101},
        pages = {101},
          doi = {10.1088/0004-637X/733/2/101},
archivePrefix = {arXiv},
       eprint = {1011.5360},
 primaryClass = {astro-ph.CO},
       adsurl = {https://ui.adsabs.harvard.edu/abs/2011ApJ...733..101G}
}

@ARTICLE{Rizzo2024_dispersion,
       author = {{Rizzo}, F. and {Bacchini}, C. and {Kohandel}, M. and {Di Mascolo}, L. and {Fraternali}, F. and {Roman-Oliveira}, F. and {Zanella}, A. and {Popping}, G. and {Valentino}, F. and {Magdis}, G. and {Whitaker}, K.},
        title = "{The ALMA-ALPAKA survey: II. Evolution of turbulence in galaxy disks across cosmic time: Difference between cold and warm gas}",
      journal = {\aap},
         year = 2024,
        month = sep,
       volume = {689},
          eid = {A273},
        pages = {A273},
          doi = {10.1051/0004-6361/202450455},
archivePrefix = {arXiv},
       eprint = {2407.06261},
 primaryClass = {astro-ph.GA},
       adsurl = {https://ui.adsabs.harvard.edu/abs/2024A&A...689A.273R}
}

@ARTICLE{deGraff2024_dispersion,
       author = {{de Graaff}, Anna and {Rix}, Hans-Walter and {Carniani}, Stefano and {Suess}, Katherine A. and {Charlot}, St{\'e}phane and {Curtis-Lake}, Emma and {Arribas}, Santiago and {Baker}, William M. and {Boyett}, Kristan and {Bunker}, Andrew J. and {Cameron}, Alex J. and {Chevallard}, Jacopo and {Curti}, Mirko and {Eisenstein}, Daniel J. and {Franx}, Marijn and {Hainline}, Kevin and {Hausen}, Ryan and {Ji}, Zhiyuan and {Johnson}, Benjamin D. and {Jones}, Gareth C. and {Maiolino}, Roberto and {Maseda}, Michael V. and {Nelson}, Erica and {Parlanti}, Eleonora and {Rawle}, Tim and {Robertson}, Brant and {Tacchella}, Sandro and {{\"U}bler}, Hannah and {Williams}, Christina C. and {Willmer}, Christopher N.~A. and {Willott}, Chris},
        title = "{Ionised gas kinematics and dynamical masses of z {\ensuremath{\gtrsim}} 6 galaxies from JADES/NIRSpec high-resolution spectroscopy}",
      journal = {\aap},
         year = 2024,
        month = apr,
       volume = {684},
          eid = {A87},
        pages = {A87},
          doi = {10.1051/0004-6361/202347755},
archivePrefix = {arXiv},
       eprint = {2308.09742},
 primaryClass = {astro-ph.GA},
       adsurl = {https://ui.adsabs.harvard.edu/abs/2024A&A...684A..87D}
}

@ARTICLE{deBlok2010_corecusp,
       author = {{de Blok}, W.~J.~G.},
        title = "{The Core-Cusp Problem}",
      journal = {Advances in Astronomy},
         year = 2010,
        month = jan,
       volume = {2010},
          eid = {789293},
        pages = {789293},
          doi = {10.1155/2010/789293},
archivePrefix = {arXiv},
       eprint = {0910.3538},
 primaryClass = {astro-ph.CO},
       adsurl = {https://ui.adsabs.harvard.edu/abs/2010AdAst2010E...5D}
}

@ARTICLE{Einasto1965,
       author = {{Einasto}, J.},
        title = "{On the Construction of a Composite Model for the Galaxy and on the Determination of the System of Galactic Parameters}",
      journal = {Trudy Astrofizicheskogo Instituta Alma-Ata},
         year = 1965,
        month = jan,
       volume = {5},
        pages = {87-100},
       adsurl = {https://ui.adsabs.harvard.edu/abs/1965TrAlm...5...87E}
}

@ARTICLE{Zhao1996,
       author = {{Zhao}, Hongsheng},
        title = "{Analytical models for galactic nuclei}",
      journal = {\mnras},
         year = 1996,
        month = jan,
       volume = {278},
       number = {2},
        pages = {488-496},
          doi = {10.1093/mnras/278.2.488},
archivePrefix = {arXiv},
       eprint = {astro-ph/9509122},
 primaryClass = {astro-ph},
       adsurl = {https://ui.adsabs.harvard.edu/abs/1996MNRAS.278..488Z}
}

@ARTICLE{Read2019_dmcores,
       author = {{Read}, J.~I. and {Walker}, M.~G. and {Steger}, P.},
        title = "{Dark matter heats up in dwarf galaxies}",
      journal = {\mnras},
         year = 2019,
        month = mar,
       volume = {484},
       number = {1},
        pages = {1401-1420},
          doi = {10.1093/mnras/sty3404},
archivePrefix = {arXiv},
       eprint = {1808.06634},
 primaryClass = {astro-ph.GA},
       adsurl = {https://ui.adsabs.harvard.edu/abs/2019MNRAS.484.1401R}
}

@ARTICLE{Navarro2004,
       author = {{Navarro}, J.~F. and {Hayashi}, E. and {Power}, C. and {Jenkins}, A.~R. and {Frenk}, C.~S. and {White}, S.~D.~M. and {Springel}, V. and {Stadel}, J. and {Quinn}, T.~R.},
        title = "{The inner structure of {\ensuremath{\Lambda}}CDM haloes - III. Universality and asymptotic slopes}",
      journal = {\mnras},
         year = 2004,
        month = apr,
       volume = {349},
       number = {3},
        pages = {1039-1051},
          doi = {10.1111/j.1365-2966.2004.07586.x},
archivePrefix = {arXiv},
       eprint = {astro-ph/0311231},
 primaryClass = {astro-ph},
       adsurl = {https://ui.adsabs.harvard.edu/abs/2004MNRAS.349.1039N}
}

@ARTICLE{Burkert2015_cores,
       author = {{Burkert}, A.},
        title = "{The Structure and Dark Halo Core Properties of Dwarf Spheroidal Galaxies}",
      journal = {\apj},
         year = 2015,
        month = aug,
       volume = {808},
       number = {2},
          eid = {158},
        pages = {158},
          doi = {10.1088/0004-637X/808/2/158},
archivePrefix = {arXiv},
       eprint = {1501.06604},
 primaryClass = {astro-ph.GA},
       adsurl = {https://ui.adsabs.harvard.edu/abs/2015ApJ...808..158B}
}

@ARTICLE{Oh2011_little_things,
       author = {{Oh}, Se-Heon and {de Blok}, W.~J.~G. and {Brinks}, Elias and {Walter}, Fabian and {Kennicutt}, Jr., Robert C.},
        title = "{Dark and Luminous Matter in THINGS Dwarf Galaxies}",
      journal = {\aj},
         year = 2011,
        month = jun,
       volume = {141},
       number = {6},
          eid = {193},
        pages = {193},
          doi = {10.1088/0004-6256/141/6/193},
archivePrefix = {arXiv},
       eprint = {1011.0899},
 primaryClass = {astro-ph.CO},
       adsurl = {https://ui.adsabs.harvard.edu/abs/2011AJ....141..193O}
}

@ARTICLE{Oh2015_little_things,
       author = {{Oh}, Se-Heon and {Hunter}, Deidre A. and {Brinks}, Elias and {Elmegreen}, Bruce G. and {Schruba}, Andreas and {Walter}, Fabian and {Rupen}, Michael P. and {Young}, Lisa M. and {Simpson}, Caroline E. and {Johnson}, Megan C. and {Herrmann}, Kimberly A. and {Ficut-Vicas}, Dana and {Cigan}, Phil and {Heesen}, Volker and {Ashley}, Trisha and {Zhang}, Hong-Xin},
        title = "{High-resolution Mass Models of Dwarf Galaxies from LITTLE THINGS}",
      journal = {\aj},
         year = 2015,
        month = jun,
       volume = {149},
       number = {6},
          eid = {180},
        pages = {180},
          doi = {10.1088/0004-6256/149/6/180},
archivePrefix = {arXiv},
       eprint = {1502.01281},
 primaryClass = {astro-ph.GA},
       adsurl = {https://ui.adsabs.harvard.edu/abs/2015AJ....149..180O}
}

@ARTICLE{Stott2016_kross,
       author = {{Stott}, John P. and {Swinbank}, A.~M. and {Johnson}, Helen L. and {Tiley}, Alfie and {Magdis}, Georgios and {Bower}, Richard and {Bunker}, Andrew J. and {Bureau}, Martin and {Harrison}, Chris M. and {Jarvis}, Matt J. and {Sharples}, Ray and {Smail}, Ian and {Sobral}, David and {Best}, Philip and {Cirasuolo}, Michele},
        title = "{The KMOS Redshift One Spectroscopic Survey (KROSS): dynamical properties, gas and dark matter fractions of typical z {\ensuremath{\sim}} 1 star-forming galaxies}",
      journal = {\mnras},
         year = 2016,
        month = apr,
       volume = {457},
       number = {2},
        pages = {1888-1904},
          doi = {10.1093/mnras/stw129},
archivePrefix = {arXiv},
       eprint = {1601.03400},
 primaryClass = {astro-ph.GA},
       adsurl = {https://ui.adsabs.harvard.edu/abs/2016MNRAS.457.1888S}
}

@ARTICLE{Kuhn2024_jwst_disks,
       author = {{Kuhn}, Vicki and {Guo}, Yicheng and {Martin}, Alec and {Bayless}, Julianna and {Gates}, Ellie and {Puleo}, AJ},
        title = "{JWST Reveals a Surprisingly High Fraction of Galaxies Being Spiral-like at 0.5 {\ensuremath{\leq}} z {\ensuremath{\leq}} 4}",
      journal = {\apjl},
         year = 2024,
        month = jun,
       volume = {968},
       number = {2},
          eid = {L15},
        pages = {L15},
          doi = {10.3847/2041-8213/ad43eb},
archivePrefix = {arXiv},
       eprint = {2312.12389},
 primaryClass = {astro-ph.GA},
       adsurl = {https://ui.adsabs.harvard.edu/abs/2024ApJ...968L..15K}
}

@ARTICLE{Chugunov2025_jwst_features,
       author = {{Chugunov}, Ilia V. and {Marchuk}, Alexander A. and {Mosenkov}, Aleksandr V.},
        title = "{Less wound and more asymmetric: JWST confirms the evolution of spiral structure in galaxies at z {\ensuremath{\lesssim}} 3}",
      journal = {\pasa},
         year = 2025,
        month = jan,
       volume = {42},
          eid = {e029},
        pages = {e029},
          doi = {10.1017/pasa.2025.6},
archivePrefix = {arXiv},
       eprint = {2501.11670},
 primaryClass = {astro-ph.GA},
       adsurl = {https://ui.adsabs.harvard.edu/abs/2025PASA...42...29C}
}

@ARTICLE{Geron2025_jwst_features,
       author = {{G{\'e}ron}, Tobias and {Smethurst}, R.~J. and {Dickinson}, Hugh and {Fortson}, L.~F. and {Garland}, Izzy L. and {Kruk}, Sandor and {Lintott}, Chris and {Makechemu}, Jason Shingirai and {Mantha}, Kameswara Bharadwaj and {Masters}, Karen L. and {O'Ryan}, David and {Roberts}, Hayley and {Simmons}, B.~D. and {Walmsley}, Mike and {Calabr{\`o}}, Antonello and {Chiba}, Rimpei and {Costantin}, Luca and {Drout}, Maria R. and {Fragkoudi}, Francesca and {Guo}, Yuchen and {Holwerda}, B.~W. and {Jogee}, Shardha and {Koekemoer}, Anton M. and {Lucas}, Ray A. and {Pacucci}, Fabio},
        title = "{Galaxy Zoo CEERS: Bar Fractions Up to z {\ensuremath{\sim}} 4.0}",
      journal = {\apj},
         year = 2025,
        month = jul,
       volume = {987},
       number = {1},
          eid = {74},
        pages = {74},
          doi = {10.3847/1538-4357/add7d0},
archivePrefix = {arXiv},
       eprint = {2505.01421},
 primaryClass = {astro-ph.GA},
       adsurl = {https://ui.adsabs.harvard.edu/abs/2025ApJ...987...74G}
}

@ARTICLE{Kartaltepe2023_jwst_disks,
       author = {{Kartaltepe}, Jeyhan S. and {Rose}, Caitlin and {Vanderhoof}, Brittany N. and {McGrath}, Elizabeth J. and {Costantin}, Luca and {Cox}, Isabella G. and {Yung}, L.~Y. Aaron and {Kocevski}, Dale D. and {Wuyts}, Stijn and {Ferguson}, Henry C. and {Bagley}, Micaela B. and {Finkelstein}, Steven L. and {Amor{\'\i}n}, Ricardo O. and {Andrews}, Brett H. and {Arrabal Haro}, Pablo and {Backhaus}, Bren E. and {Behroozi}, Peter and {Bisigello}, Laura and {Calabr{\`o}}, Antonello and {Casey}, Caitlin M. and {Coogan}, Rosemary T. and {Cooper}, M.~C. and {Croton}, Darren and {de la Vega}, Alexander and {Dickinson}, Mark and {Fontana}, Adriano and {Franco}, Maximilien and {Grazian}, Andrea and {Grogin}, Norman A. and {Hathi}, Nimish P. and {Holwerda}, Benne W. and {Huertas-Company}, Marc and {Iyer}, Kartheik G. and {Jogee}, Shardha and {Jung}, Intae and {Kewley}, Lisa J. and {Kirkpatrick}, Allison and {Koekemoer}, Anton M. and {Liu}, James and {Lotz}, Jennifer M. and {Lucas}, Ray A. and {Newman}, Jeffrey A. and {Pacifici}, Camilla and {Pandya}, Viraj and {Papovich}, Casey and {Pentericci}, Laura and {P{\'e}rez-Gonz{\'a}lez}, Pablo G. and {Petersen}, Jayse and {Pirzkal}, Nor and {Rafelski}, Marc and {Ravindranath}, Swara and {Simons}, Raymond C. and {Snyder}, Gregory F. and {Somerville}, Rachel S. and {Stanway}, Elizabeth R. and {Straughn}, Amber N. and {Tacchella}, Sandro and {Trump}, Jonathan R. and {Vega-Ferrero}, Jes{\'u}s and {Wilkins}, Stephen M. and {Yang}, Guang and {Zavala}, Jorge A.},
        title = "{CEERS Key Paper. III. The Diversity of Galaxy Structure and Morphology at z = 3-9 with JWST}",
      journal = {\apjl},
         year = 2023,
        month = mar,
       volume = {946},
       number = {1},
          eid = {L15},
        pages = {L15},
          doi = {10.3847/2041-8213/acad01},
archivePrefix = {arXiv},
       eprint = {2210.14713},
 primaryClass = {astro-ph.GA},
       adsurl = {https://ui.adsabs.harvard.edu/abs/2023ApJ...946L..15K}
}

@ARTICLE{Huertas-Company2024_jwst_disks,
       author = {{Huertas-Company}, M. and {Iyer}, K.~G. and {Angeloudi}, E. and {Bagley}, M.~B. and {Finkelstein}, S.~L. and {Kartaltepe}, J. and {McGrath}, E.~J. and {Sarmiento}, R. and {Vega-Ferrero}, J. and {Arrabal Haro}, P. and {Behroozi}, P. and {Buitrago}, F. and {Cheng}, Y. and {Costantin}, L. and {Dekel}, A. and {Dickinson}, M. and {Elbaz}, D. and {Grogin}, N.~A. and {Hathi}, N.~P. and {Holwerda}, B.~W. and {Koekemoer}, A.~M. and {Lucas}, R.~A. and {Papovich}, C. and {P{\'e}rez-Gonz{\'a}lez}, P.~G. and {Pirzkal}, N. and {Seill{\'e}}, L. -M. and {de la Vega}, A. and {Wuyts}, S. and {Yang}, G. and {Yung}, L.~Y.~A.},
        title = "{Galaxy morphology from z {\ensuremath{\sim}} 6 through the lens of JWST}",
      journal = {\aap},
         year = 2024,
        month = may,
       volume = {685},
          eid = {A48},
        pages = {A48},
          doi = {10.1051/0004-6361/202346800},
archivePrefix = {arXiv},
       eprint = {2305.02478},
 primaryClass = {astro-ph.GA},
       adsurl = {https://ui.adsabs.harvard.edu/abs/2024A&A...685A..48H}
}

@ARTICLE{Tohill2024_jwst_disks,
       author = {{Tohill}, C. and {Bamford}, S.~P. and {Conselice}, C.~J. and {Ferreira}, L. and {Harvey}, T. and {Adams}, N. and {Austin}, D.},
        title = "{A Robust Study of High-redshift Galaxies: Unsupervised Machine Learning for Characterizing Morphology with JWST up to z {\ensuremath{\sim}} 8}",
      journal = {\apj},
         year = 2024,
        month = feb,
       volume = {962},
       number = {2},
          eid = {164},
        pages = {164},
          doi = {10.3847/1538-4357/ad17b8},
archivePrefix = {arXiv},
       eprint = {2306.17225},
 primaryClass = {astro-ph.GA},
       adsurl = {https://ui.adsabs.harvard.edu/abs/2024ApJ...962..164T}
}

@ARTICLE{Smethurst2025_jwst_disks,
       author = {{Smethurst}, R.~J. and {Simmons}, B.~D. and {G{\'e}ron}, T. and {Dickinson}, H. and {Fortson}, L. and {Garland}, I.~L. and {Kruk}, S. and {Jewell}, S.~M. and {Lintott}, C.~J. and {Makechemu}, J.~S. and {Mantha}, K.~B. and {Masters}, K.~L. and {O'Ryan}, D. and {Roberts}, H. and {Thorne}, M.~R. and {Walmsley}, M. and {Calabr{\`o}}, M. and {Holwerda}, B. and {Kartaltepe}, J.~S. and {Koekemoer}, A.~M. and {Lyu}, Y. and {Lucas}, R. and {Pacucci}, F. and {Tarrasse}, M.},
        title = "{Galaxy Zoo JWST: up to 75 per cent of discs are featureless at 3 < z < 7}",
      journal = {\mnras},
         year = 2025,
        month = may,
       volume = {539},
       number = {2},
        pages = {1359-1371},
          doi = {10.1093/mnras/staf506},
archivePrefix = {arXiv},
       eprint = {2503.21869},
 primaryClass = {astro-ph.GA},
       adsurl = {https://ui.adsabs.harvard.edu/abs/2025MNRAS.539.1359S}
}

@ARTICLE{Ferreira2023_jwst_disks,
       author = {{Ferreira}, Leonardo and {Conselice}, Christopher J. and {Sazonova}, Elizaveta and {Ferrari}, Fabricio and {Caruana}, Joseph and {Tohill}, Cl{\'a}r-Br{\'\i}d and {Lucatelli}, Geferson and {Adams}, Nathan and {Irodotou}, Dimitrios and {Marshall}, Madeline A. and {Roper}, Will J. and {Lovell}, Christopher C. and {Verma}, Aprajita and {Austin}, Duncan and {Trussler}, James and {Wilkins}, Stephen M.},
        title = "{The JWST Hubble Sequence: The Rest-frame Optical Evolution of Galaxy Structure at 1.5 < z < 6.5}",
      journal = {\apj},
         year = 2023,
        month = oct,
       volume = {955},
       number = {2},
          eid = {94},
        pages = {94},
          doi = {10.3847/1538-4357/acec76},
archivePrefix = {arXiv},
       eprint = {2210.01110},
 primaryClass = {astro-ph.GA},
       adsurl = {https://ui.adsabs.harvard.edu/abs/2023ApJ...955...94F}
}

@ARTICLE{EspejoSalcedo2025_morphology,
       author = {{Espejo Salcedo}, J.~M. and {Pastras}, S. and {V{\'a}cha}, J. and {Pulsoni}, C. and {Genzel}, R. and {F{\"o}rster Schreiber}, N.~M. and {Jolly}, J. -B. and {Barfety}, C. and {Chen}, J. and {Tozzi}, G. and {Liu}, D. and {Lee}, L.~L. and {Wuyts}, S. and {Tacconi}, L.~J. and {Davies}, R. and {{\"U}bler}, H. and {Lutz}, D. and {Wisnioski}, E. and {Shangguan}, J. and {Lee}, M. and {Price}, S.~H. and {Eisenhauer}, F. and {Renzini}, A. and {Nestor Shachar}, A. and {Herrera-Camus}, R.},
        title = "{Galaxy morphologies at cosmic noon with JWST: A foundation for exploring gas transport with bars and spiral arms}",
      journal = {\aap},
         year = 2025,
        month = aug,
       volume = {700},
          eid = {A42},
        pages = {A42},
          doi = {10.1051/0004-6361/202554725},
archivePrefix = {arXiv},
       eprint = {2503.21738},
 primaryClass = {astro-ph.GA},
       adsurl = {https://ui.adsabs.harvard.edu/abs/2025A&A...700A..42E}
}

@ARTICLE{Lelli2016_TFR,
       author = {{Lelli}, Federico and {McGaugh}, Stacy S. and {Schombert}, James M.},
        title = "{The Small Scatter of the Baryonic Tully-Fisher Relation}",
      journal = {\apjl},
         year = 2016,
        month = jan,
       volume = {816},
       number = {1},
          eid = {L14},
        pages = {L14},
          doi = {10.3847/2041-8205/816/1/L14},
archivePrefix = {arXiv},
       eprint = {1512.04543},
 primaryClass = {astro-ph.GA},
       adsurl = {https://ui.adsabs.harvard.edu/abs/2016ApJ...816L..14L}
}

@ARTICLE{Lelli2019_TFR,
       author = {{Lelli}, Federico and {McGaugh}, Stacy S. and {Schombert}, James M. and {Desmond}, Harry and {Katz}, Harley},
        title = "{The baryonic Tully-Fisher relation for different velocity definitions and implications for galaxy angular momentum}",
      journal = {\mnras},
         year = 2019,
        month = apr,
       volume = {484},
       number = {3},
        pages = {3267-3278},
          doi = {10.1093/mnras/stz205},
archivePrefix = {arXiv},
       eprint = {1901.05966},
 primaryClass = {astro-ph.GA},
       adsurl = {https://ui.adsabs.harvard.edu/abs/2019MNRAS.484.3267L}
}

@ARTICLE{Boubel2024_TFH0,
       author = {{Boubel}, Paula and {Colless}, Matthew and {Said}, Khaled and {Staveley-Smith}, Lister},
        title = "{An improved Tully-Fisher estimate of H$_{0}$}",
      journal = {\mnras},
         year = 2024,
        month = sep,
       volume = {533},
       number = {2},
        pages = {1550-1559},
          doi = {10.1093/mnras/stae1925},
archivePrefix = {arXiv},
       eprint = {2408.03660},
 primaryClass = {astro-ph.CO},
       adsurl = {https://ui.adsabs.harvard.edu/abs/2024MNRAS.533.1550B}
}

@ARTICLE{Schombert2020_TFH0,
       author = {{Schombert}, James and {McGaugh}, Stacy and {Lelli}, Federico},
        title = "{Using the Baryonic Tully-Fisher Relation to Measure H$_{o}$}",
      journal = {\aj},
         year = 2020,
        month = aug,
       volume = {160},
       number = {2},
          eid = {71},
        pages = {71},
          doi = {10.3847/1538-3881/ab9d88},
archivePrefix = {arXiv},
       eprint = {2006.08615},
 primaryClass = {astro-ph.CO},
       adsurl = {https://ui.adsabs.harvard.edu/abs/2020AJ....160...71S}
}

@ARTICLE{Akins2025_jwst,
       author = {{Akins}, Hollis B. and {Casey}, Caitlin M. and {Champagne}, Jaclyn B. and {Cooper}, Olivia and {Franco}, Maximilien and {Fujimoto}, Seiji and {Knudsen}, Kirsten K. and {Koekemoer}, Anton M. and {Long}, Arianna S. and {Man}, Allison and {Manning}, Sinclaire M. and {McKinney}, Jed and {Zavala}, Jorge and {Arrabal Haro}, Pablo and {Dickinson}, Mark and {Kokorev}, Vasily and {Taylor}, Anthony J.},
        title = "{JWST+ALMA reveal the ISM kinematics and stellar structure of MAMBO-9, a merging pair of DSFGs in an overdense environment at $z=5.85$}",
      journal = {arXiv e-prints},
         year = 2025,
        month = aug,
          eid = {arXiv:2508.06607},
        pages = {arXiv:2508.06607},
          doi = {10.48550/arXiv.2508.06607},
archivePrefix = {arXiv},
       eprint = {2508.06607},
 primaryClass = {astro-ph.GA},
       adsurl = {https://ui.adsabs.harvard.edu/abs/2025arXiv250806607A}
}

@ARTICLE{Fujimoto2025_cosmicgrapes,
       author = {{Fujimoto}, S. and {Ouchi}, M. and {Kohno}, K. and {Valentino}, F. and {Gim{\'e}nez-Arteaga}, C. and {Brammer}, G.~B. and {Furtak}, L.~J. and {Kohandel}, M. and {Oguri}, M. and {Pallottini}, A. and {Richard}, J. and {Zitrin}, A. and {Bauer}, F.~E. and {Boylan-Kolchin}, M. and {Dessauges-Zavadsky}, M. and {Egami}, E. and {Finkelstein}, S.~L. and {Ma}, Z. and {Smail}, I. and {Watson}, D. and {Hutchison}, T.~A. and {Rigby}, J.~R. and {Welch}, B.~D. and {Ao}, Y. and {Bradley}, L.~D. and {Caminha}, G.~B. and {Caputi}, K.~I. and {Espada}, D. and {Endsley}, R. and {Fudamoto}, Y. and {Gonz{\'a}lez-L{\'o}pez}, J. and {Hatsukade}, B. and {Koekemoer}, A.~M. and {Kokorev}, V. and {Laporte}, N. and {Lee}, M. and {Magdis}, G.~E. and {Ono}, Y. and {Rizzo}, F. and {Shibuya}, T. and {Shimasaku}, K. and {Sun}, F. and {Toft}, S. and {Umehata}, H. and {Wang}, T. and {Yajima}, H.},
        title = "{Primordial rotating disk composed of at least 15 dense star-forming clumps at cosmic dawn}",
      journal = {Nature Astronomy},
         year = 2025,
        month = aug,
          doi = {10.1038/s41550-025-02592-w},
archivePrefix = {arXiv},
       eprint = {2402.18543},
 primaryClass = {astro-ph.GA},
       adsurl = {https://ui.adsabs.harvard.edu/abs/2025NatAs.tmp..157F}
}

@ARTICLE{Rowland2024_rebels,
       author = {{Rowland}, Lucie E. and {Hodge}, Jacqueline and {Bouwens}, Rychard and {Mancera Pi{\~n}a}, Pavel E. and {Hygate}, Alexander and {Algera}, Hiddo and {Aravena}, Manuel and {Bowler}, Rebecca and {da Cunha}, Elisabete and {Dayal}, Pratika and {Ferrara}, Andrea and {Herard-Demanche}, Thomas and {Inami}, Hanae and {van Leeuwen}, Ivana and {de Looze}, Ilse and {Oesch}, Pascal and {Pallottini}, Andrea and {Phillips}, Si{\^a}n and {Rybak}, Matus and {Schouws}, Sander and {Smit}, Renske and {Sommovigo}, Laura and {Stefanon}, Mauro and {van der Werf}, Paul},
        title = "{REBELS-25: discovery of a dynamically cold disc galaxy at z = 7.31}",
      journal = {\mnras},
         year = 2024,
        month = dec,
       volume = {535},
       number = {3},
        pages = {2068-2091},
          doi = {10.1093/mnras/stae2217},
archivePrefix = {arXiv},
       eprint = {2405.06025},
 primaryClass = {astro-ph.GA},
       adsurl = {https://ui.adsabs.harvard.edu/abs/2024MNRAS.535.2068R}
}

@ARTICLE{Pastras2025_noema3d,
       author = {{Pastras}, Stavros and {Genzel}, Reinhard and {Tacconi}, Linda J. and {Schuster}, Karl and {Neri}, Roberto and {F{\"o}rster Schreiber}, Natascha M. and {Naab}, Thorsten and {Barfety}, Capucine and {Burkert}, Andreas and {Cao}, Yixian and {Chen}, Jianhang and {Combes}, Fran{\c{c}}oise and {Davies}, Ric and {Eisenhauer}, Frank and {Espejo Salcedo}, Juan M. and {Garc{\'\i}a-Burillo}, Santiago and {Herrera-Camus}, Rodrigo and {Jolly}, Jean-Baptiste and {Lee}, Lilian L. and {Lee}, Minju M. and {Liu}, Daizhong and {Lutz}, Dieter and {Nestor Shachar}, Amit and {Parlanti}, Eleonora and {Price}, Sedona H. and {Pulsoni}, Claudia and {Renzini}, Alvio and {Scaloni}, Letizia and {Shimizu}, Taro T. and {Springel}, Volker and {Sternberg}, Amiel and {Sturm}, Eckhard and {Tozzi}, Giulia and {Wuyts}, Stijn and {{\"U}bler}, Hannah},
        title = "{NOEMA$^{\rm 3D}$: A first kpc resolution study of a $z\sim1.5$ main sequence barred galaxy channeling gas into a growing bulge}",
      journal = {arXiv e-prints},
         year = 2025,
        month = may,
          eid = {arXiv:2505.07925},
        pages = {arXiv:2505.07925},
          doi = {10.48550/arXiv.2505.07925},
archivePrefix = {arXiv},
       eprint = {2505.07925},
 primaryClass = {astro-ph.GA},
       adsurl = {https://ui.adsabs.harvard.edu/abs/2025arXiv250507925P}
}

@ARTICLE{Birkin2024_KAOSS,
       author = {{Birkin}, Jack E. and {Puglisi}, A. and {Swinbank}, A.~M. and {Smail}, Ian and {An}, Fang Xia and {Chapman}, S.~C. and {Chen}, Chian-Chou and {Conselice}, C.~J. and {Dudzevi{\v{c}}i{\={u}}t{\.{e}}}, U. and {Farrah}, D. and {Gullberg}, B. and {Matsuda}, Y. and {Schinnerer}, E. and {Scott}, D. and {Wardlow}, J.~L. and {van der Werf}, P.},
        title = "{KAOSS: turbulent, but disc-like kinematics in dust-obscured star-forming galaxies at z   1.3-2.6}",
      journal = {\mnras},
         year = 2024,
        month = jun,
       volume = {531},
       number = {1},
        pages = {61-83},
          doi = {10.1093/mnras/stae1089},
archivePrefix = {arXiv},
       eprint = {2301.05720},
 primaryClass = {astro-ph.GA},
       adsurl = {https://ui.adsabs.harvard.edu/abs/2024MNRAS.531...61B}
}

@ARTICLE{Lang2020_alma_phangs,
       author = {{Lang}, Philipp and {Meidt}, Sharon E. and {Rosolowsky}, Erik and {Nofech}, Joseph and {Schinnerer}, Eva and {Leroy}, Adam K. and {Emsellem}, Eric and {Pessa}, Ismael and {Glover}, Simon C.~O. and {Groves}, Brent and {Hughes}, Annie and {Kruijssen}, J.~M. Diederik and {Querejeta}, Miguel and {Schruba}, Andreas and {Bigiel}, Frank and {Blanc}, Guillermo A. and {Chevance}, M{\'e}lanie and {Colombo}, Dario and {Faesi}, Christopher and {Henshaw}, Jonathan D. and {Herrera}, Cinthya N. and {Liu}, Daizhong and {Pety}, J{\'e}r{\^o}me and {Puschnig}, Johannes and {Saito}, Toshiki and {Sun}, Jiayi and {Usero}, Antonio},
        title = "{PHANGS CO Kinematics: Disk Orientations and Rotation Curves at 150 pc Resolution}",
      journal = {\apj},
         year = 2020,
        month = jul,
       volume = {897},
       number = {2},
          eid = {122},
        pages = {122},
          doi = {10.3847/1538-4357/ab9953},
archivePrefix = {arXiv},
       eprint = {2005.11709},
 primaryClass = {astro-ph.GA},
       adsurl = {https://ui.adsabs.harvard.edu/abs/2020ApJ...897..122L}
}

@ARTICLE{Jones2021_alpine_alma,
       author = {{Jones}, G.~C. and {Vergani}, D. and {Romano}, M. and {Ginolfi}, M. and {Fudamoto}, Y. and {B{\'e}thermin}, M. and {Fujimoto}, S. and {Lemaux}, B.~C. and {Morselli}, L. and {Capak}, P. and {Cassata}, P. and {Faisst}, A. and {Le F{\`e}vre}, O. and {Schaerer}, D. and {Silverman}, J.~D. and {Yan}, Lin and {Boquien}, M. and {Cimatti}, A. and {Dessauges-Zavadsky}, M. and {Ibar}, E. and {Maiolino}, R. and {Rizzo}, F. and {Talia}, M. and {Zamorani}, G.},
        title = "{The ALPINE-ALMA [C II] Survey: kinematic diversity and rotation in massive star-forming galaxies at z 4.4-5.9}",
      journal = {\mnras},
         year = 2021,
        month = nov,
       volume = {507},
       number = {3},
        pages = {3540-3563},
          doi = {10.1093/mnras/stab2226},
archivePrefix = {arXiv},
       eprint = {2104.03099},
 primaryClass = {astro-ph.GA},
       adsurl = {https://ui.adsabs.harvard.edu/abs/2021MNRAS.507.3540J}
}

@ARTICLE{Rizzo2021_alma,
       author = {{Rizzo}, Francesca and {Vegetti}, Simona and {Fraternali}, Filippo and {Stacey}, Hannah R. and {Powell}, Devon},
        title = "{Dynamical properties of z  4.5 dusty star-forming galaxies and their connection with local early-type galaxies}",
      journal = {\mnras},
         year = 2021,
        month = nov,
       volume = {507},
       number = {3},
        pages = {3952-3984},
          doi = {10.1093/mnras/stab2295},
archivePrefix = {arXiv},
       eprint = {2102.05671},
 primaryClass = {astro-ph.GA},
       adsurl = {https://ui.adsabs.harvard.edu/abs/2021MNRAS.507.3952R}
}

@ARTICLE{Tiley2021_kges,
       author = {{Tiley}, Alfred L. and {Gillman}, S. and {Cortese}, L. and {Swinbank}, A.~M. and {Dudzevi{\v{c}}i{\={u}}t{\.{e}}}, U. and {Harrison}, C.~M. and {Smail}, I. and {Obreschkow}, D. and {Croom}, S.~M. and {Sharples}, R.~M. and {Puglisi}, A.},
        title = "{The KMOS galaxy evolution survey (KGES): the angular momentum of star-forming galaxies over the last {\ensuremath{\approx}}10 Gyr}",
      journal = {\mnras},
         year = 2021,
        month = sep,
       volume = {506},
       number = {1},
        pages = {323-342},
          doi = {10.1093/mnras/stab1692},
archivePrefix = {arXiv},
       eprint = {2106.05511},
 primaryClass = {astro-ph.GA},
       adsurl = {https://ui.adsabs.harvard.edu/abs/2021MNRAS.506..323T}
}

@ARTICLE{Wisnioski2019_kmos3d,
       author = {{Wisnioski}, E. and {F{\"o}rster Schreiber}, N.~M. and {Fossati}, M. and {Mendel}, J.~T. and {Wilman}, D. and {Genzel}, R. and {Bender}, R. and {Wuyts}, S. and {Davies}, R.~L. and {{\"U}bler}, H. and {Bandara}, K. and {Beifiori}, A. and {Belli}, S. and {Brammer}, G. and {Chan}, J. and {Davies}, R.~I. and {Fabricius}, M. and {Galametz}, A. and {Lang}, P. and {Lutz}, D. and {Nelson}, E.~J. and {Momcheva}, I. and {Price}, S. and {Rosario}, D. and {Saglia}, R. and {Seitz}, S. and {Shimizu}, T. and {Tacconi}, L.~J. and {Tadaki}, K. and {van Dokkum}, P.~G. and {Wuyts}, E.},
        title = "{The KMOS$^{3D}$ Survey: Data Release and Final Survey Paper}",
      journal = {\apj},
         year = 2019,
        month = dec,
       volume = {886},
       number = {2},
          eid = {124},
        pages = {124},
          doi = {10.3847/1538-4357/ab4db8},
archivePrefix = {arXiv},
       eprint = {1909.11096},
 primaryClass = {astro-ph.GA},
       adsurl = {https://ui.adsabs.harvard.edu/abs/2019ApJ...886..124W}
}

@ARTICLE{Richards2016_edges,
       author = {{Richards}, Emily E. and {van Zee}, L. and {Barnes}, K.~L. and {Staudaher}, S. and {Dale}, D.~A. and {Braun}, T.~T. and {Wavle}, D.~C. and {Dalcanton}, J.~J. and {Bullock}, J.~S. and {Chandar}, R.},
        title = "{Baryonic distributions in galaxy dark matter haloes - I. New observations of neutral and ionized gas kinematics}",
      journal = {\mnras},
         year = 2016,
        month = jul,
       volume = {460},
       number = {1},
        pages = {689-728},
          doi = {10.1093/mnras/stw1016},
archivePrefix = {arXiv},
       eprint = {1605.01638},
 primaryClass = {astro-ph.GA},
       adsurl = {https://ui.adsabs.harvard.edu/abs/2016MNRAS.460..689R}
}

@ARTICLE{Cappellari2013_atlas3d,
       author = {{Cappellari}, Michele and {Scott}, Nicholas and {Alatalo}, Katherine and {Blitz}, Leo and {Bois}, Maxime and {Bournaud}, Fr{\'e}d{\'e}ric and {Bureau}, M. and {Crocker}, Alison F. and {Davies}, Roger L. and {Davis}, Timothy A. and {de Zeeuw}, P.~T. and {Duc}, Pierre-Alain and {Emsellem}, Eric and {Khochfar}, Sadegh and {Krajnovi{\'c}}, Davor and {Kuntschner}, Harald and {McDermid}, Richard M. and {Morganti}, Raffaella and {Naab}, Thorsten and {Oosterloo}, Tom and {Sarzi}, Marc and {Serra}, Paolo and {Weijmans}, Anne-Marie and {Young}, Lisa M.},
        title = "{The ATLAS$^{3D}$ project - XV. Benchmark for early-type galaxies scaling relations from 260 dynamical models: mass-to-light ratio, dark matter, Fundamental Plane and Mass Plane}",
      journal = {\mnras},
         year = 2013,
        month = jul,
       volume = {432},
       number = {3},
        pages = {1709-1741},
          doi = {10.1093/mnras/stt562},
archivePrefix = {arXiv},
       eprint = {1208.3522},
 primaryClass = {astro-ph.CO},
       adsurl = {https://ui.adsabs.harvard.edu/abs/2013MNRAS.432.1709C}
}

@ARTICLE{Martinsson2013_diskmass,
       author = {{Martinsson}, Thomas P.~K. and {Verheijen}, Marc A.~W. and {Westfall}, Kyle B. and {Bershady}, Matthew A. and {Andersen}, David R. and {Swaters}, Rob A.},
        title = "{The DiskMass Survey. VII. The distribution of luminous and dark matter in spiral galaxies}",
      journal = {\aap},
         year = 2013,
        month = sep,
       volume = {557},
          eid = {A131},
        pages = {A131},
          doi = {10.1051/0004-6361/201321390},
archivePrefix = {arXiv},
       eprint = {1308.0336},
 primaryClass = {astro-ph.CO},
       adsurl = {https://ui.adsabs.harvard.edu/abs/2013A&A...557A.131M}
}

@ARTICLE{Gomez-lopez2019_rc_herschel,
       author = {{G{\'o}mez-L{\'o}pez}, J.~A. and {Amram}, P. and {Epinat}, B. and {Boselli}, A. and {Rosado}, M. and {Marcelin}, M. and {Boissier}, S. and {Gach}, J. -L. and {S{\'a}nchez-Cruces}, M. and {Sardaneta}, M.~M.},
        title = "{An H{\ensuremath{\alpha}} kinematic survey of the Herschel Reference Survey. I. Fabry-Perot observations with the 1.93 m telescope at OHP}",
      journal = {\aap},
         year = 2019,
        month = nov,
       volume = {631},
          eid = {A71},
        pages = {A71},
          doi = {10.1051/0004-6361/201935869},
archivePrefix = {arXiv},
       eprint = {1908.10295},
 primaryClass = {astro-ph.GA},
       adsurl = {https://ui.adsabs.harvard.edu/abs/2019A&A...631A..71G}
}

@ARTICLE{Yoon2021_manga_rc,
       author = {{Yoon}, Yongmin and {Park}, Changbom and {Chung}, Haeun and {Zhang}, Kai},
        title = "{Rotation Curves of Galaxies and Their Dependence on Morphology and Stellar Mass}",
      journal = {\apj},
         year = 2021,
        month = dec,
       volume = {922},
       number = {2},
          eid = {249},
        pages = {249},
          doi = {10.3847/1538-4357/ac2302},
archivePrefix = {arXiv},
       eprint = {2110.06033},
 primaryClass = {astro-ph.GA},
       adsurl = {https://ui.adsabs.harvard.edu/abs/2021ApJ...922..249Y}
}

@ARTICLE{DiTeodoro2021,
       author = {{Di Teodoro}, Enrico M. and {Posti}, Lorenzo and {Ogle}, Patrick M. and {Fall}, S. Michael and {Jarrett}, Thomas},
        title = "{Rotation curves and scaling relations of extremely massive spiral galaxies}",
      journal = {\mnras},
         year = 2021,
        month = nov,
       volume = {507},
       number = {4},
        pages = {5820-5831},
          doi = {10.1093/mnras/stab2549},
archivePrefix = {arXiv},
       eprint = {2109.03828},
 primaryClass = {astro-ph.GA},
       adsurl = {https://ui.adsabs.harvard.edu/abs/2021MNRAS.507.5820D}
}

@ARTICLE{DiTeodoro2023,
       author = {{Di Teodoro}, Enrico M. and {Posti}, Lorenzo and {Fall}, S. Michael and {Ogle}, Patrick M. and {Jarrett}, Thomas and {Appleton}, Philip N. and {Cluver}, Michelle E. and {Haynes}, Martha P. and {Lisenfeld}, Ute},
        title = "{Dark matter halos and scaling relations of extremely massive spiral galaxies from extended H I rotation curves}",
      journal = {\mnras},
         year = 2023,
        month = feb,
       volume = {518},
       number = {4},
        pages = {6340-6354},
          doi = {10.1093/mnras/stac3424},
archivePrefix = {arXiv},
       eprint = {2207.02906},
 primaryClass = {astro-ph.GA},
       adsurl = {https://ui.adsabs.harvard.edu/abs/2023MNRAS.518.6340D}
}

@ARTICLE{Lelli2017_sparc,
       author = {{Lelli}, Federico and {McGaugh}, Stacy S. and {Schombert}, James M. and {Pawlowski}, Marcel S.},
        title = "{One Law to Rule Them All: The Radial Acceleration Relation of Galaxies}",
      journal = {\apj},
         year = 2017,
        month = feb,
       volume = {836},
       number = {2},
          eid = {152},
        pages = {152},
          doi = {10.3847/1538-4357/836/2/152},
archivePrefix = {arXiv},
       eprint = {1610.08981},
 primaryClass = {astro-ph.GA},
       adsurl = {https://ui.adsabs.harvard.edu/abs/2017ApJ...836..152L}
}

@ARTICLE{Lelli2016_sparc,
       author = {{Lelli}, Federico and {McGaugh}, Stacy S. and {Schombert}, James M.},
        title = "{SPARC: Mass Models for 175 Disk Galaxies with Spitzer Photometry and Accurate Rotation Curves}",
      journal = {\aj},
         year = 2016,
        month = dec,
       volume = {152},
       number = {6},
          eid = {157},
        pages = {157},
          doi = {10.3847/0004-6256/152/6/157},
archivePrefix = {arXiv},
       eprint = {1606.09251},
 primaryClass = {astro-ph.GA},
       adsurl = {https://ui.adsabs.harvard.edu/abs/2016AJ....152..157L}
}

@ARTICLE{Abdurrouf2022_manga,
       author = {{Abdurro'uf} and {Accetta}, Katherine and {Aerts}, Conny and {Silva Aguirre}, V{\'\i}ctor and {Ahumada}, Romina and {Ajgaonkar}, Nikhil and {Filiz Ak}, N. and {Alam}, Shadab and {Allende Prieto}, Carlos and {Almeida}, Andr{\'e}s and {Anders}, Friedrich and {Anderson}, Scott F. and {Andrews}, Brett H. and {Anguiano}, Borja and {Aquino-Ort{\'\i}z}, Erik and {Arag{\'o}n-Salamanca}, Alfonso and {Argudo-Fern{\'a}ndez}, Maria and {Ata}, Metin and {Aubert}, Marie and {Avila-Reese}, Vladimir and {Badenes}, Carles and {Barb{\'a}}, Rodolfo H. and {Barger}, Kat and {Barrera-Ballesteros}, Jorge K. and {Beaton}, Rachael L. and {Beers}, Timothy C. and {Belfiore}, Francesco and {Bender}, Chad F. and {Bernardi}, Mariangela and {Bershady}, Matthew A. and {Beutler}, Florian and {Bidin}, Christian Moni and {Bird}, Jonathan C. and {Bizyaev}, Dmitry and {Blanc}, Guillermo A. and {Blanton}, Michael R. and {Boardman}, Nicholas Fraser and {Bolton}, Adam S. and {Boquien}, M{\'e}d{\'e}ric and {Borissova}, Jura and {Bovy}, Jo and {Brandt}, W.~N. and {Brown}, Jordan and {Brownstein}, Joel R. and {Brusa}, Marcella and {Buchner}, Johannes and {Bundy}, Kevin and {Burchett}, Joseph N. and {Bureau}, Martin and {Burgasser}, Adam and {Cabang}, Tuesday K. and {Campbell}, Stephanie and {Cappellari}, Michele and {Carlberg}, Joleen K. and {Wanderley}, F{\'a}bio Carneiro and {Carrera}, Ricardo and {Cash}, Jennifer and {Chen}, Yan-Ping and {Chen}, Wei-Huai and {Cherinka}, Brian and {Chiappini}, Cristina and {Choi}, Peter Doohyun and {Chojnowski}, S. Drew and {Chung}, Haeun and {Clerc}, Nicolas and {Cohen}, Roger E. and {Comerford}, Julia M. and {Comparat}, Johan and {da Costa}, Luiz and {Covey}, Kevin and {Crane}, Jeffrey D. and {Cruz-Gonzalez}, Irene and {Culhane}, Connor and {Cunha}, Katia and {Dai}, Y. Sophia and {Damke}, Guillermo and {Darling}, Jeremy and {Davidson}, Jr., James W. and {Davies}, Roger and {Dawson}, Kyle and {De Lee}, Nathan and {Diamond-Stanic}, Aleksandar M. and {Cano-D{\'\i}az}, Mariana and {S{\'a}nchez}, Helena Dom{\'\i}nguez and {Donor}, John and {Duckworth}, Chris and {Dwelly}, Tom and {Eisenstein}, Daniel J. and {Elsworth}, Yvonne P. and {Emsellem}, Eric and {Eracleous}, Mike and {Escoffier}, Stephanie and {Fan}, Xiaohui and {Farr}, Emily and {Feng}, Shuai and {Fern{\'a}ndez-Trincado}, Jos{\'e} G. and {Feuillet}, Diane and {Filipp}, Andreas and {Fillingham}, Sean P. and {Frinchaboy}, Peter M. and {Fromenteau}, Sebastien and {Galbany}, Llu{\'\i}s and {Garc{\'\i}a}, Rafael A. and {Garc{\'\i}a-Hern{\'a}ndez}, D.~A. and {Ge}, Junqiang and {Geisler}, Doug and {Gelfand}, Joseph and {G{\'e}ron}, Tobias and {Gibson}, Benjamin J. and {Goddy}, Julian and {Godoy-Rivera}, Diego and {Grabowski}, Kathleen and {Green}, Paul J. and {Greener}, Michael and {Grier}, Catherine J. and {Griffith}, Emily and {Guo}, Hong and {Guy}, Julien and {Hadjara}, Massinissa and {Harding}, Paul and {Hasselquist}, Sten and {Hayes}, Christian R. and {Hearty}, Fred and {Hern{\'a}ndez}, Jes{\'u}s and {Hill}, Lewis and {Hogg}, David W. and {Holtzman}, Jon A. and {Horta}, Danny and {Hsieh}, Bau-Ching and {Hsu}, Chin-Hao and {Hsu}, Yun-Hsin and {Huber}, Daniel and {Huertas-Company}, Marc and {Hutchinson}, Brian and {Hwang}, Ho Seong and {Ibarra-Medel}, H{\'e}ctor J. and {Chitham}, Jacob Ider and {Ilha}, Gabriele S. and {Imig}, Julie and {Jaekle}, Will and {Jayasinghe}, Tharindu and {Ji}, Xihan and {Johnson}, Jennifer A. and {Jones}, Amy and {J{\"o}nsson}, Henrik and {Katkov}, Ivan and {Khalatyan}, Dr., Arman and {Kinemuchi}, Karen and {Kisku}, Shobhit and {Knapen}, Johan H. and {Kneib}, Jean-Paul and {Kollmeier}, Juna A. and {Kong}, Miranda and {Kounkel}, Marina and {Kreckel}, Kathryn and {Krishnarao}, Dhanesh and {Lacerna}, Ivan and {Lane}, Richard R. and {Langgin}, Rachel and {Lavender}, Ramon and {Law}, David R. and {Lazarz}, Daniel and {Leung}, Henry W. and {Leung}, Ho-Hin and {Lewis}, Hannah M. and {Li}, Cheng and {Li}, Ran and {Lian}, Jianhui and {Liang}, Fu-Heng and {Lin}, Lihwai and {Lin}, Yen-Ting and {Lin}, Sicheng and {Lintott}, Chris and {Long}, Dan and {Longa-Pe{\~n}a}, Pen{\'e}lope and {L{\'o}pez-Cob{\'a}}, Carlos and {Lu}, Shengdong and {Lundgren}, Britt F. and {Luo}, Yuanze and {Mackereth}, J. Ted and {de la Macorra}, Axel and {Mahadevan}, Suvrath and {Majewski}, Steven R. and {Manchado}, Arturo and {Mandeville}, Travis and {Maraston}, Claudia and {Margalef-Bentabol}, Berta and {Masseron}, Thomas and {Masters}, Karen L. and {Mathur}, Savita and {McDermid}, Richard M. and {Mckay}, Myles and {Merloni}, Andrea and {Merrifield}, Michael and {Meszaros}, Szabolcs and {Miglio}, Andrea and {Di Mille}, Francesco and {Minniti}, Dante and {Minsley}, Rebecca and {Monachesi}, Antonela},
        title = "{The Seventeenth Data Release of the Sloan Digital Sky Surveys: Complete Release of MaNGA, MaStar, and APOGEE-2 Data}",
      journal = {\apjs},
         year = 2022,
        month = apr,
       volume = {259},
       number = {2},
          eid = {35},
        pages = {35},
          doi = {10.3847/1538-4365/ac4414},
archivePrefix = {arXiv},
       eprint = {2112.02026},
 primaryClass = {astro-ph.GA},
       adsurl = {https://ui.adsabs.harvard.edu/abs/2022ApJS..259...35A}
}

@ARTICLE{Bundy2015_manga,
       author = {{Bundy}, Kevin and {Bershady}, Matthew A. and {Law}, David R. and {Yan}, Renbin and {Drory}, Niv and {MacDonald}, Nicholas and {Wake}, David A. and {Cherinka}, Brian and {S{\'a}nchez-Gallego}, Jos{\'e} R. and {Weijmans}, Anne-Marie and {Thomas}, Daniel and {Tremonti}, Christy and {Masters}, Karen and {Coccato}, Lodovico and {Diamond-Stanic}, Aleksandar M. and {Arag{\'o}n-Salamanca}, Alfonso and {Avila-Reese}, Vladimir and {Badenes}, Carles and {Falc{\'o}n-Barroso}, J{\'e}sus and {Belfiore}, Francesco and {Bizyaev}, Dmitry and {Blanc}, Guillermo A. and {Bland-Hawthorn}, Joss and {Blanton}, Michael R. and {Brownstein}, Joel R. and {Byler}, Nell and {Cappellari}, Michele and {Conroy}, Charlie and {Dutton}, Aaron A. and {Emsellem}, Eric and {Etherington}, James and {Frinchaboy}, Peter M. and {Fu}, Hai and {Gunn}, James E. and {Harding}, Paul and {Johnston}, Evelyn J. and {Kauffmann}, Guinevere and {Kinemuchi}, Karen and {Klaene}, Mark A. and {Knapen}, Johan H. and {Leauthaud}, Alexie and {Li}, Cheng and {Lin}, Lihwai and {Maiolino}, Roberto and {Malanushenko}, Viktor and {Malanushenko}, Elena and {Mao}, Shude and {Maraston}, Claudia and {McDermid}, Richard M. and {Merrifield}, Michael R. and {Nichol}, Robert C. and {Oravetz}, Daniel and {Pan}, Kaike and {Parejko}, John K. and {Sanchez}, Sebastian F. and {Schlegel}, David and {Simmons}, Audrey and {Steele}, Oliver and {Steinmetz}, Matthias and {Thanjavur}, Karun and {Thompson}, Benjamin A. and {Tinker}, Jeremy L. and {van den Bosch}, Remco C.~E. and {Westfall}, Kyle B. and {Wilkinson}, David and {Wright}, Shelley and {Xiao}, Ting and {Zhang}, Kai},
        title = "{Overview of the SDSS-IV MaNGA Survey: Mapping nearby Galaxies at Apache Point Observatory}",
      journal = {\apj},
         year = 2015,
        month = jan,
       volume = {798},
       number = {1},
          eid = {7},
        pages = {7},
          doi = {10.1088/0004-637X/798/1/7},
archivePrefix = {arXiv},
       eprint = {1412.1482},
 primaryClass = {astro-ph.GA},
       adsurl = {https://ui.adsabs.harvard.edu/abs/2015ApJ...798....7B}
}

@ARTICLE{ForsterSchreiber2018_sinf,
       author = {{F{\"o}rster Schreiber}, N.~M. and {Renzini}, A. and {Mancini}, C. and {Genzel}, R. and {Bouch{\'e}}, N. and {Cresci}, G. and {Hicks}, E.~K.~S. and {Lilly}, S.~J. and {Peng}, Y. and {Burkert}, A. and {Carollo}, C.~M. and {Cimatti}, A. and {Daddi}, E. and {Davies}, R.~I. and {Genel}, S. and {Kurk}, J.~D. and {Lang}, P. and {Lutz}, D. and {Mainieri}, V. and {McCracken}, H.~J. and {Mignoli}, M. and {Naab}, T. and {Oesch}, P. and {Pozzetti}, L. and {Scodeggio}, M. and {Shapiro Griffin}, K. and {Shapley}, A.~E. and {Sternberg}, A. and {Tacchella}, S. and {Tacconi}, L.~J. and {Wuyts}, S. and {Zamorani}, G.},
        title = "{The SINS/zC-SINF Survey of z {\ensuremath{\sim}} 2 Galaxy Kinematics: SINFONI Adaptive Optics-assisted Data and Kiloparsec-scale Emission-line Properties}",
      journal = {\apjs},
         year = 2018,
        month = oct,
       volume = {238},
       number = {2},
          eid = {21},
        pages = {21},
          doi = {10.3847/1538-4365/aadd49},
archivePrefix = {arXiv},
       eprint = {1802.07276},
 primaryClass = {astro-ph.GA},
       adsurl = {https://ui.adsabs.harvard.edu/abs/2018ApJS..238...21F}
}

@ARTICLE{Hunter2012AJ_little_things,
       author = {{Hunter}, Deidre A. and {Ficut-Vicas}, Dana and {Ashley}, Trisha and {Brinks}, Elias and {Cigan}, Phil and {Elmegreen}, Bruce G. and {Heesen}, Volker and {Herrmann}, Kimberly A. and {Johnson}, Megan and {Oh}, Se-Heon and {Rupen}, Michael P. and {Schruba}, Andreas and {Simpson}, Caroline E. and {Walter}, Fabian and {Westpfahl}, David J. and {Young}, Lisa M. and {Zhang}, Hong-Xin},
        title = "{Little Things}",
      journal = {\aj},
         year = 2012,
        month = nov,
       volume = {144},
       number = {5},
          eid = {134},
        pages = {134},
          doi = {10.1088/0004-6256/144/5/134},
archivePrefix = {arXiv},
       eprint = {1208.5834},
 primaryClass = {astro-ph.GA},
       adsurl = {https://ui.adsabs.harvard.edu/abs/2012AJ....144..134H}
}

@ARTICLE{Walter2008_things,
       author = {{Walter}, Fabian and {Brinks}, Elias and {de Blok}, W.~J.~G. and {Bigiel}, Frank and {Kennicutt}, Jr., Robert C. and {Thornley}, Michele D. and {Leroy}, Adam},
        title = "{THINGS: The H I Nearby Galaxy Survey}",
      journal = {\aj},
         year = 2008,
        month = dec,
       volume = {136},
       number = {6},
        pages = {2563-2647},
          doi = {10.1088/0004-6256/136/6/2563},
archivePrefix = {arXiv},
       eprint = {0810.2125},
 primaryClass = {astro-ph},
       adsurl = {https://ui.adsabs.harvard.edu/abs/2008AJ....136.2563W}
}

@ARTICLE{Blumenthal1986,
       author = {{Blumenthal}, G.~R. and {Faber}, S.~M. and {Flores}, R. and {Primack}, J.~R.},
        title = "{Contraction of Dark Matter Galactic Halos Due to Baryonic Infall}",
      journal = {\apj},
         year = 1986,
        month = feb,
       volume = {301},
        pages = {27},
          doi = {10.1086/163867},
       adsurl = {https://ui.adsabs.harvard.edu/abs/1986ApJ...301...27B}
}

@ARTICLE{Mo1998,
       author = {{Mo}, H.~J. and {Mao}, Shude and {White}, Simon D.~M.},
        title = "{The formation of galactic discs}",
      journal = {\mnras},
         year = 1998,
        month = apr,
       volume = {295},
       number = {2},
        pages = {319-336},
          doi = {10.1046/j.1365-8711.1998.01227.x},
archivePrefix = {arXiv},
       eprint = {astro-ph/9707093},
 primaryClass = {astro-ph},
       adsurl = {https://ui.adsabs.harvard.edu/abs/1998MNRAS.295..319M}
}

@ARTICLE{Governato2012,
       author = {{Governato}, F. and {Zolotov}, A. and {Pontzen}, A. and {Christensen}, C. and {Oh}, S.~H. and {Brooks}, A.~M. and {Quinn}, T. and {Shen}, S. and {Wadsley}, J.},
        title = "{Cuspy no more: how outflows affect the central dark matter and baryon distribution in {\ensuremath{\Lambda}} cold dark matter galaxies}",
      journal = {\mnras},
         year = 2012,
        month = may,
       volume = {422},
       number = {2},
        pages = {1231-1240},
          doi = {10.1111/j.1365-2966.2012.20696.x},
archivePrefix = {arXiv},
       eprint = {1202.0554},
 primaryClass = {astro-ph.CO},
       adsurl = {https://ui.adsabs.harvard.edu/abs/2012MNRAS.422.1231G}
}

@ARTICLE{Navarro1997_nfw,
       author = {{Navarro}, Julio F. and {Frenk}, Carlos S. and {White}, Simon D.~M.},
        title = "{A Universal Density Profile from Hierarchical Clustering}",
      journal = {\apj},
         year = 1997,
        month = dec,
       volume = {490},
       number = {2},
        pages = {493-508},
          doi = {10.1086/304888},
archivePrefix = {arXiv},
       eprint = {astro-ph/9611107},
 primaryClass = {astro-ph},
       adsurl = {https://ui.adsabs.harvard.edu/abs/1997ApJ...490..493N}
}

@ARTICLE{Vogelsberger2014_illustris,
       author = {{Vogelsberger}, Mark and {Genel}, Shy and {Springel}, Volker and {Torrey}, Paul and {Sijacki}, Debora and {Xu}, Dandan and {Snyder}, Greg and {Nelson}, Dylan and {Hernquist}, Lars},
        title = "{Introducing the Illustris Project: simulating the coevolution of dark and visible matter in the Universe}",
      journal = {\mnras},
         year = 2014,
        month = oct,
       volume = {444},
       number = {2},
        pages = {1518-1547},
          doi = {10.1093/mnras/stu1536},
archivePrefix = {arXiv},
       eprint = {1405.2921},
 primaryClass = {astro-ph.CO},
       adsurl = {https://ui.adsabs.harvard.edu/abs/2014MNRAS.444.1518V}
}

@ARTICLE{Sommerville2015,
       author = {{Somerville}, Rachel S. and {Dav{\'e}}, Romeel},
        title = "{Physical Models of Galaxy Formation in a Cosmological Framework}",
      journal = {\araa},
         year = 2015,
        month = aug,
       volume = {53},
        pages = {51-113},
          doi = {10.1146/annurev-astro-082812-140951},
archivePrefix = {arXiv},
       eprint = {1412.2712},
 primaryClass = {astro-ph.GA},
       adsurl = {https://ui.adsabs.harvard.edu/abs/2015ARA&A..53...51S}
}

@ARTICLE{Schaye2015_eagle,
       author = {{Schaye}, Joop and {Crain}, Robert A. and {Bower}, Richard G. and {Furlong}, Michelle and {Schaller}, Matthieu and {Theuns}, Tom and {Dalla Vecchia}, Claudio and {Frenk}, Carlos S. and {McCarthy}, I.~G. and {Helly}, John C. and {Jenkins}, Adrian and {Rosas-Guevara}, Y.~M. and {White}, Simon D.~M. and {Baes}, Maarten and {Booth}, C.~M. and {Camps}, Peter and {Navarro}, Julio F. and {Qu}, Yan and {Rahmati}, Alireza and {Sawala}, Till and {Thomas}, Peter A. and {Trayford}, James},
        title = "{The EAGLE project: simulating the evolution and assembly of galaxies and their environments}",
      journal = {\mnras},
         year = 2015,
        month = jan,
       volume = {446},
       number = {1},
        pages = {521-554},
          doi = {10.1093/mnras/stu2058},
archivePrefix = {arXiv},
       eprint = {1407.7040},
 primaryClass = {astro-ph.GA},
       adsurl = {https://ui.adsabs.harvard.edu/abs/2015MNRAS.446..521S}
}

@BOOK{Binney_and_Tremaine2008,
       author = {{Binney}, James and {Tremaine}, Scott},
        title = "{Galactic Dynamics: Second Edition}",
         year = 2008,
       adsurl = {https://ui.adsabs.harvard.edu/abs/2008gady.book.....B}
}


\bsp	
\label{lastpage}
\end{document}